\documentclass[preprint,amsmath,amssymb,aps,prd,nofootinbib]{revtex4}
\usepackage{epsfig,graphicx,color,appendix,fge}
\begin{document}
\def\CP{{\it CP}~}
\def\cp{{\it CP}}
\title{\mbox{}\\[15pt]
 QCD axion and Neutrino physics\\ induced by Hidden flavor structure}

\author{Y. H. Ahn$^{1,2}$\footnote{Email: axionahn@naver.com},
Xiaojun Bi$^{2,3}$\footnote{Email: bixj@ihep.ac.cn}
}
\affiliation{$^{1}$Department of Physics, Chung-Ang University, Seoul 06974, Korea.}
\affiliation{$^2$Key Laboratory of Particle Astrophysics, Institute of High Energy Physics,
Chinese Academy of Sciences, Beijing, 100049, China}
\affiliation{$^3$School of Physics, University of Chinese Academy of Sciences, Beijing 100049, China}



\begin{abstract}
We study the reasonable requirements of two anomalous $U(1)$s in a flavored-axion framework for the anomaly cancellations of both $U(1)$-mixed gravity and $U(1)_Y\times[U(1)]^2$ which in turn determine the $U(1)_Y$ charges where $U(1)_Y$ is the hypercharge gauge symmetry of the standard model.
We argue that, with a flavor symmetry group, axion-induced topology in symmetry-broken phases plays crucial roles in describing how quarks and leptons are organized at a fundamental level and make deep connections with each other.
A unified model, as an example, is then proposed in a simple way to describe a whole spectrum of particles where both flavored-axion interactions with normal matter and the masses and mixings of fermions emerge from the spontaneous breaking of a given symmetry group. Once a scale of active neutrino mass defined at a seesaw scale is fixed by the commensurate $U(1)$ flavored-PQ charge of fermions, that of QCD axion decay constant $F_A$ is determined. In turn, fundamental physical parameters complementary to each other are predicted with the help of precision flavor experiments.
Model predictions are extracted on the characteristics of neutrino and flavored-axion: $F_A=3.57^{\,+1.52}_{\,-1.53}\times10^{10}$ GeV (consequently, QCD axion mass $m_a=1.52^{+1.14}_{-0.46}\times10^{-4}$ eV, axion to photon coupling $|g_{a\gamma\gamma}|=2.15^{+1.61}_{-0.64}\times10^{-14}\,\text{GeV}^{-1}$, axion to electron coupling $g_{Aee}=3.29^{+2.47}_{-0.98}\times10^{-14}$, etc.);  atmospheric mixing angle $\theta_{23}$, Dirac CP phase $\delta_{CP}$, and $0\nu\beta\beta${\it-decay rate} for normal mass ordering and inverted one by taking quantum corrections into account.
\end{abstract}

\maketitle 
\section{Introduction}
 Symmetries have played a pivotal role in physics in general and particularly in quantum field theory. The standard model (SM) as a low-energy effective theory has been very predictive and well tested, due to the absence of one-loop $SU(N)$ and $U(1)$ gauge anomalies and the symmetries satisfied by the theory - Lorentz invariance plus the $G_{\rm SM}\equiv SU(3)_C\times SU(2)_L\times U(1)_Y$ gauge symmetry in addition to the discrete space-time symmetries like $P$ (parity) and $CP$ (charge-conjugation and parity). For all the success of the SM, however, it leaves many open questions for the theoretical and cosmological issues that have not been solved yet. 
 It is widely believed that the SM should be extended to a more fundamental underlying theory. 
 
If nature is stringy, string theory, the only framework we have for a consistent theory with both quantum mechanics and gravity, should give insight into all such fundamental issues. String theory when compactified to four dimensions can generically contain {\it gauged} $U(1)$ plus non-Abelian symmetries. 
Furthermore, since the gauge invariance of the SM gauge theory does not restrict the flavor structure of quark and lepton, additional (non)-Abelian symmetries might be required\,\cite{Ahn:2014gva, Ahn:2016hbn, Ahn:2018nfb, Ahn:2018cau}. Such additional symmetries may be hidden in our world, but they should manifest themselves if we look at nature at a more fundamental level.
From that perspective, with an undisputed ansatz we can start: 
\begin{eqnarray}
&\text{\it All elementary fields that make up the Universe are charged under the gauge group}\nonumber\\
 &G_0=G_{\rm SM}\times G_f\,,
 \label{ansa}
\end{eqnarray}
where $G_f$ stands for a newly introduced gauge group containing (non)-Abelian symmetries as a copy of the SM gauge system. A theory based on $G_f$ can naturally involve several, but restricted by a flavor theory, symmetry breaking fields which play a messenger role in links between elementary fields. We assume that from all the elementary fermions that form a chiral set, reflecting none of them has a gauge-invariant mass term. Once we know the quantum numbers of all elementary fields on $G_0$ it is almost the same as knowing the physics itself since there is a connection from nature to the specific numbers.
Together with gauge anomaly cancellation for the consistency of low-energy effective theory, an extended theory under the ansatz\,(\ref{ansa}) has the $CPT$ (charge-conjugation, parity, and time-reversal) invariance and general coordinate invariance.
Since chiral fermions are the main ingredient of the extended theory, the gauged- and gravitational anomalies of gauged $U(1)_{X_i}$ (with $i=1,2,...$) and {\it non-Abelian} $G_{na}$ symmetries are generically present, making the theory inconsistent. 
Such anomalies, if present, must cancel when adding the contributions, due to the various chiral fermions. Hence, finding missing pieces about symmetry and anomaly may be the key to understanding the unsolved theoretical and cosmological issues\,\cite{Ahn:2018cau, Ahn:2018nfb, Ahn:2017dpf, Ahn:2016hbn, Ahn:2016typ, Ahn:2014gva}. 

Hence, two requirements needed for the extended theory under the ansatz\,(\ref{ansa}) are there: (i) For the presence of gauged $U(1)$ symmetry, the anomaly of $U(1)_Y\times[U(1)_X]^2$ whose cancellation is not automatically guaranteed should be free, while the mixed anomalies $U(1)_{X}\times[G^{\rm SM}_{i}]^2$ could be canceled by the Green-Schwarz (GS) mechanism\,\cite{Green:1984sg} where $G^{\rm SM}_i$ is one of the factors of $G_{\rm SM}$. Once gauged $U(1)$s are introduced in an extended theory its mixed gravitational anomaly should be free\,\cite{Kim:2012wc}, or else it leads to an explicit breaking of the gauge symmetry by gravitational interactions. It may require at least two $U(1)$s to make its linear combination cancel each other in the presence of a chiral Majorana fermion as shown in Refs.\,\cite{Ahn:2018nfb, Ahn:2016hbn}, and then the global $U(1)$-mixed gravitational anomaly is also canceled by the same diagrams due to the same tracing over $U(1)$ representations. 
(ii) For gauged non-Abelian symmetry, the mixed anomalies $G_{na}\times[G^{\rm SM}_{i}]^2$ and $G^{\rm SM}_{i}\times[G_{na}]^2$ should be free, such that one of candidates for $G_{na}$ could be a new $SU(2)$ that is supposed to be spontaneously broken at a high energy limit and whose discrete subgroup can be used in describing flavor theory at low energy, where such anomalies including $G_{na}$-mixed gravitational anomaly can be free by matching the representations of particle content under $SU(2)$ to those under its discrete subgroup (with the Witten's anomaly-free condition\,\cite{Witten:1982fp}). 
We assume that the (non)-Abelian gauge symmetry in $G_f$ is spontaneously broken at some ultraviolet (UV) scale, leaving behind its global subgroup.
That is, at low energy, we have a surviving gauge theory based on $G_{\rm SM}$, and besides the theory has a {\it global} symmetry group $G_F=U(1)\times${\it non-Abelian finite symmetries} (which is a subset of the gauge group $G_f$). The charge of field content under $G_F$ is hidden at low energy, which is indirectly inferred from physical observables since the global symmetry acts on the physical observables.
Then at low energy, it eventually gives deep connections among different types of elementary fields due to the ansatz\,(\ref{ansa}), illuminating the hidden flavor structures with $N_f$ generations well defined in the non-Abelian discrete symmetry and thereby identifying the masses and mixings of the SM fermions and the cosmological dark matter as an unexpected bonus, etc.

 One goal of this paper is to investigate the reasonable requirements of two anomalous $U(1)$s for the anomaly cancellations of both $U(1)$-mixed gravity and $U(1)_Y\times[U(1)]^2$ providing unique $U(1)_Y$ charges. Such anomaly cancellations require the presence of additional fermions that make the seesaw mechanism\,\cite{Minkowski:1977sc} natural. 
 At high energy limit of the theory one can consider gauged $U(1)_X$ and $U(1)_Y$ symmetries since the global $U(1)_X$ originates from gauged one. Then there exists a GS counter term to the $U(1)_Y\times[U(1)_X]^2$ anomaly. The 4D version of GS mechanism\,\cite{Ferrara:2011dz, string_book} gives extra masses to the gauge bosons involved, which is unacceptable for hypercharge. So, in order not to induce a mass term for the $U(1)_Y$ gauge field its counter term has to vanish by assuming that the hypercharge field does not couple to a closed string sector, and therefore the $U(1)_Y\times[U(1)_X]^2$ anomaly should vanish at high energy limit.
Unlike Refs.\,\cite{Ahn:2018cau, Ahn:2018nfb} where $U(1)_Y\times[U(1)_X]^2$ anomaly cancellation is not considered such that several choices of $U(1)_X$ quantum numbers for SM fields spectra are there, in the present model the associated quantum numbers of the quark and lepton fields can be uniquely determined by the anomaly cancellations of both $U(1)_Y\times[U(1)_X]^2$ and $U(1)_X\times[gravity]^2$. This will give rise to unique textures in the Yukawa matrices, illuminating  deep connection between the SM quark and charged-lepton flavor structures. Along this line, the other goal, therefore, aims to establish model predictions on the QCD axion mass and its photon coupling, and the unsolved observables in neutrino sector like the atmospheric neutrino mixing angle and Dirac CP phase together with the mass orderings of neutrino mass spectra, through constructing an explicit model based on the ansatz\,(\ref{ansa}) with those anomaly cancellations.

Under $G_{\rm SM}\times G_F$ considering a general flavored-axion framework where flavored-PQ $U(1)$ symmetry is embedded in non-Abelian finite group\,\cite{Ahn:2014gva, Ahn:2016hbn, Ahn:2018nfb, Ahn:2018cau}. The $U(1)_X$ invariance embedded in a non-Abelian finite symmetry forbids renormalizable Yukawa interactions for the light families except for top quark and some new additional SM gauge singlet fermions, but would allow them through effective non-renormalizable Yukawa couplings suppressed by $({\cal F}/\Lambda)^n$. Here $n$ is the positive integer and ${\cal F}$ called flavon stands for an SM gauge singlet scalar field charged under $G_F$.  
Then a set of flavon fields ${\cal F}$ act on dimension-four(three) operators well-sewed by $G_F\times G_{\rm SM}$ with different orders\,\cite{Ahn:2014zja, Ahn:2014gva, Ahn:2016hbn, Ahn:2018cau, Ahn:2018nfb}
\begin{eqnarray}
 \tilde{c}_{1}\,{\cal OP}_{3}\,({\cal F})^{1}+{\cal OP}_{4}\sum^{\rm finite}_{n=0} c_{n}\,\left(\frac{{\cal F}}{\Lambda}\right)^n\,,\quad\text{with}~\frac{1}{\sqrt{10}}\lesssim|\tilde{c}_{1}|,|c_n|\lesssim\sqrt{10}
  \label{AFN}
\end{eqnarray}
where ${\cal OP}_{4(3)}$ is a dimension-$4(3)$ operator, and all the coefficients $c_{n}$ and $\tilde{c}_{1}$ are complex numbers with absolute value of order unity. What the meaning of `finite' in Eq.\,(\ref{AFN}) is that the number of operators is finite due to the entry fields charged under the $G_F\times G_{\rm SM}$, while there could be an infinite number of higher-dimensional operators proportional to the leading order terms, that yet gives tiny shifts in measurables below error bars.
After a set of flavon fields ${\cal F}$ that transform nontrivially under a non-Abelian finite symmetry get VEVs, breaking spontaneously the so-called flavored-PQ symmetry $U(1)_X$ simultaneously, leads to all Yukawa couplings being suppressed by $(\langle{\cal F}\rangle/\Lambda)^n$ and naturally triggers the seesaw mechanism\,\cite{Minkowski:1977sc} and Peccei-Quinn (PQ) mechanism\,\cite{Peccei-Quinn}. Here the scale of flavor dynamics $\Lambda$, so-called Froggatt-Nielsen (FN) scale\,\cite{Froggatt:1978nt}, above which there exists unknown physics, is the UV cutoff of the new gauge symmetry $G_{0}$-invariant effective theory. Such a fundamental scale may come from where some string moduli are stabilized\,\cite{Ahn:2016typ, Ahn:2016hbn, Ahn:2018nfb, Ahn:2018cau}. Moreover, in the flavored-axion framework of Eq.\,(\ref{AFN}), there remains no residual symmetry since the non-Abelian finite symmetry is explicitly broken by higher-order effects in a given model. Hence there is no room for a spontaneously broken discrete symmetry to give rise to a domain-wall problem, as well as it can be protected from quantum-gravitational effects.

The rest of this paper is organized as follows. In the following section, we discuss the model setup showing how flavor symmetry including the SM gauge symmetry originates and study the reasonable requirements of two anomalous $U(1)$s in a flavored-axion framework for the anomaly cancellations of both $U(1)$-mixed gravity and $U(1)_Y\times[U(1)]^2$. In Sec.\,\ref{exam} we construct an explicit model sewed by the flavor symmetry group $G_F$ together with the SM gauge symmetry by arguing that axion-induced topology in symmetry-broken phases plays crucial roles in describing how quarks and leptons are organized at a fundamental level and make deep connections with each other. Consecutively, in Sec.\,\ref{visu}, we show a whole spectrum of quarks and leptons in a new parameterization way. And in Sec.\,\ref{nu_sec} we show how neutrino sector could be tightly connected to the quark and charged-lepton sectors, exploring numerically what values of the Dirac CP phase and atmospheric mixing angle in the low-energy neutrino oscillation and the effective Majorana neutrino mass ($0\nu\beta\beta${\it-decay rate}) can be predicted depending on the neutrino mass spectra through quantum corrections. 
In Sec.\,\ref{fl-qcd}, we show predictions on the characteristics of flavored-axions.
In the final section, we summarize our conclusions. In the Appendix, a desired vacuum configuration is considered.

\section{The model setup: Anomaly and Symmetry}
\label{sec_m}
Consider a gauge theory, as an example\,\footnote{There are some other examples in Refs.\,\cite{oth_ex} realized in conventional $SO(10)$ or $SU(5)$ grand unified theory, without considering the flavor hierarchy puzzles.} for the origin of flavor symmetry including the SM gauge symmetry, in a four-stack model $G_0=U(3)\times [U(2)]^2\times U(1)$ on $D$ branes\,\cite{string_book} where four gauged $U(1)$s are generically anomalous.
Since $U(2)$ is isomorphic to $SU(2)\times U(1)$ there are two $SU(2)$ symmetries: one is responsible for the electroweak gauge symmetry at low energy limit, while the other one for the flavor structure of quarks and leptons at some high energy scale. If a $U(2)$ symmetry is broken spontaneously at a high energy limit, a flavor global symmetry $G_F$ is obtained as its subgroup when their heavy gauge bosons corresponding to $SU(2)_H$ and $U(1)_R$ decouple at energy scale $\Lambda_{SU(2)_H, U(1)_R}\equiv\Lambda$, 
so that $G_0\rightarrow SU(3)_C\times SU(2)_L\times U(1)_Y\times G_F$ where the $SU(2)_L$ responsible for the SM gauge symmetry that remains unbroken at this level and the global symmetry $U(1)_R$ in $G_F$ originates from the gauged one. The symmetry $G_{\rm SM}\times G_F$ is eventually broken spontaneously down to $G_{\rm SM}$ when a set of flavon fields develop VEVs, {\it that is} $G_{\rm SM}\times G_F\rightarrow G_{\rm SM}$, where the non-Abelian finite symmetry in $G_F$ is explicitly broken by higher-order effects in the flavored-axion framework of Eq.\,(\ref{AFN}).

{\bf a) Anomalies associated with the flavored $SU(2)\times U(1)$s:}\\
When the continuous gauge symmetry $SU(2)_H$ is spontaneously broken by a heavy scalar VEV that leaves its subgroup unbroken, the flavor global symmetry $G_F$ contains the product of a (discrete) subgroup of $SU(2)_H$, for example\,\footnote{The double tetrahedral group, $SL_2(F_3)$, is defined as the group of all 24 proper rotations in three dimensions leaving a regular tetrahedron invariant in the $SU(2)$ double covering of $SO(3)$. We do not review the group theory of $SL_2(F_3)$, which is discussed in Ref.\,\cite{sl2f3}.}, $SL_2(F_3)$:
\begin{eqnarray}
 G_F= SL_2(F_3)\times U(1)_X\times U(1)_R\,.
 \label{flsy}
\end{eqnarray} 
Since there is the continuous gauge symmetry in $G_0$ our consideration must satisfy the anomaly cancellation-conditions as mentioned in Introduction, once the $SL_2(F_3)$ is embedded in the $SU(2)_H$ the anomaly diagrams linear in this $SU(2)_H$, $SL_2(F_3)\times[G^{\rm SM}_i]^2$ and $G^{\rm SM}_i\times[SL_2(F_3)]^2$ anomalies, vanish due to $SL_2(F_3)\subset SU(2)_H$ if the particle content of a given model satisfies complete $SU(2)_H$ representations\,\footnote{For example, see Table-\ref{reps_q} and -\ref{reps_l} where the singlet, doublet, and triplet representations of $SL_2(F_3)$ correspond to those of $SU(2)_H$.}.
The group $SL_2(F_3)$ is the smallest with ${\bf 1}$-, ${\bf 2}$-, and ${\bf 3}$-dimensional representations and the multiplication rule ${\bf 2}\otimes{\bf 2}={\bf 3}\oplus{\bf 1}$, allowing for flavons that transform nontrivial singlet, doublet, and triplet, which can well describe the present SM flavor puzzle associated with the leptonic and quark mixing angles in the flavored-axion framework of Eq.\,(\ref{AFN})\,\cite{Ahn:2018cau, Ahn:2018nfb}. Phenomenologically, ${\bf 2}\oplus{\bf 1}$ and ${\bf 3}$ representation structures for quark and lepton fields are strongly supported by the Cabbibo angle and large leptonic mixing angles, respectively\,\cite{Ahn:2018cau, Ahn:2018nfb, Ahn:2013ema}.

With respect to the global $U(1)_R$ in Eq.\,(\ref{flsy}) we set a chiral set to become vectorial (non-anomalous): all elementary fields participating in a given theory could be arranged according to the $U(1)_R$ and divided into groups according to their supercharge configurations in a theory with supersymmetry, {\it e.g.} see Table-\ref{DrivingRef}, -\ref{reps_q}, and -\ref{reps_l}.

A linear combination of three $U(1)$s contributes to two gauged $U(1)$s\,\footnote{Even the gauged $U(1)$s have Landau poles, since the elementary fermions charged under the $U(1)$s are simultaneously charged under the $G_F$ flavor group $\subset G_f$, a cut-off (see Eq.\,(\ref{lamScale})) well below the Landau poles can be achieved in the flavored-axion framework of Eq.\,(\ref{AFN}) where the scale of the Landau pole is defined by the blowup of a coupling constant, preventing the poles from having observable physical consequences.}: For the anomaly cancellation of $U(1)\times[gravity]^2$ since a Majorana sterile fermion with a single $U(1)$ does not meet, its anomaly could be canceled by another $U(1)$ contribution\,\cite{Ahn:2016hbn} as mentioned in Introduction. Hence, the flavored-PQ symmetry $U(1)_X$ could be composed of two anomalous symmetries generated by the charges $X_1$ and $X_2$\,\cite{Ahn:2014gva, Ahn:2016typ, Ahn:2016hbn, Ahn:2017dpf, Ahn:2018nfb, Ahn:2018cau}:
\begin{eqnarray}
U(1)_X\equiv U(1)_{X_1}\times U(1)_{X_2}\,.
\label{fl_ax}
\end{eqnarray}
According to the well-defined Kahler potential based on type-IIB string theory in Refs.\,\cite{Ahn:2016typ, Ahn:2017dpf} where three
size moduli and one axion are stabilized with positive masses while leaving two closed-string axions massless, there will be two closed-strings $(a_{T_i})$ and two open-string axions ($A_i$) with $i=1,2$. Two linear combinations of the $a_{T_i}$ and $A_i$ fields are eaten by two massive gauge bosons associated with the gauged $U(1)_{X_i}$, and then after decoupling the two massive gauge bosons the other two axionic directions survive to low energies as flavored axions, leaving behind low energy symmetries which are {\it anomalous global} $U(1)_{X_i}$ with $i=1,2$, see the details in Ref.\,\cite{Ahn:2016hbn}.
The global $U(1)_X$ charges of the field contents of a given theory must be commensurate through quantum anomalies such as $U(1)\times[gravity]^2$, $U(1)\times[G^{\rm SM}_{i}]^2$, and $U(1)_Y\times[U(1)]^2$. The spontaneous breaking of $U(1)_X$ realizes the existence of the Nambu-Goldstone (NG) modes (flavored-axions) and provides an elegant solution to the strong CP problem\,\cite{fl_ax}. Moreover, all forms of the mass hierarchy of quarks and leptons are generated from the flavored-PQ symmetry\,\cite{Ahn:2014gva}.

Some flavons charged under $SL_2(F_3)$ are simultaneously charged under the $U(1)_X$ group.
Once a flavon charged under $SL_2(F_3)\times U(1)_X$ acquires a VEV, the flavor group $G_F$ is broken and hidden, illuminating the flavor structure of mixing patterns and mass hierarchies of quarks and leptons. Hence it is very crucial to find a vacuum configuration of flavon fields in a given theory for a desirable flavor structure under $G_F$, see example in Sec.\,\ref{vac_a}.

{\bf b) Anomalies associated with the electroweak $SU(2)_L\times U(1)_Y$:}\\
At low energy theory, the hypercharge $U(1)_Y$ is the unique anomaly-free linear combination of three $U(1)$s, which should not induce a Stuckelberg mass for $U(1)_Y$ gauge field\,\cite{Ahn:2017dpf, Ahn:2016typ}. The $U(1)_Y$ charge of the particle contents of a given theory is uniquely determined in a way that a mixed anomaly between the hypercharge and $U(1)_X$ gauge fields $U(1)_Y\times[U(1)]^2$ and the mixed anomalies $U(1)_Y\times[SU(2)_L]^2$ and $U(1)_Y\times[SU(3)_C]^2$ are canceled together with the anomaly cancellation of $U(1)\times[gravity]^2$, which will be discussed.

Considering a Lagrangian for massless flavors which contains two anomalous axial $U(1)$ symmetries generated by the charges $X_1$ and $X_2$ in Eq.\,(\ref{fl_ax}). A chiral set with respect to each of the individual $U(1)$'s could become axial and vectorial for $U(1)_{X}$, that is, $U(1)_X= U(1)_{\tilde{X}}\times U(1)_V$. Under the QCD instanton background, there remains one axial-vector symmetry $U(1)_{\tilde{X}}$, while instead of the vector symmetry $U(1)_{V}$ a physical quantity between $U(1)_{X_i}$ is used, see {\it e.g.} Eq.\,(\ref{m_para}). Hence the $U(1)_X$ becomes pure axial symmetry. This corresponds to the transformations\,\cite{Ahn:2014gva, Ahn:2016hbn}
\begin{eqnarray}
 U(1)_{\tilde{X}}:\quad\psi_f\rightarrow e^{i\gamma_5\tilde{X}_{\psi_f}\alpha/2}\psi_f\qquad\text{with}~\tilde{X}_{\psi_f}=\delta^{\rm G}_2\,X_{1\psi_f}+\delta^{\rm G}_1\,X_{2\psi_f}\,,
 \label{axial_transf}
\end{eqnarray}
where $\alpha$ is a transformation parameter and $\tilde{X}_{\psi_f}$ is the charge of the representation under the $U(1)_{\tilde{X}}$. 
Here $\delta^{\rm G}_k$ is the QCD anomaly coefficient of the $U(1)_{X_k}\times[SU(3)_C]^2$ defined as $\delta^{\rm G}_k\delta^{ab}=2\sum_{\psi_f}X_{k\psi_f}{\rm Tr}(T^aT^b)$ in the QCD instanton backgrounds where the $T^a$ are the generators of the representation of $SU(3)_C$ to which Dirac fermion  belongs with $X$-charge.
If $U(1)_{\tilde{X}}$ was spontaneously broken one would expect an NG boson (QCD axion).
The axial part of $U(1)_X$ which is a linear combination of $U(1)_{X_i}$ has a QCD anomaly $U(1)_{\tilde{X}}\times[SU(3)_C]^2$ (with its color coefficient $N_C$) towards the QCD axion direction from the phases of scalar fields since each $U(1)_{X_i}$ is broken by its corresponding scalar field attaining VEV.
Since the SM fermions are at last massive after electroweak symmetry breaking, the $U(1)_{\tilde{X}}$ could be interpreted as the QCD axion direction. 
Then the SM gauge group $G_{\rm SM}$ becomes anomalous when the background field associated with $U(1)_{\tilde{X}}$ is turned on: 
\begin{eqnarray}
U(1)_Y\times[U(1)_{\tilde{X}}]^2\,, \quad  U(1)_{\tilde{X}}\times\big\{[SU(3)_C]^2, ~[SU(2)_L]^2, ~[U(1)_Y]^2\big\}\,.
\label{six_chi}
\end{eqnarray}
In addition to these, there are two more anomalies associated with the global $U(1)_{\tilde{X}}$:
\begin{eqnarray}
U(1)_{\tilde{X}}\times[gravity]^2\,,\qquad[U(1)_{\tilde{X}}]^3\,.
\label{six_chi2}
\end{eqnarray}
Thus the $U(1)_{\tilde{X}}$ charge of the SM fermions leads in general to six different anomalies. The SM fermions are massive due to their interactions with the Higgs field that acquires a VEV and is responsible for the electroweak symmetry breaking. After the electroweak symmetry breaking the symmetry $SU(3)_C\times SU(2)_L\times U(1)_Y$ is broken down to the low-energy gauge group $SU(3)_C\times U(1)_{EM}$. 
Below the weak scale, the anomalies $[SU(2)_{L}]^2\times U(1)_{\tilde{X}}$ and $[U(1)_{Y}]^2\times U(1)_{\tilde{X}}$ merge to give the electromagnetic anomaly $[U(1)_{EM}]^2\times U(1)_{\tilde{X}}$ in the gauge boson mass eigenstate basis.
Hence, there remain five anomalies among six:   
\begin{eqnarray}
U(1)_{\tilde{X}}\times\big\{[SU(3)_C]^2, ~[U(1)_{EM}]^2\big\}\,,~ U(1)_{EM}\times[U(1)_{\tilde{X}}]^2\,,~ U(1)_{\tilde{X}}\times[gravity]^2\,,~[U(1)_{\tilde{X}}]^3\,.
\label{four_chi2}
\end{eqnarray}
In what follows we will be discussing why the mixed anomaly $U(1)_Y\times[U(1)_{\tilde{X}}]^2$, the $U(1)$-mixed gravitational anomaly, and the cubic anomaly $[U(1)_{\tilde{X}}]^3$ should vanish at low-energy theory, while the mixed $U(1)_{\tilde{X}}\times[G_i^{\rm SM}]^2$ anomalies are not necessarily required to vanish.

Firstly, there are two main reasons why the $U(1)$-mixed gravitational anomaly should vanish in that the global $U(1)_{\tilde{X}}$ originates from the gauged one as addressed in Ref.\,\cite{Ahn:2016hbn}. (i) To consistently couple gravity to matter, it should vanish. (ii) Non-perturbative quantum gravitational anomaly effects\,\cite{Kamionkowski:1992mf}, leading to a non-conservation of the corresponding current $\partial_\mu J^\mu_{\tilde{X}}\propto {\cal R}\tilde{\cal R}$ where ${\cal R}$ is the Riemann tensor and $\tilde{\cal R}$ is its dual, spoil the axion solution to the strong CP problem. Thus in order for the breaking effects of the axionic shift symmetry by gravity to disappear, the $U(1)$-mixed gravitational anomaly should vanish.
The $U(1)$-mixed gravitational anomaly $U(1)_{\tilde{X}}\times[gravity]^2$ must be equivalent to the sum of each of the individual anomaly $U(1)_{X_i}\times[gravity]^2$ ($i=1,2$) proportional to the trace of a single charge since gravity does not experience any symmetry broken phase of the SM gauge theory. The $U(1)$-mixed gravitational anomaly-free condition can be decomposed into quark and lepton part, that is, $\big\{U(1)_{\tilde{X}}|_{\rm quark}+U(1)_{\tilde{X}}|_{\rm lepton}\big\}\times[SO(4)]^2$. 
As shown in Refs.\,\cite{Ahn:2016hbn, Ahn:2018cau, Ahn:2018nfb}, the non-vanishing anomaly coefficient of the quark sector $\big\{U(1)_{X_i}\times[gravity]^2\big\}_{\rm quark}$ constrains the quantity $\sum_i^{N_f}X_{\psi_i}$ in the gravitational instanton backgrounds (with $N_f$ generations well-defined in a non-Abelian finite group, {\it e.g.}$N_f=3$ in $SL_2(F_3)$), and in turn whose quantity is congruent to the $U(1)_{X_i}\times[SU(3)_C]^2$ anomaly coefficient $\delta^{\rm G}_i$. 
The $\delta^{\rm G}_i$ with respect to each of the individual $U(1)$'s is determined in a way that the anomaly coefficient of $U(1)_{X_1}\times[graviy]^2\big|_{\rm quark}$ is equal to that of $U(1)_{X_2}\times[graviy]^2\big|_{\rm quark}$: for example, once $U(1)_{\tilde{X}}\times[graviy]^2\big|_{\rm quark}=3[aX_1\delta^{\rm G}_2+bX_2\delta^{\rm G}_1]$ with $a, b$ being integers is given, $\delta^{\rm G}_1$ and $\delta^{\rm G}_2$ are determined to be $aX_1$ and $bX_2$, respectively.
Then, the $U(1)$-mixed gravitational anomaly-free condition can be expressed in terms of $\delta^{\rm G}_i$ 
\begin{eqnarray}
 -6\,\delta^{\rm G}_1\delta^{\rm G}_2=\delta^{\rm G}_2\delta^{\rm L}_1+\delta^{\rm G}_1\delta^{\rm L}_2\,,
\label{gr_re}
\end{eqnarray}
 where $\delta^{\rm L}_k\equiv\sum_{\psi_i={\rm lepton}}X_{k\psi_i}$ for lepton fields. The left and right of the above equation come from the quark and lepton sector, respectively; the left part can be written in terms of the color anomaly coefficient as ``$-3N_C$" (see Eq.\,(\ref{cano})). Hence we refer to its anomaly-free condition under the ansatz\,(\ref{ansa}) as ``{\it flavor quantum number conservation}". Eq.\,(\ref{gr_re}) indicates that the SM quarks can be deeply connected to the SM leptons as well as new fermions which are SM gauge singlets.

Secondly, we discuss the anomaly cancellation of $U(1)_Y\times[U(1)_{\tilde{X}}]^2$.
In order to see this, one can weakly gauge the $U(1)_{X_i}$ symmetries at high-energy limit of the theory since each global $U(1)_{X_i}$ originates from the gauged one\,\cite{Ahn:2016hbn, Ahn:2016typ}. Then the anomalies between hypercharge and two $U(1)$ gauge fields appear already before $SL_2(F_3)\times U(1)_{\tilde{X}}$ symmetry breaking. The anomaly coefficients of the mixed $U(1)_Y\times[U(1)_{X_i}]^2$ and $U(1)_Y\times U(1)_{X_i}\times U(1)_{X_j}$ with $j\neq i=1,2$ are given at high-energy limit of the theory, respectively, by
\begin{eqnarray}
\delta^X_{i}=2{\rm Tr}[Q^Y_f(X_{i\psi_f})^2]\,,\qquad\delta^X_{ij}=2{\rm Tr}[Q^Y_fX_{i\psi_f}X_{j\psi_f}]\,.
\label{AY1}
\end{eqnarray}
Considering, for example, the $U(1)_Y$ and $U(1)_{X_i}$ combined covariant derivative $D_\mu\supset\partial_\mu+iYB_\mu+iXA_{\mu i}$ in a UV theory, where $B_\mu$ and $A_{\mu i}$ correspond to the $U(1)_Y$ and $U(1)_{X_i}$ gauge fields, respectively, and their gauge couplings are absorbed into the associated gauge fields. Then there exist GS counterterms to the $U(1)_Y\times[U(1)_{X_i}]^2$ anomalies, and the hypercharge field appears via a Stuckelberg mass coupling. In order not to induce a Stuckelberg mass for the $U(1)_Y$ gauge field (see, for example, the effective action of Refs.\,\cite{Ahn:2017dpf, Ahn:2016typ}), their counter terms must vanish at high-energy limit.    
Hence, assuming that the hypercharge field does not couple to the closed string sector which would induce a Stuckelberg mass for the hypercharge, it can not modify the GS anomaly cancellation mechanism. So, the $U(1)_Y\times[U(1)_{\tilde{X}}]^2$ anomaly in the spectrum after $SL_2(F_3)\times U(1)_{\tilde{X}}$ symmetry breaking should be removed by the same diagrams as before $SL_2(F_3)\times U(1)_{\tilde{X}}$ symmetry breaking:
in other words, the $U(1)_Y\times [U(1)_{\tilde{X}}]^2$ anomaly has to vanish, whose condition 
is equivalently written as its anomaly coefficient 
\begin{eqnarray}
A_Y\equiv2\sum_{\psi_f}(\tilde{X}_{\psi_f})^2Q^Y_f=2\sum_{i,j}\alpha_{ij}{\rm Tr}[X_{i\psi_f}X_{j\psi_f}Q^Y_f]=0\,,
\label{AY}
\end{eqnarray}
where $\alpha_{ij}$ can be expressed in terms of the QCD anomaly coefficient as $\alpha_{11}=(\delta^{\rm G}_2)^2$, $\alpha_{12}=2\delta^G_1\delta^{\rm G}_2$, and $\alpha_{22}=(\delta^{\rm G}_1)^2$. Here $Q^{\rm EM}_f=I^f_3+Q^Y_f$ (the electric charge $Q^{\rm EM}_f$ of a $f$-particle is related to its isospin $I^f_3$ and hypercharge $Q^Y_f$).
In effect, such anomaly-free condition severely constrains the electromagnetic anomaly (see Eq.\,(\ref{ema})) with its anomaly coefficient
\begin{eqnarray}
E=2\sum_{\psi_f}\tilde{X}_{\psi_f}(Q^{\rm EM}_f)^2\,.
\label{Ae}
\end{eqnarray}
The anomaly-free condition of $U(1)_Y\times [U(1)_{\tilde{X}}]^2$ is associated with only the SM quarks and charged-leptons, eventually illuminating a deep connection between the SM quark and charged-lepton flavor structures.

Finally, we comment on that there may be no reason for the cubic anomaly $[U(1)_{\tilde{X}}]^3$ calculated for the global symmetry $U(1)_{\tilde{X}}$ itself to vanish at low energy limit. Since the global $U(1)_{\tilde{X}}$ originates from a UV gauge symmetry, its cubic anomaly should vanish at high energy limit. Hence we assume that, by weakly gauging the two $U(1)_{X_i}$ their cubic anomalies (consequently the cubic $[U(1)_{\tilde{X}}]^3$ anomaly) should be cancelled by the GS mechanism\,\cite{Green:1984sg} at high energy limit (see also its related Ref.\,\cite{Ahn:2016typ}), and after decoupling their gauge bosons and their related fields there remains the cubic anomaly $[U(1)_{\tilde{X}}]^3$ calculated for the global symmetry $U(1)_{\tilde{X}}$ itself at low energy limit.

Considering two anomalous $U(1)$s where the linear combination of two $U(1)$s becomes a pure axial symmetry as in Eq.\,(\ref{axial_transf}) while instead of the vectorial part of the other combination a physical quantity between $U(1)_{X_i}$ is obtained. Then the scalar (flavon) fields ${\cal F}_1$ and ${\cal F}_2$ charged under $U(1)_{X_1}$ and $U(1)_{X_2}$ have VEVs $\langle{\cal F}_1\rangle$ and $\langle{\cal F}_2\rangle$, respectively, and 
their ratios to the QCD axion direction and to the $U(1)_{X_i}$ flat direction could be expressed, respectively, as\,\cite{Ahn:2018cau, Ahn:2018nfb}
\begin{eqnarray}
\nabla_{{\cal F}_1}=\Big|\frac{X_2\,\delta^{\rm G}_1}{X_1\,\delta^{\rm G}_2}\Big|\nabla_{{\cal F}_2}\,,\qquad\nabla_{{\cal F}'_i}=\text{const.}\times\nabla_{{\cal F}_i}\,,
\label{m_para}
\end{eqnarray}
with $\nabla_{{\cal F}_i}\equiv\langle{\cal F}_i\rangle/\Lambda$, meaning that the set of parameters expressed as the VEVs of flavons charged under $U(1)_{X_i}$ are linearly dependent.
Then, all the SM fermions including new fermions are connected with each other through the relations Eqs.\,(\ref{gr_re}) and (\ref{m_para}) in the flavored-axion framework of Eq.\,(\ref{AFN}). 
If the particle content of low energy theory is constrained under the flavor symmetry group $G_F$ together with the SM gauge symmetry $G_{\rm SM}$ in a flavored-axion framework, according to the above discussions the anomalies of both $U(1)_Y\times[U(1)]^2$ and $U(1)\times[gravity]^2$ should be canceled by enlarging particle content with another $U(1)$ contribution.

Using both the {\it updated} Weinberg value from Particle Data Group (PDG)\,\cite{PDG} and the bound from the ADMX (Axion Dark Matter eXperiment) experiment\,\cite{ADMX}, respectively,
  \begin{eqnarray}
 z\equiv\frac{m^{\overline{\rm MS}}_u(2{\rm GeV})}{m^{\overline{\rm MS}}_d(2{\rm GeV})}=0.47^{+0.06}_{-0.07}\,,\qquad (g_{a\gamma\gamma}/m_a)^2\leq1.44\times10^{-19}\,{\rm GeV}^{-2}{\rm eV}^{-2}\,,
 \label{axipra}
 \end{eqnarray}
 where $g_{a\gamma\gamma}$ and $m_a$ stand for QCD axion-photon coupling and QCD axion mass, respectively, and $z$ is the value in the $\overline{\rm MS}$ scheme at a scale $\approx2$ GeV, a crucial  criteria for modeling can be extracted in terms of the ratio of electromagnetic anomaly coefficient $E$ to color one $N_C$ (see Eq.\,(\ref{gagg}))
   \begin{eqnarray}
 0<E/N_C<4\,.
 \label{b_a1}
 \end{eqnarray}
  Eq.\,(\ref{b_a1}) directly indicates that some models giving $E=0$ would be excluded, see Sec.\,\ref{cpvf} and also footnote\,7.

\section{The model as an example}
\label{exam}
We set up an explicit model with Eq.\,(\ref{AFN}) sewed by $G_F$ in Eq.\,(\ref{flsy}) and which generically contains several, but restricted by the flavor model, scalar fields under the ansatz\,(\ref{ansa}).
Modeling based on $G_F\times G_{\rm SM}$ considers the followings:
 {\bf(a)} the flavor symmetry $G_F$ is spontaneously broken at energies much above the electroweak scale, which requires a desired vacuum configuration to depict the hierarchy of  quark and charged-lepton masses and mixings, as well as to implement a flavored seesaw\,\footnote{ Here the flavored-seesaw is referred to as the seesaw in the flavored-axion framework of Eq.\,(\ref{AFN}) that makes several associated parameters compact.} mechanism for the active neutrino masses and mixings,
 {\bf(b)} in the flavored-axion framework of Eq.\,(\ref{AFN}),  the flavored-PQ symmetry $U(1)_X$ embedded in $SL_2(F_3)$ provides flavored-axion interactions with quarks and leptons as well as a neat solution to the strong CP problem and its resulting QCD axion as a dark matter candidate, and
{\bf(c)} both $U(1)$-mixed gravitational and $U(1)_Y\times[U(1)]^2$ anomalies are canceled by Eqs.\,(\ref{gr_re}) and (\ref{AY}).

\subsection{Vacuum configuration}
\label{vac_a}
We need to account for the vacuum configuration of flavon fields, which are responsible for the spontaneous breaking of a new flavor symmetry $G_F$ in Eq.\,(\ref{flsy}).
After the flavor symmetry breaking, there appear associated dimensionless parameters $\langle {\cal F}\rangle/\Lambda$ in Eq.\,(\ref{AFN}) in the effective Yukawa structures of the low-energy theory. 
Then the observed hierarchy of fermion masses and mixing angles (in the Cabbibo-Kobayashi-Maskawa (CKM) for quark mixings and Pontecorvo-Maki-Nakagawa-Sakata (PMNS) for leptonic mixings) should be explained by the effective Yukawa structures where the VEVs of flavon fields having $U(1)_X$ charges make a connection in each Yukawa structure through Eq.\,(\ref{m_para}). The details on flavon and driving fields configurations, which are overlapped with Refs.\,\cite{Ahn:2018cau, Ahn:2018nfb},  and the vacuum configuration of flavon fields are presented in Appendix\,\ref{vacu_0}.

 The global minima of the potential of Eq.\,(\ref{super_d}) reads from Eqs.\,(\ref{vevdirection1}, \ref{vevdirection2}, \ref{vevdirection3}) at leading order
{\small\begin{eqnarray}
 ~\langle\Phi_T\rangle=\frac{v_{T}}{\sqrt{2}}(1,0,0)\,,\qquad\quad\langle\Phi_S\rangle=\frac{v_{S}}{\sqrt{2}}(1,1,1)\,,\qquad\quad\langle\eta\rangle=\frac{v_{\eta}}{\sqrt{2}}(1,0)\,,\nonumber\\
 \langle\Psi\rangle=\langle\tilde{\Psi}\rangle=\frac{v_\Psi}{\sqrt{2}}\,, \qquad\qquad\quad\langle\Theta\rangle=\frac{v_\Theta}{\sqrt{2}}\,,\qquad\qquad\qquad \langle\tilde{\Theta}\rangle=0\,,
  \label{vev}
\end{eqnarray}}
where $\langle\eta\rangle\sim(+1,0)$ has been taken. The VEVs of Eq.\,(\ref{vev}) can be corrected by higher-dimensional operators, see Eqs.\,(\ref{Npo4}, \ref{Npo5}) in Appendix\,\ref{vacu_l}. Since the $\Phi_S, \Theta, \Psi(\tilde{\Psi})$ fields are charged under $U(1)_X$ we set the decomposition of complex scalar fields\,\cite{Ahn:2014gva, Ahn:2016hbn, Ahn:2018nfb, Ahn:2018cau} as follows
 \begin{eqnarray}
  &\Phi_{Si}=\frac{e^{i\frac{\phi_{S}}{v_{S}}}}{\sqrt{2}}\left(v_{S}+h_{S}\right)\,,\qquad\qquad\quad~\Theta=\frac{e^{i\frac{\phi_{\theta}}{v_{\Theta}}}}{\sqrt{2}}\left(v_{\Theta}+h_{\Theta}\right)\,,\nonumber\\
&\Psi=\frac{v_{\Psi}}{\sqrt{2}}e^{i\frac{\phi_{\Psi}}{v_{g}}}\left(1+\frac{h_{\Psi}}{v_{g}}\right)\,,\qquad\qquad\,\tilde{\Psi}=\frac{v_{\tilde{\Psi}}}{\sqrt{2}}e^{-i\frac{\phi_{\Psi}}{v_{g}}}\left(1+\frac{h_{\tilde{\Psi}}}{v_{g}}\right)\,,
  \label{NGboson}
 \end{eqnarray}
in which we have set $\Phi_{S1}=\Phi_{S2}=\Phi_{S3}\equiv\Phi_{Si}$ and $h_{\Psi}=h_{\tilde{\Psi}}$ in the SUSY limit, and $v_{g}=\sqrt{v^2_{\Psi}+v^2_{\tilde{\Psi}}}$. And the NG modes $A_1$ and $A_2$ are expressed as
 \begin{eqnarray}
  A_1=\frac{v_{S}\,\phi_{S}+v_{\Theta}\,\phi_{\theta}}{v_{\cal F}}\,,\qquad A_{2}=\phi_{\Psi}
  \label{NGboson1}
 \end{eqnarray}
with the angular fields $\phi_{S}$, $\phi_{\theta}$ and $\phi_{\Psi}$, where $v_{\cal F}=v_\Theta\sqrt{1+\kappa^2}$. All the VEVs of Eq.\,(\ref{vev}) breaking $G_F$ are correlated with one another and they can be determined together with FN scale in a simple way to describe flavor structures which will be seen in the following.
\subsection{Yukawa superpotentials and Anomaly coefficients}
Now let us impose $SL_2(F_3)\times U(1)_X$ quantum numbers to the SM quarks and leptons including SM gauge singlet Majorana neutrinos with $U(1)_R=+1$, summarized in Table-\ref{reps_q} and -\ref{reps_l}\,\footnote{All fields appearing in Table-\ref{reps_q} and -\ref{reps_l} are left-handed particles/antiparticles.}, in a way that their quantum numbers satisfy both conditions of Eqs\,(\ref{gr_re}) and (\ref{AY}), which are different from those of Refs.\,\cite{Ahn:2018cau, Ahn:2018nfb}, as well as their masses and mixings are well described.

\begin{table}[h]
\caption{\label{reps_q} Representations of the quark fields under $G_{\rm SM}\times SL_2(F_3)\times U(1)_{X}$ with $U(1)_R=+1$ and $U(1)_X\equiv U(1)_{X_1}\times U(1)_{X_2}$. In $({\cal Q}_1, {\cal Q}_2)_ Y$ of $G_{\rm SM}$, ${\cal Q}_1$ and ${\cal Q}_2$ are the representations under $SU(3)_C$ and $SU(2)_L$ respectively, and the script $Y$ denotes the $U(1)$ hypercharge.}
\begin{ruledtabular}
\begin{tabular}{cccccccc}
Field &$Q_{1}$& $Q_{2}$& $Q_{3}$&${\cal D}^c$ & $b^c$&${\cal U}^c$ & $t^c$\\
\hline
$G_{\rm SM}$&$(3,2)_{1/6}$&$(3,2)_{1/6}$&$(3,2)_{1/6}$&$(3,1)_{1/3}$&$(3,1)_{1/3}$&$(3,1)_{-2/3}$&$(3,1)_{-2/3}$\\
$SL_2(F_3)$&$\mathbf{1}$ &$\mathbf{1}'$ & $\mathbf{1^{\prime\prime}}$ & $\mathbf{2}'$&$\mathbf{1}'$&$\mathbf{2}'$&$\mathbf{1^{\prime}}$\\
$U(1)_{X}$&$-3X_1-2X_2$&$-2X_1-3X_2$&$0$ &$2X_1-2X_2$&$-2X_2$&$2X_1+10X_2$&$0$\\
\end{tabular}
\end{ruledtabular}
\end{table}
Under $SL_2(F_3)\times U(1)_X$, we assign the left-handed quark $SU(2)_{L}$ doublets denoted as $Q_{1}$, $Q_{2}$ and $Q_{3}$ to the $({\bf 1}, -3X_1-2X_2)$, $({\bf 1}', -2X_1-3X_2)$ and $({\bf 1}'', 0)$, respectively, while the right-handed up-type quark $SU(2)_{L}$ singlets are assigned as ${\cal U}^{c}=\{u^{c}, c^{c}\}$ and $t^{c}$ to the $({\bf 2}', 2X_1+10X_2)$ and $({\bf 1}', 0)$, respectively, and the right-handed down-type quarks ${\cal D}^{c}=\{d^{c}, s^{c}\}$ and $b^c$ to the $({\bf 2}', 2X_1-2X_2)$ and $({\bf 1}', -2X_2)$, respectively. Note that the first and second generation of right-handed quarks are assigned as $SL_2(F_3)$ doublet fields so as to give a relatively large Cabbibo angle $\lambda\equiv\sin\theta_C\approx0.225$.
Then the $U(1)_{X_i}\times[SU(3)_C]^2$ anomaly coefficient reads
 \begin{eqnarray}
  \delta^{\rm G}_i=2\Big(X_{iQ_1}+X_{iQ_2}+X_{iQ_3}+X_{iD^c}+X_{iU^c}\Big)+X_{ib^c}+X_{it^c}\,,
 \label{dGi}
 \end{eqnarray}
leading to 
 \begin{eqnarray}
 \delta^{\rm G}_1=-2X_1\,,\qquad\delta^{\rm G}_2=4X_2\,.
 \label{dGi1}
  \end{eqnarray}
 Under $G_{\rm SM}\times SL_2(F_3)\times U(1)_{X}$ with $U(1)_R=+1$, quark fields are sewed by the four representations ${\bf 1}$, ${\bf 1}'$, ${\bf 1}''$, and ${\bf 2}'$ of $SL_2(F_3)$ whose singlets and doublet match up with the singlet and doublet of $SU(2)_H$ representations, respectively, as in Table-\ref{reps_q}. The Yukawa superpotential for up-type and down-type quark sectors invariant under $G_{\rm SM}\times G_F$ is sewed through Eq.\,(\ref{AFN}), respectively, as
{\begin{eqnarray}
 W^u_q &=&
  \hat{y}_{t}\,t^cQ_{3}H_u
  +y_{c}\,(\eta {\cal U}^c)_{{\bf 1}''}Q_2\frac{H_u}{\Lambda}+\tilde{y}_{c}\,[(\eta {\cal U}^c)_{{\bf 3}}\Phi_T]_{{\bf 1}''}Q_2\frac{H_u}{\Lambda^2}\nonumber\\
  &+&\bar{y}_{u}\,[(\eta {\cal U}^c)_{\bf 3}\eta\eta]_{\bf 1}Q_1\frac{H_u}{\Lambda^3}+y_{u}\,[(\eta {\cal U}^c)_{\bf 3}\Phi_S]_{\bf 1}Q_1\frac{H_u}{\Lambda^2}\nonumber\\
  &+&{\small\text{a term replaced by}\,( y_u\rightarrow\tilde{y}_u~\text{with}~\Phi_S\rightarrow\Phi_T)}\,,\label{lagrangian_qu}\\
 W^d_q &=& y_{b}\,b^cQ_{3}H_d
  +y_{s}\,(\eta {\cal D}^c)_{{\bf 1}''}Q_2\frac{H_d}{\Lambda}+Y_{s}\,b^cQ_2(\Phi_S\Phi_S)_{{\bf 1}'}\frac{H_d}{\Lambda^2}\nonumber\\
    &+& Y_{d}\,b^cQ_1(\Phi_S\Phi_S)_{{\bf 1}''}\frac{H_d}{\Lambda^2}+y_{d}\,[(\eta {\cal D}^c)_{\bf 3}\Phi_S]_{{\bf 1}}Q_{1}\frac{H_d}{\Lambda^2}\nonumber\\
    &+&{\small\text{three terms replaced by} \,( Y_s\rightarrow\tilde{Y}_s;Y_d\rightarrow\tilde{Y}_d;\,y_d\rightarrow\tilde{y}_d~\text{with}~\Phi_S\rightarrow\Phi_T)}\,.
 \label{lagrangian_qd}
 \end{eqnarray}}
In the above superpotentials the top quark operator is only renormalizable, leading to the top quark mass as the pole mass. 
The other higher-dimensional operators driven by $\Phi_T$ and $\eta$ fields can be absorbed into the leading terms and redefined. According to the quantum numbers of the matter field contents as in Table-\ref{reps_q}, the Yukawa couplings of quarks are visualized as function of the SM gauge-singlet flavon fields $\Psi(\tilde{\Psi})$ and/or $\Theta(\Phi_S)$, except for the top Yukawa coupling, see the details in Sec.\,\ref{visu}.

Below the $U(1)_X$ group breaking scale, the effective interaction of QCD axion with the gluon field strength tensor $G^a_{\mu\nu}$ (here $a$ is an $SU(3)$-adjoint index) and its dual $\tilde{G}^a_{\mu\nu}$ are expressed via the chiral rotation of Eq.\,(\ref{axial_transf}) as\,\cite{Ahn:2014gva, Ahn:2016hbn, Ahn:2018nfb, Ahn:2018cau}
\begin{eqnarray}
{\cal L}_{Ag}=\Big(\frac{A_1}{F_{a_1}}+\frac{A_2}{F_{a_2}}\Big)\frac{g^2_s}{32\pi^2}\,G^{\mu\nu a}\tilde{G}^a_{\mu\nu}\equiv\frac{A}{F_A}\frac{g^2_s}{32\pi^2}\,G^{\mu\nu a}\tilde{G}^a_{\mu\nu}\,,
\label{agc}
\end{eqnarray}
where $F_{a_i}=f_{a_i}/\delta^{\rm G}_i$ ($i=1,2$), $F_A$ is the QCD axion decay constant, $A$ is the QCD axion field, and $g_s$ stands for the gauge coupling constant of $SU(3)_C$. Here the color anomaly coefficient of $U(1)_{\tilde{X}}\times[SU(3)_C]^2$ defined as $N_C\equiv2{\rm Tr}[\tilde{X}_{\psi_f}T^2_{SU(3)_C}]=2\delta^{\rm G}_1\delta^{\rm G}_2$ reads from Eq.\,(\ref{dGi1}) as
\begin{eqnarray}
N_C=-16X_1X_2\,.
\label{cano}
\end{eqnarray}
Note that $U(n)$ generators ($n\geq2$) are normalized according to ${\rm Tr}[T^aT^b]=\delta^{ab}/2$.
And the QCD axion decay constant is given by 
\begin{eqnarray}
F_A=\frac{f_A}{N_C}=\frac{f_{a_i}}{\sqrt{2}\,\delta^{\rm G}_i}\qquad\text{with}~i=1,2\,,
\label{adeca}
\end{eqnarray}
where $f_A=\sqrt{2}\,\delta^{\rm G}_2f_{a_1}=\sqrt{2}\,\delta^{\rm G}_1f_{a_2}$ with $f_{a_1}=|X_1|v_{\cal F}$ and $f_{a_2}=|X_2|v_g$ is used.
The $U(1)_{X_i}$ is broken down to its discrete subgroup $Z_{N_i}$ in the backgrounds of QCD instanton, and the quantities $N_i$ (nonzero integers) associated
to the axionic domain-wall are given by\,\cite{Ahn:2018nfb, Ahn:2018cau}
\begin{eqnarray}
N_1=\Big|\frac{\delta^{\rm G}_1}{X_1/2}\Big|\,,\qquad N_2=\Big|\frac{\delta^{\rm G}_2}{X_2}\Big|\,.
\end{eqnarray}
Taking $X_1=1$ and $X_2=1$ gives $N_1=2|\delta^{\rm G}_1|=4$ and $N_2=|\delta^{\rm G}_2|=4$. There exists $Z_4$ discrete symmetry and therefore four domain-walls (that are not topologically stable) at the QCD phase transition, meaning that the scale of $G_F$ breakdown should occur during (or before) cosmic inflation\,\footnote{Two massless axions coupled to $G^{\mu\nu a}\tilde{G}^a_{\mu\nu}$ arise when $U(1)_{X_i}$ global symmetries ($i=1,2$) are spontaneously broken, which have variations on cosmological distances. Quantum fluctuations during inflation are imprinted into two massless axions, which can lead to observable signatures in cosmic microwave background radiation temperature fluctuation as the form of density perturbation when the quantum fluctuations are crossing back inside the Hubble radius long after inflation has been completed.}, see Eq.\,(\ref{k_bound}). 

Next we turn to the lepton sector fields charged under $G_{\rm SM}\times SL_2(F_3)\times U(1)_X$ with $U(1)_R=+1$. Taking into account the active neutrino mass is generated by the seesaw mechanism\,\cite{Minkowski:1977sc} and the large leptonic mixing pattern\,\cite{PDG} is generated by the given $G_F$ symmetry. Under $SL_2(F_3)\times U(1)_X$, we assign the left-handed charged lepton $SU(2)_{L}$ doublets denoted as $L$  to the $({\bf 3},  X_{1L}+X_{2L})$, while the right-handed charged leptons denoted as $e^c,\,\mu^c$ and $\tau^c$, the electron flavor to the $({\bf 1}, X_{1e^c}+X_{2e^c})$, the muon flavor to the $({\bf 1}'', X_{1\mu^c}+X_{2\mu^c})$, and the tau flavor to the $({\bf 1}', X_{1\tau^c}+X_{2\tau^c})$. 
Then the requirement of the $U(1)_Y\times[U(1)_{\tilde{X}}]^2$ anomaly-free condition (\ref{AY}) reads 
 \begin{eqnarray}
  \delta^X_1+(\delta^{\rm G}_1/\delta^{\rm G}_2)^2\delta^X_2+2(\delta^{\rm G}_1/\delta^{\rm G}_2)\delta^X_{12}=0\,,
   \label{uyxf}
 \end{eqnarray}
where
 \begin{eqnarray}
  \delta^X_1&=&2\Big\{(X_{1Q_1})^2+(X_{1Q_2})^2+2(X_{1D^c})^2+(X_{1b^c})^2\nonumber\\
  &&~-4(X_{1U^c})^2-3(X_{1L})^2+(X_{1e^c})^2+(X_{1\mu^c})^2+(X_{1\tau^c})^2\Big\}\,,\nonumber
     \end{eqnarray}
   \begin{eqnarray}
    \delta^X_2&=&2\Big\{(X_{2Q_1})^2+(X_{2Q_2})^2+2(X_{2D^c})^2+(X_{2b^c})^2\nonumber\\
  &&~-4(X_{2U^c})^2-3(X_{2L})^2+(X_{2e^c})^2+(X_{2\mu^c})^2+(X_{2\tau^c})^2\Big\}\,,\nonumber\\
    \delta^X_{12}&=&2\Big\{X_{1Q_1}X_{2Q_1}+X_{1Q_2}X_{2Q_2}+2X_{1D^c}X_{2D^c}+X_{1b^c}X_{2b^c}\nonumber\\
  &&~-4X_{1U^c}X_{2U^c}-3X_{1L}X_{2L}+X_{1e^c}X_{2e^c}+X_{1\mu^c}X_{2\mu^c}+X_{1\tau^c}X_{2\tau^c}\Big\}\,.\nonumber
 \label{}
 \end{eqnarray}
Meanwhile, in order to satisfy the anomaly-free condition (\ref{gr_re}) of $U(1)_X\times[gravity]^2$ we introduce sterile neutrinos which would interact with particles through the gravitational force -- a force that acts on mass alone -- and their masses are heavy and/or light (see Refs.\,\cite{Ahn:2016hbn, Ahn:2018cau, Ahn:2018nfb}). The heavy Majorana right-handed neutrino mass is responsible for the seesaw mechanism\,\cite{Minkowski:1977sc}. Under $SL_2(F_3)\times U(1)_X$ we assign the right-handed heavy neutrinos $SU(2)_L$ singlets denoted as $N^c$ to the $({\bf 3}, -X_1/2)$ while additional right-handed light neutrino $SU(2)_L$ singlet denoted as $S^c$ to the $({\bf 1}, X_{2S^c})$, which guarantees the absence of Yukawa terms $S^cN^c{\cal F}$ and $LS^cH_u({\cal F}/\Lambda)^n$ ($n=0,1,2,..$), such as $ L\eta\eta S^c H_u/\Lambda^2$, due to $X_{1S^c}=0$.
Then the $U(1)_{\tilde{X}}\times[gravity]^2$ anomaly-free condition reads 
  \begin{eqnarray}
  \frac{N_C}{2}\Big(6+\frac{\delta^{\rm L}_1}{\delta^{\rm G}_1}+\frac{\delta^{\rm L}_2}{\delta^{\rm G}_2}\Big)=0\,,
 \label{grf}
 \end{eqnarray}
where the quantities $\delta^{\rm L}_i$ associated with leptons 
  \begin{eqnarray}
  \delta^{\rm L}_i=6(X_{iL})+X_{ie^c}+X_{i\mu^c}+X_{i\tau^c}+3X_{iN^c}+X_{iS^c}\quad(i=1,2)\,.\nonumber
 \label{}
 \end{eqnarray}

Together with the given quantum number of quarks in Table-\ref{reps_q}, taking into account the anomaly cancellations of both $U(1)_Y\times[U(1)_X]^2$ and $U(1)_X\times[gravity]^2$ of Eqs.\,(\ref{uyxf}, \ref{grf}) and the observed mass hierarchies of charged leptons, we find a $U(1)_X$ quantum number of lepton fields $L$: $-X_1/2+6X_2$, $e^c$: $X_1/2-20X_2$, $\mu^c$: $-X_1/2-10X_2$, $\tau^c$: $X_1/2-7X_2$, and $S^c$: $-43X_2$, whose quantum number is uniquely determined differently from those of Refs.\,\cite{Ahn:2018cau, Ahn:2018nfb}. And the quantum number of lepton fields is summarized in Table-\ref{reps_l}.
\begin{table}[h]
\caption{\label{reps_l} Representations of the lepton fields under $G_{\rm SM}\times SL_2(F_3)\times U(1)_{X}$ with $U(1)_R=+1$ and $U(1)_X\equiv U(1)_{X_1}\times U(1)_{X_2}$. In $({\cal Q}_1, {\cal Q}_2)_ Y$ of $G_{\rm SM}$, ${\cal Q}_1$ and ${\cal Q}_2$ are the representations under $SU(3)_C$ and $SU(2)_L$ respectively, and the script $Y$ denotes the $U(1)$ hypercharge.}
\begin{ruledtabular}
\begin{tabular}{ccccccc}
Field &$L$&$e^c$&$\mu^c$ & $\tau^c$&$N^c$&$S^c$\\
\hline
$G_{\rm SM}$&$(1,2)_{-1/2}$&$(1,1)_1$&$(1,1)_1$&$(1,1)_1$&$(1,1)_0$&$(1,1)_0$\\
$SL_2(F_3)$&$\mathbf{3}$&$\mathbf{1}$ & $\mathbf{1^{\prime\prime}}$ & $\mathbf{1^\prime}$&$\mathbf{3}$&$\mathbf{1}$\\
$U(1)_{X}$& $ \frac{1}{2}X_1-7X_2$ & $-\frac{1}{2}X_1-6X_2$ & $-\frac{1}{2}X_1+12X_2$ & $-\frac{1}{2}X_1+6X_2$& $-\frac{1}{2}X_1$&$6X_2$
\end{tabular}
\end{ruledtabular}
\end{table}
Under $G_{\rm SM}\times SL_2(F_3)\times U(1)_{X}$ with $U(1)_R=+1$, lepton fields are sewed by the four representations ${\bf 1}$, ${\bf 1}'$, ${\bf 1}''$, and ${\bf 3}$ of $SL_2(F_3)$ whose singlets and triplet match up with the singlet and triplet of $SU(2)_H$ adjoint representations, respectively, as in Table-\ref{reps_l}. As discussed in Refs.\,\cite{Ahn:2018nfb, Ahn:2016hbn, Ahn:2014gva}, with the condition of $U(1)_X$-$[gravity]^2$ anomaly cancellation new additional Majorana fermion $S^{c}$ besides the heavy Majorana neutrinos can be introduced in the lepton sector. 
In compliance with Table-\ref{reps_l} via Eq.\,(\ref{AFN}), the Yukawa superpotential for charged-lepton and neutrino sectors invariant under $G_{\rm SM}\times G_F$ reads at leading order
\begin{eqnarray}
 W_{\ell} &=&
y_\tau\,\tau^c (L\Phi_T)_{{\bf 1}''} \frac{H_d}{\Lambda}+y_\mu\,\mu^c (L\Phi_T)_{{\bf 1}'} \frac{H_d}{\Lambda}+y_e\,e^c (L\Phi_T)_{{\bf 1}} \frac{H_d}{\Lambda}\,,\label{lagrangian_chL}\\
 W_{\nu} &=& y_{\nu}(LN^c)_{{\bf 1}}H_{u}+\frac{1}{2}(\hat{y}_\Theta\Theta+\hat{y}_{\tilde{\Theta}}\tilde{\Theta})(N^{c}N^{c})_{{\bf 1}}+\frac{\hat{y}_R}{2}(N^{c}N^{c})_{{\bf 3}} \Phi_S\nonumber\\
 &+&\frac{1}{2}y_{s}\,S^c S^c\,\Phi_T\frac{\Phi_T}{\Lambda}\nonumber\\
 &+&\tilde{y}_{\nu}(LN^c\Phi_T)_{{\bf 1}}\frac{H_{u}}{\Lambda}+\frac{1}{2}\frac{\hat{y}_{r}}{\Lambda}\,(N^{c}N^{c}\Phi_S\,\Phi_T)_{{\bf 1}}+\frac{1}{2}\frac{\hat{y}_{\chi}}{\Lambda}(N^{c}N^{c})_{{\bf 3}} \Phi_T+...
 \label{lagrangian2}
 \end{eqnarray}
In the above charged-lepton Yukawa superpotential $W_{\ell}$, it has three independent Yukawa terms at the leading: apart from the Yukawa couplings, each term involves flavon field $\Phi_T$ and does not allow the interchange between $\Phi_T$ and $\Phi_S$. The next leading order contributions such as $\ell^cLH_d\,\eta\eta/\Lambda^2$ ($\ell=e, \mu, \tau$) to the charged-lepton mass matrix are negligible due to the VEV alignment Eq.\,(\ref{vev}). 

In the neutrino Yukawa superpotential $W_{\nu}$, there are two renormalizable terms for the right-handed neutrino $N^c$, which implement the seesaw mechanism by making the VEV $\langle\Theta\rangle$ $(=\langle\Phi_S\rangle/\kappa)$ large. In Eq.\,(\ref{lagrangian2}) dots represent higher-order terms triggered by the fields $\Phi_{T}$ and/or $\eta$, and especially the contribution induced by the $\eta$ field such as  $(L\,N^c\eta\eta)_{{\bf 1}}H_u/\Lambda^2$ is neglected due to the VEV alignment (\ref{vev}).
As addressed in above Eq.\,(\ref{grf}), the $SL_2(F_3)\times U(1)_X$ quantum number of Table-\ref{reps_l} guarantees the absence of the Yukawa terms $S^cN^c{\cal F}$ and $LS^cH_u({\cal F}/\Lambda)^n$ ($n=0,1,2,...$) while the latter is allowed in Ref.\,\cite{Ahn:2018nfb}.
Consequently, one Dirac neutrino mass term associated with $N^c$ is generated, while two Majorana mass terms associated with $N^c$ and $S^c$ are generated.
Besides, the flavored PQ symmetry $U(1)_{X}$  forbids the renormalizable Yukawa term $N^{c}N^{c}\Phi_{T}$ and does not allow the interchange between $\Phi_{T}$ and $\Phi_{S}$ both of which transform differently under $U(1)_{X}$. 
Due to the higher-dimensional operators, there remain no residual symmetries\,\footnote{Clearly, the flavor symmetry $SL_2(F_3)$ is broken down to its subgroup: at the leading order, the charged-lepton mass terms and the Dirac neutrino mass terms containing $N^cL$ are invariant under the subgroup $G_T$, while the Majorana mass terms containing $N^cN^c$ are invariant under the subgroup $G_S$.}.  
 The details on how the active neutrino mass and mixing are predicted under this model will be presented in Sec.\,\ref{nu_sec}.

Below the $U(1)_X$ group breaking scale (equivalent to the seesaw scale) the effective interactions of QCD axion with the electroweak and hypercharge gauge bosons and with the photon are expressed through the chiral rotation of Eq.\,(\ref{axial_transf}), respectively, as
\begin{eqnarray}
{\cal L}^{WY}_{A}&=&\frac{A}{f_A}\frac{1}{32\pi^2}\Big\{g^2_W\,N_W\,W^{\mu\nu}\tilde{W}_{\mu\nu}+g^2_Y\,N_Y\,Y^{\mu\nu}\tilde{Y}_{\mu\nu}\Big\}\,,\label{}\\
{\cal L}^{\gamma}_{A}&=&\frac{A}{f_A}\frac{e^2}{32\pi^2}\,E\,F^{\mu\nu}\tilde{F}_{\mu\nu}\,,
\label{ema}
\end{eqnarray}
where $g_{W}$, $g_Y$, and $e$ stand for the gauge coupling constant of $SU(2)_L$, $U(1)_Y$, and $U(1)_{EM}$, respectively; their corresponding gauge field strengths $W^{\mu\nu}, Y^{\mu\nu}$, and $F^{\mu\nu}$ with their dual forms $\tilde{W}_{\mu\nu}, \tilde{Y}_{\mu\nu}$, and $\tilde{F}_{\mu\nu}$, respectively. Here $N_W\equiv2{\rm Tr}[\tilde{X}_{\psi_f}T^2_{SU(2)}]$ and $N_Y\equiv2{\rm Tr}[\tilde{X}_{\psi_f}(Q_f^Y)^2]$ are the anomaly coefficients of $U(1)_{\tilde{X}}\times[SU(2)_L]^2$ and $U(1)_{\tilde{X}}\times[U(1)_Y]^2$, respectively. And the electromagnetic anomaly coefficient $E$ of $U(1)_{\tilde{X}}\times[U(1)_{EM}]^2$ defined in Eq.\,(\ref{Ae}) is explicitly expressed as 
  \begin{eqnarray}
  E&=&N_W+N_Y\nonumber\\
  &=&\frac{10}{3}(\tilde{X}_{Q_1}+\tilde{X}_{Q_2})+\frac{16}{3}\tilde{X}_{U^c}+\frac{4}{3}\tilde{X}_{D^c}+\frac{2}{3}\tilde{X}_{b^c}+6\tilde{X}_{L}+2(\tilde{X}_{e^c}+\tilde{X}_{\mu^c}+\tilde{X}_{\tau^c})\,.
 \label{eano}
 \end{eqnarray}
Recalling that $\tilde{X}_{\psi_f}=\delta^{\rm G}_2\,X_{1\psi_f}+\delta^{\rm G}_1\,X_{2\psi_f}$ in Eq.\,(\ref{axial_transf}), Eq\,(\ref{eano}) leads to $E=-128X_1X_2/3$.
 
With the given field contents in Table-\ref{reps_q} and -\ref{reps_l} satisfying Eqs.\,(\ref{gr_re}, \ref{AY}, \ref{b_a1}) a unique prediction\,\footnote{Another possible case is there: in lepton sector $L$: $X_1/2+6X_2$, $e^c$: $-X_1/2-20X_2$, $\mu^c$: $-X_1/2-13X_2$, $\tau^c$: $-X_1/2-4X_2$, and $S^c$: $-35X_2$, and in quark sector $Q_1$: $-2X_1-6X_2$, $Q_2$: $-X_1-7X_2$, $Q_3$: $0$, ${\cal U}^c$: $X_1+13X_2$, $t^c$: $0$, ${\cal D}^c$: $X_1+2X_2$, and $b^c$: $-X_1+2X_2$ . This case clearly gives $E/N_C=0$, which is consistent with the prediction of the Kim-Shifman-Vainshtein-Zakharov (KSVZ) axion models\,\,\cite{KSVZ}. According to Eq.\,(\ref{axipra}), however, this case would be excluded.} is derived from Eqs.\,(\ref{cano}, \ref{eano}) as 
 \begin{eqnarray}
  \frac{E}{N_C}&=&\frac{8}{3}\,,
 \label{cas}
 \end{eqnarray}
which is consistent with the predictions of the simplest Dine-Fischler-Srednicki-Zhitnitsky (DFSZ) axion models\,\cite{DFSZ}, making it different from those in Refs.\,\cite{Ahn:2018cau, Ahn:2018nfb}.
This predictive value in the flavored axion model is very crucial to determine the QCD axion to photon coupling of Eq.\,(\ref{gagg}), if QCD axion mass is fixed via the masses of active neutrinos with the help of hints from astro-particle constraints, {\it e.g.} Eq.\,(\ref{emg}) and Refs.\,\cite{Ahn:2018nfb, Ahn:2014gva}.

\section{Quark and Charged-Lepton: Masses and mixings}
\label{visu}
Let us discuss a realization of quark and charged-lepton masses and mixings in which their physical mass hierarchies are directly responsible for the assignment of $U(1)_X$ quantum numbers, satisfying the anomaly cancellations of both $U(1)_Y\times[U(1)_X]^2$ and $U(1)_X\times[gravity]^2$. Since the Yukawa interactions of Eq.\,(\ref{AFN}) realized under $G_F$ are responsible for the fermion masses they must be related in a very simple way at a scale much higher than the electroweak scale via Eq.\,(\ref{m_para}).

The relevant quark and charged-lepton interactions with their chiral fermions are given by 
 \begin{eqnarray}
  -{\cal L}_q &=&
  \overline{q^{u}_{R}}\,\mathcal{M}_{u}\,q^{u}_{L}+\overline{q^{d}_{R}}\,\mathcal{M}_{d}\,q^{d}_{L}+\overline{\ell_{R}}\,{\cal M}_{\ell}\,\ell_{L}
   +\frac{g}{\sqrt{2}}W^+_\mu\overline{q^u_L}\gamma^\mu\,q^d_L+\text{h.c.}\,,
  \label{AxionLag2}
 \end{eqnarray}
where $g$ is the $SU(2)_L$ coupling constant, $q^{u}=(u,c,t)$, $q^{d}=(d,s,b)$, and $\ell=(e, \mu, \tau)$.

\subsection{Quark}
With the VEV directions in Eq.\,(\ref{vev}) the mass matrices ${\cal M}_{d}$ and ${\cal M}_{u}$ for down- and up-type quarks\,\footnote{The quark sector has VEV corrections by Eqs.\,(\ref{Npo4}, \ref{Npo5}) where the new VEV of $\langle\Phi_S\rangle$ gives corrections to the mass matrices but below the few-percent level and all other corrections by the new VEVs are absorbed into the leading order terms and redefined.} in the above Lagrangian\,(\ref{AxionLag2}) are described in terms of $\nabla_Q\equiv v_Q/(\sqrt{2}\,\Lambda)$ with $Q=\eta, S, T, \Theta, \Psi, \tilde{\Psi}$, respectively:
 \begin{eqnarray}
  &{\cal M}_{d}=\tilde{R}{\left(\begin{array}{ccc}
 (iy_d\nabla_S+\tilde{y}_{d}\nabla_T)\nabla_\eta &  0 &  0 \\
 \frac{1-i}{2}y_{d}\nabla_\eta\nabla_S  &  y_{s}\nabla_\eta &  0   \\
 (3Y_{d}\nabla_T+\tilde{Y}_d\nabla_T)\nabla_S  &  (3Y_{s}\nabla_S+\tilde{Y}_s\nabla_T)\nabla_S  & y_{b}
 \end{array}\right)}\tilde{L}\,v_{d}\,.
 \label{Ch2}\\
 &{\cal M}_{u}= {\left(\begin{array}{ccc}
 (iy_{u}\nabla_S+i\tilde{y}_u\nabla_T-\bar{y}_u\nabla^2_\eta)\nabla_\eta\,e^{i(\frac{A_1}{v_{\cal F}}-8\frac{A_{2}}{v_{g}})} &  0 &  0 \\
 \frac{1-i}{2}y_{u}\nabla_S\nabla_\eta\,e^{i(\frac{A_1}{v_{\cal F}}-8\frac{A_{2}}{v_{g}})} &  (y_{c}+\frac{1-i}{2}\,\tilde{y}_c\nabla_T)\nabla_\eta\,e^{-7i\frac{A_{2}}{v_{g}}} &  0   \\
 0 &  0  &  \hat{y}_{t}
 \end{array}\right)}v_{u}\,,  \label{Ch1}
 \end{eqnarray}
Here $v_{d}\equiv\langle H_{d}\rangle=v\cos\beta/\sqrt{2}$, $v_{u}\equiv\langle H_{u}\rangle =v\sin\beta/\sqrt{2}$ with $v\simeq246$ GeV and 
 \begin{eqnarray}
 \tilde{C}={\rm Diag.}\big(
 e^{i(C_{1d}\frac{A_1}{v_{\cal F}}+C_{2d}\frac{A_{2}}{v_{g}})} , e^{i(C_{1s}\frac{A_1}{v_{\cal F}}+C_{2s}\frac{A_{2}}{v_{g}})},
e^{i(C_{1b}\frac{A_1}{v_{\cal F}}+C_{2b}\frac{A_{2}}{v_{g}})}\big)\quad\text{with}~C={\cal L}, {\cal R}\,,
 \label{c_num}
 \end{eqnarray}
where (${\cal R}_{1d}={\cal R}_{1s}=0$, ${\cal R}_{2d}={\cal R}_{2s}=5$, ${\cal R}_{1b}=2$, ${\cal R}_{2b}=5$) and (${\cal L}_{1d}=1$, ${\cal L}_{2d}=-1$, ${\cal L}_{1s}={\cal L}_{2s}=0$, ${\cal L}_{1b}=-2$, ${\cal L}_{2b}=-3$).
The $X$-charge of quarks\,\footnote{Here the $X$-charge $X_q$ is defined in the form of axion to the SM matter field interactions $-{\cal L}^{aq\ell}=\sum_{k=1,2}\Big(\frac{A_k}{f_{a_k}}\sum_{q=u,d,...}X_{q_k}m_q\bar{q}i\gamma_5 q+\frac{A_k}{f_{a_k}}\sum_{\ell=e,\mu,..} X_{\ell_k} m_\ell\bar{\ell}i\gamma_5 \ell$\Big).} corresponds to $X_{1u}=-X_1$, $X_{1c}=0$, $X_{1t}=0$, $X_{1d}=-X_1$, $X_{1s}=0$, $X_{1b}=0$, $X_{2u}=8X_2$, $X_{2c}=7X_2$, $X_{2t}=0$, $X_{2d}=-4X_2$, $X_{2s}=-5X_2$, $X_{2b}=-2X_2$. 
Plugging the VEVs of $U(1)_X$ charged flavons in Eq.\,(\ref{vev}) into Eq.\,(\ref{m_para}) 
 one obtains the flavor structure relation
\begin{eqnarray}
 \nabla_\Theta&=&\frac{1}{\kappa}\nabla_S=\Big|\frac{X_2\delta^{\rm G}_1}{X_1\delta^{\rm G}_2}\Big|\sqrt{\frac{2}{1+\kappa^2}}\nabla_\Psi\,.
  \label{expan_1}
\end{eqnarray}
The Yukawa couplings in the above mass matrices are simply expressed through Eq.\,(\ref{expan_1}): for up-type quarks
  \begin{eqnarray}
 &&y_{u}=\hat{y}_{u}\,\nabla^8_{\Psi}\,,\quad y_{c}=\hat{y}_{c}\,\nabla^7_{\Psi}\,, \quad
 \tilde{y}_{c}=\hat{y}_{\tilde{c}}\,\nabla^7_{\Psi}\,,\quad\tilde{y}_{u}=\hat{y}_{\tilde{u}}\nabla_\Theta\,\nabla^8_{\Psi}\,,\quad\bar{y}_{u}=\hat{y}_{\bar{u}}\nabla_\Theta\,\nabla^8_{\Psi}\,,
 \label{Top1}
 \end{eqnarray}
and for down-type quarks
 \begin{eqnarray}
 && y_{d}= \hat{y}_{d}\,\nabla^4_{\tilde{\Psi}}\,,\qquad\quad~~
 y_{s}=\hat{y}_{s}\,\nabla^5_{\tilde{\Psi}}\,,\qquad\quad~\, Y_{s}=\hat{Y}_{s}\,\nabla^5_{\Psi}\,,\qquad\quad~~ y_{b}=\hat{y}_{b}\,\nabla^2_{\tilde{\Psi}}\,,\nonumber\\
 &&Y_{d}=\hat{Y}_{d}\nabla_\Theta\,\nabla^4_{\Psi}\,,\qquad\tilde{Y}_{d}=\hat{Y}_{\tilde{d}}\nabla^2_\Theta\,\nabla^4_{\Psi}\,,\qquad\tilde{y}_{d}=\hat{y}_{\tilde{d}}\nabla_\Theta\,\nabla^4_{\Psi}\,,\qquad\tilde{Y}_{s}=\hat{Y}_{\tilde{s}}\nabla_\Theta\,\nabla^5_{\Psi}\,,
 \label{Dow1}
 \end{eqnarray}
 where $\hat{Y}_{\tilde{d}}=\hat{Y}_{d1}+3\kappa^2\hat{Y}_{d2}$
and $1/\sqrt{10}\lesssim|\hat{y}|, |\hat{Y}_s|,  |\hat{Y}_{\tilde{s}}|,  |\hat{Y}_{d}|, |\hat{Y}_{d_i}|\lesssim\sqrt{10}$.
The quark mass matrices ${\cal M}_{u}$ in Eq.\,(\ref{Ch1}) and ${\cal M}_{d}$ in Eq.\,(\ref{Ch2}) generate the up- and down-type quark masses:
 \begin{eqnarray}
 \widehat{\mathcal{M}}_{u}=V^{u\dag}_{R}\,{\cal M}_{u}\,V^u_{L}
 ={\rm diag}(m_{u},m_{c},m_{t})\,,\quad \widehat{\mathcal{M}}_{d}=V^{d\dag}_{R}\,{\cal M}_{d}\,V^{d}_{L}
 ={\rm diag}(m_{d},m_{s},m_{b})\,.
 \label{Quark21}
 \end{eqnarray}
Then $V^{f}_{L}$ and $V^{f}_{R}$ can be determined by diagonalizing the matrices for ${\cal M}^{\dag}_{f}{\cal M}_{f}$ and ${\cal M}_{f}{\cal M}^{\dag}_{f}$ ($f=u,d$)\,\cite{Ahn:2011yj}, respectively, indicated by Eqs.\,(\ref{Ch1}) and (\ref{Ch2}). Then the CKM mixing matrix $V_{\rm CKM}\equiv V^u_LV^{d\dag}_L$ is generated via the charged quark-current term in Eq.\,(\ref{AxionLag2}). The physical structure of the up- and down-type quark Lagrangian should match up with the empirical results calculated from the Particle Data Group (PDG)\,\cite{PDG} and calculated from the CKMfitter\,\cite{ckm}.
The details on the quark masses and mixings are presented in Sec.\,\ref{quark_l}.

As expected in the ansatz\,(\ref{ansa}) and explicitly shown in Eqs.\,(\ref{qum}, \ref{qckm}, \ref{qm}, \ref{cpDirac0}), all parameters in the quark sector are correlated with one another. Hence it is very crucial for excavating values of the new expansion parameters in Eq.\,(\ref{expan_1}) to reproduce the empirical results of the CKM mixing angles in Eq.\,(\ref{ckmmixing}) and quark masses in PDG. Moreover, such parameters are also tightly correlated with those in the lepton sector, so finding the values of the parameters are crucial to producing the empirical results of the charged leptons and the light active neutrino masses (see below Eq.\,(\ref{ChL1}) and below Eq.\,(\ref{MR1})).
\subsection{Charged-lepton}
In the charged-lepton superpotential\,(\ref{lagrangian_chL}), the interchange between $\Phi_T$ and $\Phi_S$ is not allowed, and hence it is expected that charged-lepton mass matrix does not have off-diagonal entries, and in turn, the flavored-axion mixing with charged-leptons is negligible at leading order\,\footnote{The vacuum corrections of Eqs.\,(\ref{Npo4}, \ref{Npo5}) to the superpotential\,(\ref{lagrangian_chL}) including the next leading order terms $\ell^cLH_d\,\eta\eta/\Lambda^2$ with $\ell=e, \mu, \tau$ do not make off-diagonal entries and they are absorbed into the leading order terms and redefined.}.
With the help of the vacuum alignment (\ref{vev}) the charged-lepton mass matrix in the Lagrangian (\ref{AxionLag3}) is written as
 \begin{eqnarray}
 {\cal M}_{\ell}&=& {\left(\begin{array}{ccc}
 y_{e}\,e^{13i\frac{A_{2}}{v_{g}}} & 0 &  0 \\
 0 & y_{\mu}\,e^{-5i\frac{A_{2}}{v_{g}}} & 0 \\
 0 & 0 & y_{\tau}\,e^{i\frac{A_{2}}{v_{g}}}
 \end{array}\right)}v_{d}\,,
 \label{ChL1}
 \end{eqnarray}
where $y_{e}=\hat{y}_{e}\,\nabla^{13}_{\Psi}$, $y_{\mu}=\hat{y}_{\mu}\,\nabla^{5}_{\Psi}$, and $y_{\tau}=\hat{y}_{\tau}\,\nabla_{\Psi}$. The $X$-charge of the charged leptons reads $X_{1e}=X_{1\mu}=X_{1\tau}=0$, $X_{2e}=-13X_2$, $X_{2\mu}=5X_2$, and $X_{2\tau}=-X_2$. 
Their corresponding masses after electroweak symmetry breaking are given by
 \begin{eqnarray}
 m_{e}&=&\hat{y}_{e}\,\nabla^{13}_{\Psi}\nabla_T\,v_d\,,\qquad m_{\mu}=\hat{y}_{\mu}\,\nabla^{5}_{\Psi}\nabla_T\,v_d\,,\qquad
 m_{\tau}=\hat{y}_{\tau}\,\nabla_{\Psi}\nabla_T\,v_d\,.
 \label{cLep1}
 \end{eqnarray}
Here the hat Yukawa couplings of charged-lepton sector can be fixed by the numerical values of $\nabla_T$, $\nabla_\Psi$, and $\tan\beta$ obtained from the quark sector, and vice versa. 

\subsection{Numerical analysis for quark and charged-lepton masses and CKM mixings}
\label{num1}
All elementary fields are charged under the ansatz\,(\ref{ansa}) and they have deep connections among different types of elementary fields each other. On this account, their corresponding parameters, $\kappa$, $\tan\beta$, $\nabla_T$, $\nabla_\eta$, $\nabla_S$, $\nabla_\Psi$, $|\hat{y}_i|$, and $\arg(\hat{y}_i)$, are connected each other as already shown in Eqs.\,(\ref{qum}, \ref{qckm}, \ref{qm}, \ref{cLep1}) (and Eqs.\,(\ref{MR1}, \ref{Ynu1}) for neutrino sector) analytically. 
The Yukawa matrices in Eqs.\,(\ref{Ch2}, \ref{Ch1}, \ref{ChL1}) of SM charged-fermions are taken at the scale of $G_F$ symmetry breakdown. Hence their masses are subject to quantum corrections. Then the Yukawa matrices are run down to $m_t$ and diagonalized. But, we assume that the Yukawa matrices renormalized at the scale of $G_F$ breakdown are equal to those at scale $m_t$ since it is expected that there is no sizable one-loop RG running effect on the observables for the hierarchical mass spectra. The low-energy Yukawa couplings for experimental values are extracted from the physical masses and the mixing angles compiled by the PDG\,\cite{PDG} and the CKMfitter\,\cite{ckm}, respectively.

We perform a numerical simulation\,\footnote{In numerical calculation, we only consider the mass matrices in Eqs.\,(\ref{Ch2}, \ref{Ch1}, \ref{ChL1}), which are leading order terms, since it is expected that the corrections to the VEVs due to higher-dimensional operators are below the few-percent level. And once RG running effect is considered, the input values can be tuned.} using the linear algebra tools of Ref.\,\cite{Antusch:2005gp}. With the inputs 
\begin{eqnarray}
\tan\beta=8.530\,, \qquad\kappa=0.460\,,
  \label{tanpara}
\end{eqnarray}
and $|\hat{y}_d|=0.512$ ($\phi_d=1.647$ rad), $|\hat{y}_s|=0.453$ ($\phi_s=0$), $|\hat{y}_b|=0.763$ ($\phi_b=0$), $|\hat{y}_u|=0.330$ ($\phi_u=2.480$ rad), $|\hat{y}_{\tilde{u}}|=1.000$ ($\phi_{\tilde{u}}=6.120$ rad), $|\hat{y}_{\bar{u}}|=0.660$ ($\phi_{\bar{u}}=3.020$ rad), $|\hat{y}_c|=2.480$ ($\phi_c=0$), $|\hat{y}_{\tilde{c}}|=1.760$ ($\phi_{\tilde{c}}=1.000$ rad), $|\hat{y}_t|=1.002$ ($\phi_t=0$), $|\hat{Y}_{d}|=0.540$ ($\phi_{Y_{d}}=5.160$ rad),  $|\hat{Y}_{\tilde{d}}|=0.680$ ($\phi_{Y_{\tilde{d}}}=1.220$ rad), $|\hat{Y}_{s}|=2.900$ ($\phi_{Y_{s}}=3.340$ rad), $|\hat{Y}_{\tilde{s}}|=2.500$ ($\phi_{Y_{\tilde{s}}}=2.760$ rad), leading to 
\begin{eqnarray}
\nabla_\Psi=0.520\,, \quad\nabla_S=0.154\,, \quad\nabla_T=0.140\,, \quad \nabla_\eta=0.270\,,
 \label{quarkvalue}
 \end{eqnarray} 
we obtain the mixing angles and Dirac CP phase $\theta^q_{12}=13.054^{\circ}$, $\theta^q_{23}=2.331^{\circ}$, $\theta^q_{13}=0.222^{\circ}$, $\delta^q_{CP}=65.551^{\circ}$ compatible with the $3\sigma$ Global fit of CKMfitter\,\cite{ckm}, see Eq.\,(\ref{ckmmixing}); the quark masses $m_d=4.661$ MeV, $m_s=96.706$ MeV, $m_b=4.182$ GeV, $m_u=2.164$ MeV, $m_c=1.271$ GeV, and $m_t=173.1$ GeV compatible with the values in PDG\,\cite{PDG}. Note that as seen in Eq.\,(\ref{cpDirac}) and footnote-12, the Dirac CP phase $\delta^q_{CP}$ analytically depends on the phases ($\phi_s, \phi_{Y_d},  \phi_{Y_{\tilde{d}}},  \phi_{Y_s}, \phi_{Y_{\tilde{s}}}$, $\phi_d$) and ($\phi_c$, $\phi_{\tilde{c}}$, $\phi_u$) from down- and up-type quark sector, respectively, at leading order. And also, $\phi_d$ is fixed by both $m_d$ and $\theta^q_{12}$; ($\phi_u$, $\phi_{\tilde{u}}$, $\phi_{\bar{u}}$), ($\phi_{Y_d},  \phi_{Y_{\tilde{d}}}$), and ($ \phi_{Y_s}, \phi_{Y_{\tilde{s}}}$) are important to fix $m_u$, $\theta^q_{13}$, and $\theta^q_{23}$, respectively;  without loss of generality, we have set $\phi_s=0$ for the above numerical calculation.

For the charged-leptons Eq.\,(\ref{cLep1}), together with the above numerical results of quark sector, if we set $\hat{y}_e=1.128$, $\hat{y}_\mu=0.981$, and $\hat{y}_\tau=1.205$, the results reading $m_e=0.511$ MeV, $m_\mu=105.683$ MeV, $m_{\tau}=1776.86$ MeV lie in the empirical one in PDG\,\cite{PDG}.

\section{Neutrino}
\label{nu_sec}
The recent analysis based on global fits\,\cite{Esteban:2018azc, deSalas:2017kay, Capozzi:2018ubv} of the neutrino oscillations enters into a new phase of precise determination of mixing angles and mass squared differences, while they still show large uncertainties on the atmospheric mixing angle and Dirac-CP phase.
The seesaw mechanism does naturally work well in a framework based on {\it non-Abelian}$\times${\it flavored}-$U(1)$ symmetries. 
Then the flavored-seesaw under the ansatz\,(\ref{ansa}) that makes compact several associated parameters can give predictions on the atmospheric mixing angle and Dirac-CP phase.

The flavor symmetry $U(1)_{X}\times SL_2(F_{3})$ is spontaneously broken when the scalar fields $\Phi_{S}, \Theta, \tilde{\Theta},\Psi$ and $\tilde{\Psi}$ get VEVs, and all neutrinos obtain masses at energies below the electroweak scale, similar to the quarks and charged-leptons.
The relevant neutrino interactions with chiral fermions $\psi=\nu, S, N$ are given by 
 \begin{eqnarray}
  -{\cal L}_{\nu} &=&
 \frac{1}{2} \begin{pmatrix} \overline{\nu^c_L} & \overline{N_R} & \overline{S_R} \end{pmatrix} \begin{pmatrix} 0 & m^T_{D} & 0  \\ m_{D} & e^{i\frac{A_{1}}{v_{\cal F}}}\,M_R & 0  \\ 0 & 0 & M_S \end{pmatrix} \begin{pmatrix} \nu_L \\ N^c_R \\ S^c_R \end{pmatrix} +\text{h.c.}\,,
  \label{AxionLag3}
 \end{eqnarray}
where $\nu=(\nu_e, \nu_\mu, \nu_\tau)$ and $N=(N_e, N_\mu, N_\tau)$. Remark that the sterile neutrinos $N^c$ and $S^c$ (which interact with only gravity) have been introduced to solve the anomaly-free condition of $U(1)\times[gravity]^2$ and especially for $N^c$ to explain the small active neutrino masses as well as to provide a theoretically well-motivated PQ symmetry breaking scale. 
In flavored PQ models\,\cite{Ahn:2014gva, Ahn:2016hbn, Ahn:2018nfb, Ahn:2018cau} where the scale of $U(1)_X$ group breakdown is congruent to that of heavy right-handed Majorana neutrino, a flavored-axion is responsible for spontaneous heavy Majorana neutrino mass generation.

In order to eliminate the NG modes $A_{1,2}$ from the neutrino Yukawa Lagrangian\,(\ref{AxionLag3}) we transform the neutrino fields by chiral rotations under $U(1)_X$:
 \begin{eqnarray}
N\rightarrow e^{-iX_1\frac{A_1}{f_{a1}}\frac{\gamma_5}{2}}\,N\,,\qquad
S\rightarrow e^{i12\,X_2\frac{A_2}{f_{a2}}\frac{\gamma_5}{2}}\,S\,,\qquad
 \nu\rightarrow e^{i(X_1\frac{A_1}{f_{a1}}-14X_2\frac{A_2}{f_{a2}})\frac{\gamma_5}{2}}\,\nu\,.
 \end{eqnarray}
In addition, since the masses of Majorana neutrino $N_R$ are much larger than those of Dirac and light right-handed Majorana ones, after integrating out the right-handed heavy Majorana neutrinos, we obtain the following effective Lagrangian for neutrinos
 \begin{eqnarray}
  -{\cal L}^{a-\nu}_{W} &\simeq&\frac{1}{2}\overline{\nu^{c}_L} {\cal M}_{\nu} \nu_L +\frac{1}{2}M_S\overline{S_R}\,S^c_R+\frac{1}{2}\overline{N_R}\,M_R\,N^c_R + \frac{g}{\sqrt{2}}W^-_\mu\overline{\ell_L}\gamma^\mu\,\nu_L+\text{h.c.}\nonumber\\
  &-&\frac{1}{2}\,\overline{\nu}\,i\! \! \not\!\partial\,\nu-\frac{1}{2}\,\overline{S}\,i\! \! \not\!\partial\,S-\frac{1}{2}\,\overline{N}\,i\! \! \not\!\partial\,N-\frac{X_{1}}{2}\,\frac{A_1}{f_{a1}}\,\overline{N}\,i\gamma_{5}M_R\,N^c \nonumber\\
  &+&\left\{\frac{X_1}{2}\frac{A_1}{f_{a1}}-7X_2\frac{A_2}{f_{a2}}\right\}\,\overline{\nu}\,i\gamma_{5}{\cal M}_\nu\,\nu +6X_2\frac{A_2}{f_{a2}}\overline{S}\,i\gamma_{5}M_S\,S^c\,,
\label{Axion_nu_La}
  \label{neut1}
 \end{eqnarray}
where the effective neutrino mass matrix ${\cal M}_{\nu}$ is given at leading order by
 \begin{eqnarray}
 {\cal M}_{\nu}\simeq -m^T_DM^{-1}_Rm_D\,.
   \label{neut2}
 \end{eqnarray}
The $X$-charges of $N$, $S$, and $\nu$ correspond to $X_{N}=-X_1/2$, $X_{S}=6X_2$, $X_{1\nu}=X_1/2$, and $X_{2\nu}=-7X_2$.
 Here we used four-component Majorana spinors, ($N^c=N, S^c=S$, and $\nu^c=\nu$).
 
 Let us turn to the neutrino mixing and mass.
We perform basis rotations from weak to mass eigenstates in the leptonic sector
 \begin{eqnarray}
 \nu_L\rightarrow U^\dag_\nu P_\nu\,\nu_L\,,\qquad N_R\rightarrow U^\dag_R\,N_R\,,\qquad \ell_L\rightarrow P_\ell\,\ell_L\,,\qquad \ell_R\rightarrow P_\ell\,\ell_R\,,
   \label{bas01}
 \end{eqnarray}
where $P_\nu$ and $P_\ell$ are phase matrices of diagonal, and $U_\nu$ and $U_R$ are unitary matrices chosen so as the matrices
 \begin{eqnarray}
 \hat{{\cal M}_{\nu}}=U^T_\nu\,P^\ast_\nu{\cal M}_{\nu}P^\ast_\nu\,U_\nu= -U^T_\nu P^\ast_\nu\,m^T_DU^\ast_R\,\hat{M}^{-1}_R\,U^\dag_Rm_DP^\ast_\nu\,U_\nu\,,\quad \hat{M}_R=U^\dag_R\,M_R\,U^\ast_R
   \label{neut3}
 \end{eqnarray}
 are diagonal. Then from the charged current term in Eq.\,(\ref{Axion_nu_La}) we obtain the lepton mixing matrix
  \begin{eqnarray}
 U_{\rm PMNS}=U_\nu\,,
   \label{pmns0}
 \end{eqnarray}
where $U_\nu$ is the $3\times3$ PMNS mixing matrix $U_{\rm PMNS}$. The matrix $U_{\rm PMNS}$ is expressed in terms of three mixing angles, $\theta_{12}, \theta_{13}, \theta_{23}$, and three \cp-odd phases (one $\delta_{CP}$ for the Dirac neutrinos and two $\varphi_{1,2}$ for the Majorana neutrinos) as\,\cite{PDG}
 \begin{eqnarray}
  U_{\rm PMNS}=
  {\left(\begin{array}{ccc}
   c_{13}c_{12} & c_{13}s_{12} & s_{13}e^{-i\delta_{CP}} \\
   -c_{23}s_{12}-s_{23}c_{12}s_{13}e^{i\delta_{CP}} & c_{23}c_{12}-s_{23}s_{12}s_{13}e^{i\delta_{CP}} & s_{23}c_{13}  \\
   s_{23}s_{12}-c_{23}c_{12}s_{13}e^{i\delta_{CP}} & -s_{23}c_{12}-c_{23}s_{12}s_{13}e^{i\delta_{CP}} & c_{23}c_{13}
   \end{array}\right)}Q_{\nu}\,,
 \label{PMNS1}
 \end{eqnarray}
where $s_{ij}\equiv \sin\theta_{ij}$, $c_{ij}\equiv \cos\theta_{ij}$ and $Q_{\nu}={\rm Diag}(e^{-i\varphi_1/2}, e^{-i\varphi_2/2},1)$.
Then the neutrino masses are obtained by the transformation
 \begin{eqnarray}
  U^T_{\rm PMNS}\,{\cal M}_{\nu}\,U_{\rm PMNS}={\rm Diag}.(m_{\nu_1}, m_{\nu_2}, m_{\nu_3})\,.
 \label{nu_mas}
 \end{eqnarray}
Here $m_{\nu_i}$ ($i=1,2,3$) are the light neutrino masses.
As is well-known, because of the observed hierarchy $|\Delta m^{2}_{\rm Atm}|= |m^{2}_{\nu_3}-(m^{2}_{\nu_1}+m^{2}_{\nu_2})/2|\gg\Delta m^{2}_{\rm Sol}\equiv m^{2}_{\nu_2}-m^{2}_{\nu_1}>0$ and the requirement of a Mikheyev-Smirnov-Wolfenstein resonance for solar neutrinos\,\cite{Wolfenstein:1977ue}, there are two possible neutrino mass spectra: (i) the normal mass ordering (NO) $m^2_{\nu_1}<m^2_{\nu_2}<m^2_{\nu_3}$, and (ii) the inverted mass ordering (IO) $m^2_{\nu_3}<m^2_{\nu_1}<m^2_{\nu_2}$.  From Eqs.\,(\ref{PMNS1}) and (\ref{nu_mas}) we see that there are nine physical observables ($\theta_{23}$, $\theta_{13}$, $\theta_{12}$, $\delta_{CP}$, $\varphi_1$, $\varphi_2$; $m_{\nu_1}$, $m_{\nu_2}$, $m_{\nu_3}$). The recent analysis based on the global fits\,\cite{Esteban:2018azc, deSalas:2017kay, Capozzi:2018ubv} of the neutrino oscillations enters into a new phase of precise determination of mixing angles and mass squared differences. In particular, in the most recent analysis\,\cite{Esteban:2018azc}, global fit values and $3\sigma$ intervals for the neutrino mixing angles and the neutrino mass-squared differences are listed in Table-\ref{exp_nu}, where $\theta_{23}$ and $\delta_{CP}$ have large uncertainties at 3$\sigma$. 
\begin{table}[h]
\caption{\label{exp_nu} The global fit of three-flavor oscillation parameters at the best-fit and $3\sigma$ level with Super-Kamiokande atmospheric data\,\cite{Esteban:2018azc}. NO = normal neutrino mass ordering; IO = inverted mass ordering. And $\Delta m^{2}_{\rm Sol}\equiv m^{2}_{\nu_2}-m^{2}_{\nu_1}$, $\Delta m^{2}_{\rm Atm}\equiv m^{2}_{\nu_3}-m^{2}_{\nu_1}$ for NO, and  $\Delta m^{2}_{\rm Atm}\equiv m^{2}_{\nu_2}-m^{2}_{\nu_3}$ for IO.}
\begin{ruledtabular}
\begin{tabular}{cccccccccccc} &$\theta_{13}[^{\circ}]$&$\delta_{CP}[^{\circ}]$&$\theta_{12}[^{\circ}]$&$\theta_{23}[^{\circ}]$&$\Delta m^{2}_{\rm Sol}[10^{-5}{\rm eV}^{2}]$&$\Delta m^{2}_{\rm Atm}[10^{-3}{\rm eV}^{2}]$\\
\hline
$\begin{array}{ll}
\hbox{NO}\\
\hbox{IO}
\end{array}$&$\begin{array}{ll}
8.61^{+0.37}_{-0.39} \\
8.65^{+0.38}_{-0.38}
\end{array}$&$\begin{array}{ll}
217^{+149}_{-82} \\
280^{+71}_{-84}
\end{array}$&$33.82^{+2.45}_{-2.21}$&$\begin{array}{ll}
49.7^{+2.5}_{-8.8} \\
49.7^{+2.4}_{-8.5}
\end{array}$&$7.39^{+0.62}_{-0.60}$
 &$\begin{array}{ll}
2.525^{+0.097}_{-0.094} \\
2.512^{+0.094}_{-0.099}
\end{array}$ \\
\end{tabular}
\end{ruledtabular}
\end{table}

The current agenda of neutrino oscillation experiments are the precise measurements on the Dirac CP-violating phase $\delta_{CP}$ and atmospheric mixing angle $\theta_{23}$.
Based on the given model, we explore what values of the low energy Dirac CP phase $\delta_{CP}$ and atmospheric mixing angle $\theta_{23}$ can predict a value for the mass hierarchy of neutrino (normal or inverted mass ordering), and investigate the observables that can be tested in the current and the next generation of experiments: a recent upper bound on the rate of $0\nu\beta\beta$ decay shows
  \begin{eqnarray}
 ({\cal M}_{\nu})_{ee}<\left\{
             \begin{array}{ll}
             0.104-0.228\,{\rm eV}\,, & (^{76}\text{Ge-based experiment\,\cite{Agostini:2019hzm}})  \\
             0.076-0.234\,{\rm eV}\,, & (^{136}\text{Xe-based experiment\,\cite{KamLAND-Zen:2016pfg}})
             \end{array}
           \right.\,
 \label{nubb}
 \end{eqnarray}
at $90\%$ C.L, where $({\cal M}_{\nu})_{ee}$ stands for the effective Majorana mass (the modulus of the $ee$-entry of the effective neutrino mass matrix).
Since the $0\nu\beta\beta$ decay is a probe of lepton number violation at low energy, its measurement could be the strongest evidence for lepton number violation at high energy. In other words, the discovery of $0\nu\beta\beta$ decay would suggest the Majorana character of the neutrinos and thus the existence of heavy Majorana neutrinos (via the seesaw mechanism\,\cite{Minkowski:1977sc}).
\subsection{Neutrino Mixing and Masses}
From the superpotential (\ref{lagrangian2}) with the desired VEV alignment in Eq.\,(\ref{vev}), the Majorana neutrino mass terms in the Lagrangian\,(\ref{AxionLag3}) read
 \begin{eqnarray}
M_{R}=M{\left(\begin{array}{ccc}
  1+\frac{2\tilde{\kappa}\,e^{i\phi}}{3}+(\tilde{\kappa}_\chi+\frac{13}{9}\tilde{\kappa}_s\,e^{i\tilde{\phi}})\nabla_T & -\frac{\tilde{\kappa}\,e^{i\phi}}{3}+\frac{17}{18}\tilde{\kappa}_s\,e^{i\tilde{\phi}}\nabla_T & -\frac{\tilde{\kappa}\,e^{i\phi}}{3}+\frac{23}{18}\tilde{\kappa}_s\,e^{i\tilde{\phi}}\nabla_T  \\
  -\frac{\tilde{\kappa}\,e^{i\phi}}{3}+\frac{17}{18}\tilde{\kappa}_s\,e^{i\tilde{\phi}}\nabla_T & \frac{2\tilde{\kappa}\,e^{i\phi}}{3}+\frac{4}{9}\tilde{\kappa}_s\,e^{i\tilde{\phi}}\nabla_T & 1-\frac{\tilde{\kappa}\,e^{i\phi}}{3}+\frac{7}{9}\tilde{\kappa}_s\,e^{i\tilde{\phi}}\nabla_T  \\
 -\frac{\tilde{\kappa}\,e^{i\phi}}{3}+\frac{23}{18}\tilde{\kappa}_s\,e^{i\tilde{\phi}}\nabla_T & 1-\frac{\tilde{\kappa}\,e^{i\phi}}{3}+\frac{7}{9}\tilde{\kappa}_s\,e^{i\tilde{\phi}}\nabla_T & \frac{2\tilde{\kappa}\,e^{i\phi}}{3}+\frac{10}{9}\tilde{\kappa}_s\,e^{i\tilde{\phi}}\nabla_T
 \end{array}\right)}\,,
 \label{MR1}
 \end{eqnarray}
 where $v_S=\kappa\,v_\Theta$ is used and the $ \hat{y}_\Theta$ in $M\equiv|\hat{y}_\Theta v_\Theta/\sqrt{2}|$ is a rescaled factor by $\hat{y}_\Theta-\hat{y}_\chi\nabla_T/3$, and
   \begin{eqnarray}
\tilde{\kappa}\equiv\kappa\Big|\frac{\hat{y}_{R}}{\hat{y}_\Theta}\Big|\,,\quad \phi\equiv\arg\left(\hat{y}_{R}/\hat{y}_{\Theta}\right)\,,\quad \tilde{\kappa}_\chi\equiv\Big|\frac{\hat{y}_\chi}{\hat{y}_\Theta}\Big|\,,\quad \tilde{\kappa}_s\equiv\kappa\Big|\frac{\hat{y}_r}{\hat{y}_\Theta}\Big|\,,\quad\arg(\tilde{\kappa}_s)\equiv\tilde{\phi}\,.
  \label{MR5}
  \end{eqnarray}
Here we set, without loss of generality, $\hat{y}_\Theta=1$ and $\arg(\tilde{\kappa}_\chi)=0$.
A common factor $M$ in $M_R$ can be replaced by the QCD axion decay constant $F_A$ via Eq.\,(\ref{adeca}) (see also Eq.\,(\ref{emg})):
 \begin{eqnarray}
 M=\Big|\frac{\hat{y}_{\Theta}\,\delta^{\rm G}_1}{X_1}\Big|\frac{F_A}{\sqrt{1+\kappa^2}}\,.
 \label{MR2}
 \end{eqnarray}
The additional sterile neutrino mass term reads  $M_S=\hat{y}_s(\nabla_\Psi)^{12}\,\nabla_T\,v_T/\sqrt{2}$, leading to $M_S\simeq1.27\times10^6$ GeV for $\hat{y}_s=1$ from Eqs.\,(\ref{quarkvalue}) and (\ref{lamScale}).
And the Dirac mass term reads
  \begin{eqnarray}
m_{D}=\hat{y}_\nu\,\nabla^{7}_{\Psi}{\left(\begin{array}{ccc}
 1+2\nabla_Ta_\nu &  0 &  0 \\
 0 &  0 &  1-\frac{3}{2}\nabla_Ta_\nu   \\
 0 &  1-\frac{1}{2}\nabla_Ta_\nu  &  0
 \end{array}\right)}\,v_{u} \,,
 \label{Ynu1}
 \end{eqnarray}
 where $a_\nu\equiv\hat{y}_{\tilde{\nu}}/\hat{y}_\nu$.  From Eq.\,(\ref{neut3}), as pointed out in Ref.\,\cite{Ahn:2012cg}, complex phases in $m_D$ of Eq.\,(\ref{Ynu1}) can always be rotated away by appropriately choosing the phases of left-handed charged lepton fields since the phase matrix $P_\nu$ accompanies the Dirac-neutrino mass matrix $m_D$. The mass terms of neutrinos of Eqs.\,(\ref{MR1}, \ref{Ynu1}) could be corrected by a shift of the vacuum configuration of Eq.\,(\ref{vev}). But the VEV corrections of Eqs.\,(\ref{Npo4}, \ref{Npo5}) induced by higher-dimensional operators do not change their flavor structures. 

It is expected that at leading order ({\it i.e.} in the limit $\nabla_T\rightarrow0$) the mass matrix of Eq.\,(\ref{neut2}) with Eqs.\,(\ref{MR1}, \ref{Ynu1}) reflects an exact TBM\,\cite{Harrison:2002er} with their corresponding mass eigenvalues
  \begin{eqnarray}
 && \quad~\theta_{13}=0\,,\qquad\qquad\quad \theta_{23}=\frac{\pi}{4}=45^\circ\,,\qquad\qquad\quad\theta_{12}=\sin^{-1}\Big(\frac{1}{\sqrt{3}}\Big)\simeq35.3^\circ\,
\nonumber\\
&& m_{\nu_1}=\frac{|\hat{y}_\nu|^2\,\nabla^{14}_\Psi}{M(1+\tilde{\kappa}\,e^{i\phi})}v^2_u\,,\qquad m_{\nu_2}=\frac{|\hat{y}_\nu|^2\,\nabla^{14}_\Psi}{M}v^2_u\,,\qquad m_{\nu_3}=\frac{|\hat{y}_\nu|^2\,\nabla^{14}_\Psi}{M(1-\tilde{\kappa}\,e^{i\phi})}v^2_u\,.
\label{tbm}
 \end{eqnarray}
However, the non-zero but small value $\nabla_T$  of Eq.\,(\ref{quarkvalue}) given by both the quark and charged-lepton sectors reflects that the TBM and their mass eigenvalues in Eq.\,(\ref{tbm}) for three flavors should be corrected. 

 The precisely measured values of $\theta_{13}$, $\theta_{12}$, $\Delta m^2_{\rm Sol}$, and $\Delta m^2_{\rm Atm}$ in Table-\ref{exp_nu} can be fitted by  
the higher-dimensional operators induced in the superpotential (\ref{lagrangian2}) by the $G_F$ symmetry with the anomaly cancellations of both $U(1)_Y\times[U(1)]^2$ and $U(1)\times[gravity]^2$, and in turn their nontrivial contributions can give predictions on the neutrino masses, atmospheric mixing angle, Dirac CP phase, and mass ordering of mass spectra.
After seesawing via Eq.\,(\ref{neut2}) the effective mass matrix contains only eight physical degree of freedoms
 \begin{eqnarray}
 m_0\,,\quad|a_\nu|\,,\quad\tilde{\kappa}\,,\quad\tilde{\kappa}_\chi\,,\quad\tilde{\kappa}_s\,,\quad\phi\,,\quad\tilde{\phi}\,,\quad\arg{(a_\nu)}\,,
 \label{dof_nu}
 \end{eqnarray}
 defined at the seesaw scale $M$, where $m_0\equiv m_0(M)$ stands for the overall factor of Eq.\,(\ref{neut2}), 
  \begin{eqnarray}
 m_0(M)=|\hat{y}_\nu|^2\,\sqrt{1+\kappa^2}|X_1|\frac{\nabla^{14}_\Psi\,v^2_u}{F_A\,|\hat{y}_\Theta\,\delta^{\rm G}_1|}\,.
 \label{mnu0}
 \end{eqnarray}
The four measured quantities ($\theta_{12}$, $\theta_{13}$, $\Delta m^2_{\rm Sol}$, $\Delta m^2_{\rm Atm}$) among nine observables can be used as constraints, and the rest four degree of freedoms of Eq.\,(\ref{dof_nu}) corresponding to four measurable quantities give their measurements and the rest one quantity is predicted among five measurable quantities {\it e.g.} ($\theta_{23}, \delta_{CP}, \varphi_{1,2}, 0\nu\beta\beta$-decay rate).
\begin{figure}[h]
\begin{minipage}[h]{7.3cm}
\epsfig{figure=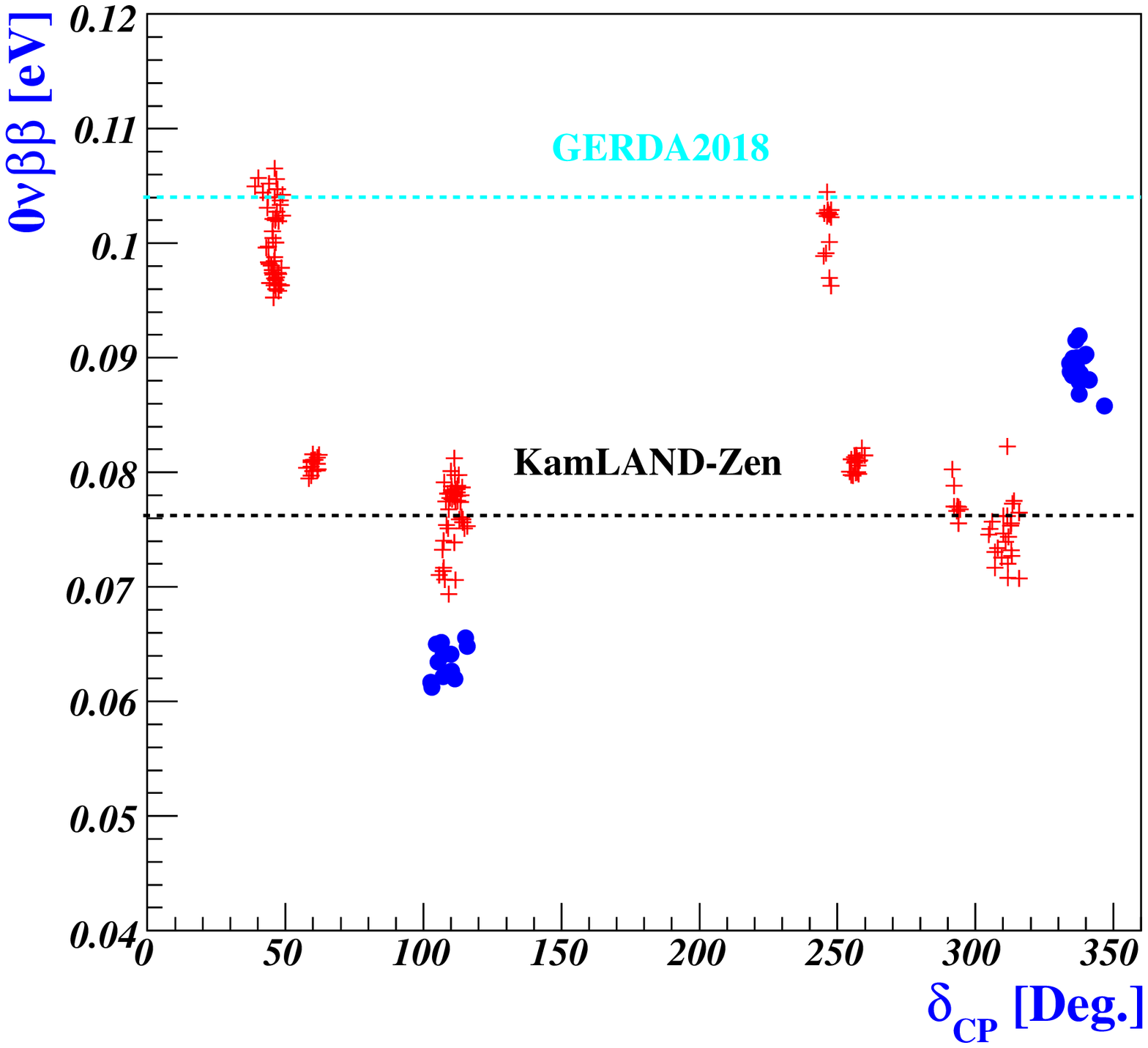,width=7.3cm,angle=0}
\end{minipage}
\hspace*{1.0cm}
\begin{minipage}[h]{7.3cm}
\epsfig{figure=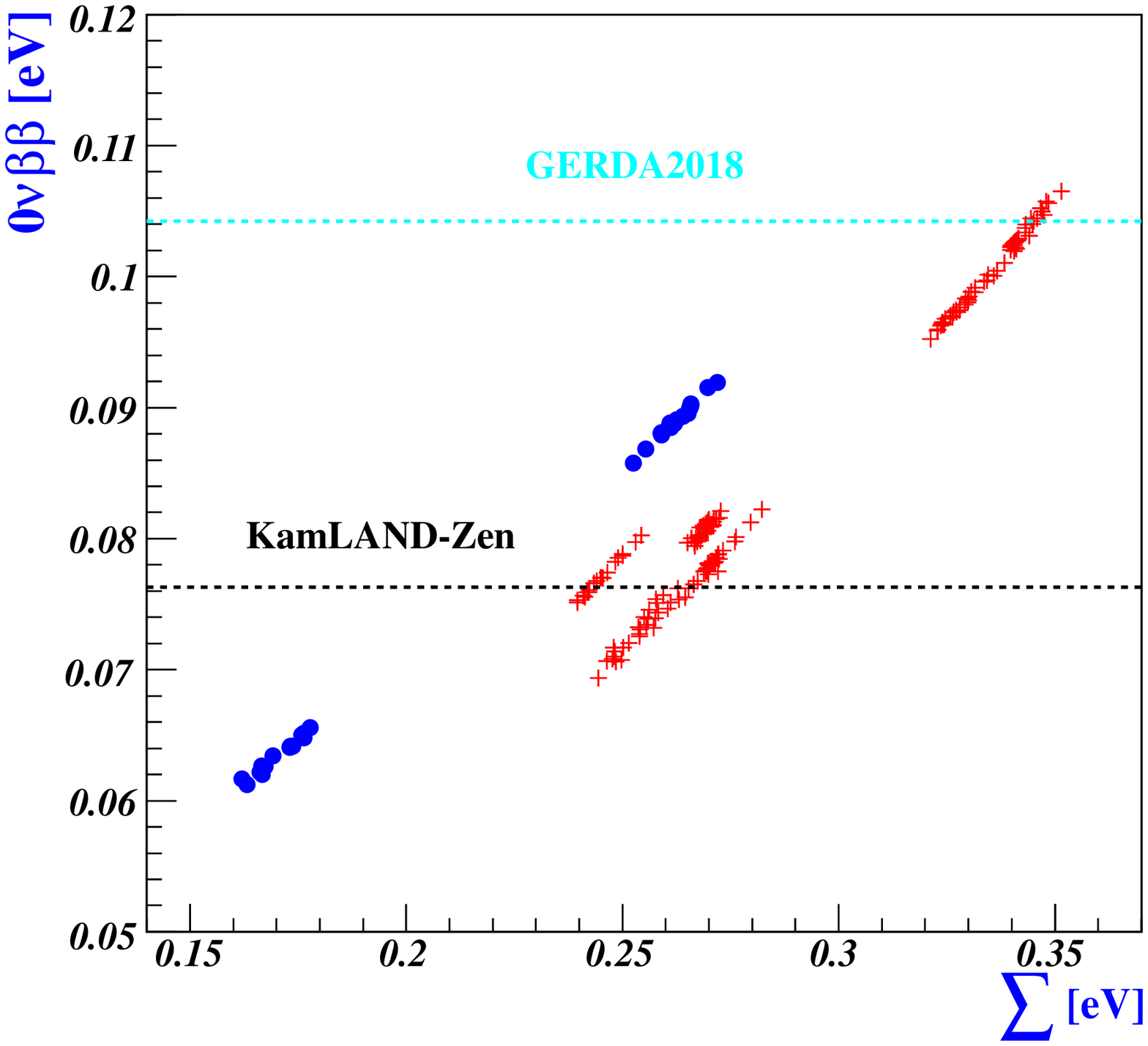,width=7.3cm,angle=0}
\end{minipage}\\
\begin{minipage}[h]{7.3cm}
\epsfig{figure=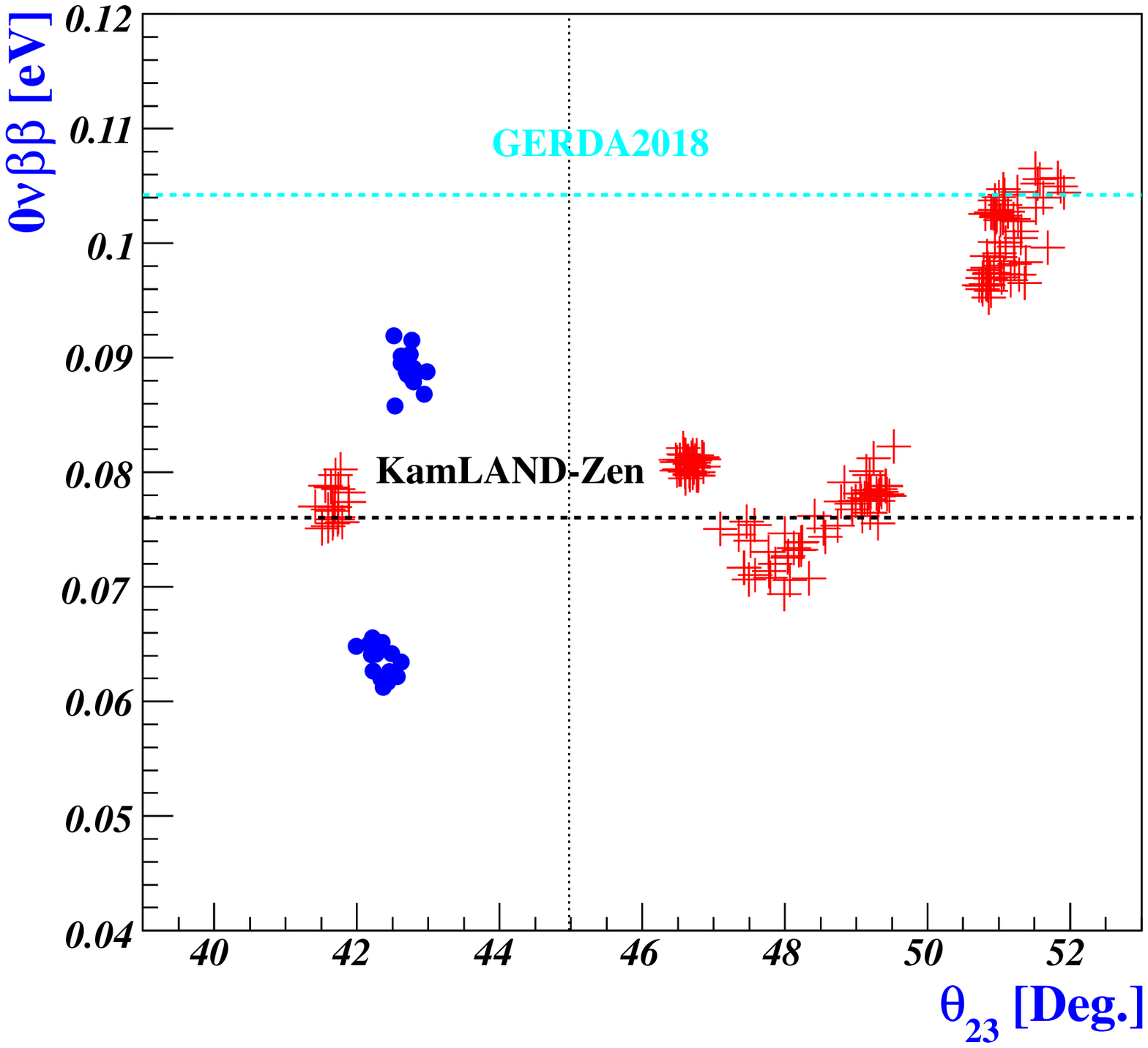,width=7.3cm,angle=0}
\end{minipage}
\hspace*{1.0cm}
\begin{minipage}[h]{7.3cm}
\epsfig{figure=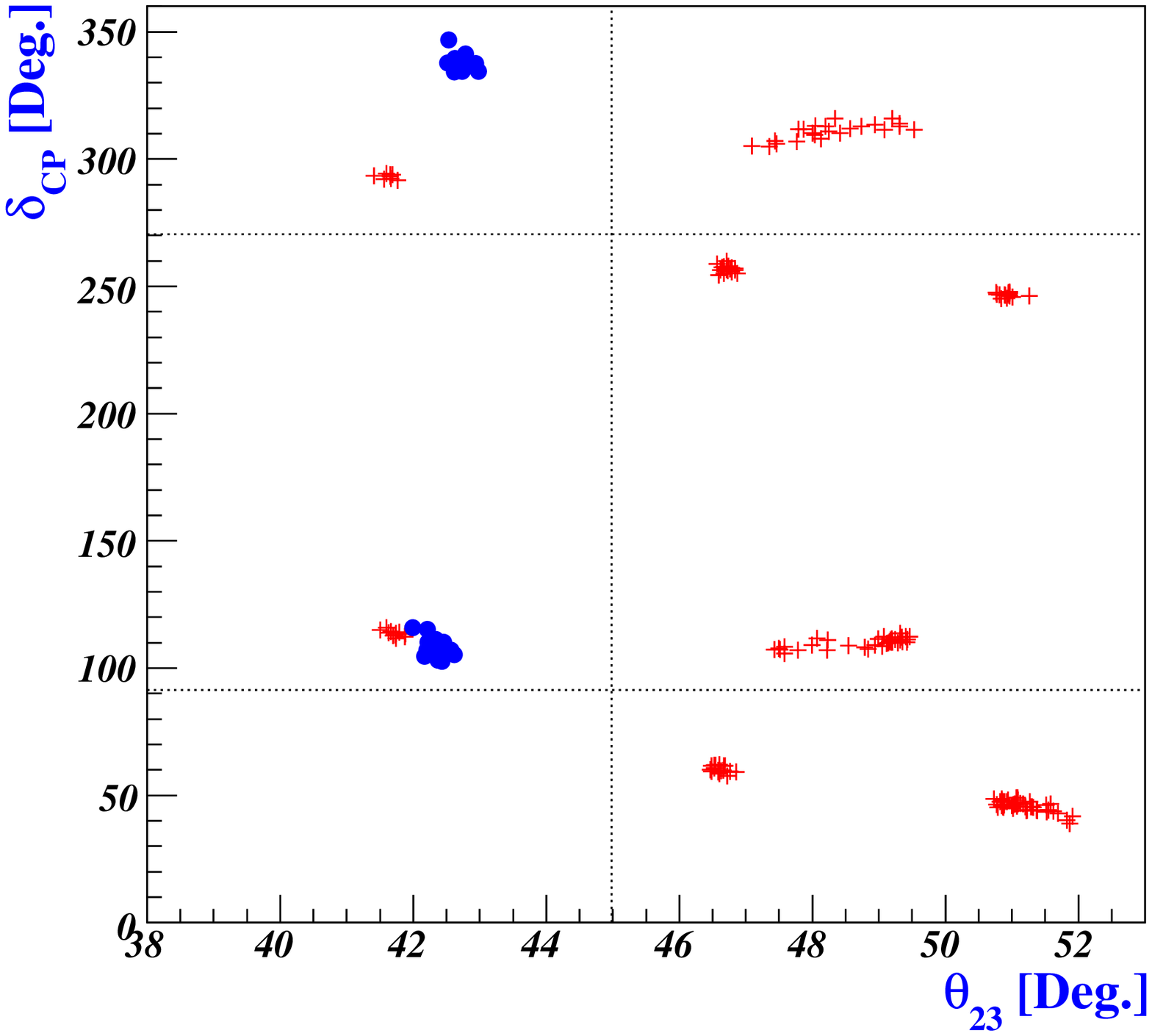,width=7.3cm,angle=0}
\end{minipage}
\caption{\label{FigA4} Predictions for $({\cal M}_\nu)_{ee}\equiv0\nu\beta\beta$ are shown, left upper, right upper, and left lower panel, as function of the Dirac CP phase $\delta_{CP}$, the sum of neutrino masses $\sum$, and the atmospheric mixing angle $\theta_{23}$, respectively, where the horizontal cyan-dashed (black-dotted) line indicates an upper bound from the GERDA\,\cite{Agostini:2019hzm} (KamLAND-Zen\,\cite{KamLAND-Zen:2016pfg}) Collaboration. Right lower panel represents predictions on $\delta_{CP}$ as a function of $\theta_{23}$. The red crosses and blue spots correspond to NO and IO, respectively.}
\end{figure}
\subsection{Numerical analysis}
The seesaw in Eq.\,(\ref{neut2}) operates at the scale of $G_F$ symmetry breakdown while its implications are measured by experiments below the electroweak scale. Therefore, quantum corrections to their masses and mixing angles could be very crucial, especially, for degenerate neutrino masses\,\cite{Antusch:2005gp}. Moreover, in the present model since the Dirac neutrino Yukawa coupling $|y_\nu|\sim\nabla^7_\Psi$ in Eq.\,(\ref{Ynu1}) is much less than the top Yukawa coupling $\hat{y}_t\simeq1$, the running of the neutrino mass eigenvalues strongly depends on the top Yukawa coupling. 
For degenerate mass spectra and $|y_\nu|\sim\nabla^7_\Psi\ll\hat{y}_t$, it is expected that the parameter spaces obtained at the tree-level satisfying the low-energy neutrino data of Table-\ref{exp_nu} are different from those by one-loop RG running effects. 

\begin{table}[h]
\caption{\label{inp_1} Input parameter spaces and predictions for NO.}
\begin{ruledtabular}
\begin{tabular}{ccccc}
inputs/space &I& II& III&IV \\
\hline
$m_{0}$&$0.227-0.237$&$0.243-0.248$&$0.236-0.278$&$0.318-0.341$\\
$|a_{\nu}|$ &$0.826-0.877$ &$0.768-0.869$ & $0.775-0.853$ & $0.737-0.774$\\
$\tilde{\kappa}$&$0.453-0.564$&$0.389-0.423$&$0.490-0.659$ &$0.586-0.646$\\
$\tilde{\kappa}_\chi$&$2.90-3.23$&$2.81-3.21$&$2.78-3.23$&$2.62-3.14$\\
$\tilde{\kappa}_s$&$0.929-1.067$&$1.012-1.141$&$1.256-1.509$&$0.888-1.090$\\
$\phi[^\circ]$ &$284.6-286.7$&$280.7-282.3$&$77.6-81.3$&$275.0-277.4$\\
$\tilde{\phi}[^\circ]$ &$284.2-288.6$&$73.2-74.4$&$273.0-280.9$&$72.0-74.6$\\
$\arg(a_\nu)[^\circ]$ &$60.6-62.6$&$58.7-64.9$&$61.6-64.7$&$60.2-64.0$\\
\hline
predictions && & & \\
\hline
$\theta_{23}[^\circ]$ & $41.4-41.9$ & $46.5-46.9$ & $47.1-49.5$ & $50.7-51.9$ \\
$\delta_{CP}[^\circ]$ & $\begin{array}{ll}
111.6-116.0 \\
291.7-294.4
\end{array}$ & $\begin{array}{ll}
57.7-62.1 \\
254.4-260.0
\end{array}$ & $\begin{array}{ll}
105.6-113.7 \\
304.9-315.8
\end{array}$ & $\begin{array}{ll}
38.9-49.0 \\
245.0-247.7
\end{array}$ \\
$ ({\cal M}_\nu)_{ee}[{\rm eV}]$ & $0.075-0.080$ & $0.080-0.082$ & $0.069-0.082$ & $0.095-0.107$ \\
$\sum[{\rm eV}]$ & $ 0.239-0.254$ & $0.265-0.273$ & $0.244-0.282$ & $0.321-0.348$ \\
\end{tabular}
\end{ruledtabular}
\end{table}
To obtain a truly meaningful parameter space at the scale of $G_F$ symmetry breakdown, which satisfies the neutrino data obtained at low energy scales in Table-\ref{exp_nu}, we run the gauge\,\footnote{Values of the gauge couplings at the scale of $G_F$ symmetry breakdown can be obtained by running with the precision values extracted at the electroweak scale.} and Yukawa couplings from the FN scale $\Lambda$ of Eq.\,(\ref{lamScale}) to the electroweak scale taken to be 100 GeV$\sim m_Z$. And in this analysis, threshold effects at either end of the spectrum are not included.
 \begin{table}[h]
\caption{\label{inp_2} Input parameter spaces and predictions for IO.}
\begin{ruledtabular}
\begin{tabular}{ccc}
parameter/space &I & II \\
\hline
$m_{0}$&$0.240-0.254$ & $0.146-0.171$\\
$|a_{\nu}|$ &$0.945-0.990$ & $0.951-1.079$\\
$\tilde{\kappa}$&$0.396-0.437$ & $0.538-0.589$\\
$\tilde{\kappa}_\chi$&$3.00-3.25$ &$1.77-2.93$\\
$\tilde{\kappa}_s$&$0.851-0.957$ &$0.543-0.727$ \\
$\phi[^\circ]$ &$74.5-76.8$ & $304.1-313.0$ \\
$\tilde{\phi}[^\circ]$ &$288.3-291.8$ & $25.7-43.1$\\
$\arg(a_\nu)[^\circ]$ &$47.4-52.6$ & $9.0-23.3$\\
\hline
predictions & &   \\
\hline
$\theta_{23}[^\circ]$ & $42.5-43.0$ & $42.0-42.6$   \\
$\delta_{CP}[^\circ]$ & $334.2-346.7$ & $102.6-115.9$   \\
$({\cal M}_\nu)_{ee}[{\rm eV}]$ & $0.088-0.092$ & $0.061-0.066$   \\
$\sum[{\rm eV}]$ & $0.253-0.272$ & $0.162-0.178$   \\
\end{tabular}
\end{ruledtabular}
\end{table}
Scanning all the parameter spaces of Eq.\,(\ref{dof_nu}) by putting the precision constraints $\{\theta_{13}$, $\theta_{12}$, $\Delta m^2_{\rm Sol}$, $\Delta m^2_{\rm Atm}\}$ at $3\sigma$ in Table-\ref{exp_nu} with the given values of $\nabla_T$, $\nabla_\Psi$, $\kappa$ in Eqs.\,(\ref{tanpara}, \ref{quarkvalue}) and taking a value of $F_A=3.57\times10^{10}$ GeV which corresponds to the central value of Eq.\,(\ref{k_bound}), we obtain the input parameter spaces at the scale of $G_F$ breakdown for both NO and IO, leading to the predictions in Table-\ref{inp_1} for NO and Table-\ref{inp_2} for IO at the electroweak scale, marked as red-crosses and blue-spots in Fig.\,\ref{FigA4}, respectively.
 In Fig.\,\ref{FigA4}, left upper, right upper, and left lower panel shows predictions for $({\cal M}_\nu)_{ee}\equiv0\nu\beta\beta$ as function of the Dirac CP phase $\delta_{CP}$, the sum of neutrino masses $\sum$, and the atmospheric mixing angle $\theta_{23}$, respectively, where the horizontal cyan-dashed (black-dotted) line indicates an upper bound in Eq.\,(\ref{nubb}) from the GERDA\,\cite{Agostini:2019hzm} (KamLAND-Zen\,\cite{KamLAND-Zen:2016pfg}) Collaboration. The right lower panel in Fig.\,\ref{FigA4} stands for the predictions of $\delta_{\rm CP}$ as a function of $\theta_{23}$. As can be seen in Fig.\,\ref{FigA4}, it is clear that the model will be excluded or strongly favored by future $0\nu\beta\beta$ decay searches and/or precise measurements of $\theta_{23}$, providing the narrow region of predictions on $\delta_{CP}$ and mass orderings.

\section{QCD axion mass from flavored-Axions properties}
\label{fl-qcd}
\noindent 
The direct interactions of flavored-axions with quarks and leptons arise through Yukawa interactions in the flavored-axion framework of Eq.\,(\ref{AFN}) as in Ref.\,\cite{Ahn:2014gva}, see Eqs.\,(\ref{AxionLag2}) and (\ref{AxionLag3}). Such flavored-axion interactions with normal matter allow looking for the QCD axion in rare decays in particle physics and stellar evolutions in astroparticle physics\,\cite{raffelt, Adler:2008zza, Bolton:1988af}. 
In the present model, such coupling strengths of the flavored-axion particles with normal matter and radiation are more strongly constrained compared with those in Refs.\,\cite{Ahn:2018cau, Ahn:2018nfb} by the anomaly cancellation conditions of both $U(1)_Y\times[U(1)_X]^2$ and $U(1)_X\times[gravity]^2$, which should satisfy observations and experimental data.

The model we propose should satisfy two tight constraints 
coming from the active neutrino mass Eq.\,(\ref{neut2}) originated from the seesaw mechanism\,\cite{Minkowski:1977sc} and the flavor-violating processes induced by the flavored-axions such as $K^+\rightarrow\pi^++A_i$\,\cite{Artamonov:2008qb} in particle physics.
Once the quantum number of the model, {\it e.g.} Table-\ref{reps_q} and -\ref{reps_l}, is satisfied with those constraints, one can derive a QCD axion decay constant $F_A$ (or QCD axion mass, or equivalently seesaw scale) and in turn, whose value should not conflict with the constraints coming from astronomical observations of the stellar evolution.

\subsection{Flavor-Changing process $K^+\rightarrow\pi^++A_i$ induced by the flavored-axions}
\noindent 
Since a direct interaction of the SM gauge singlet flavon fields charged under $U(1)_X$ 
with the SM quarks charged under $U(1)_X$ can arise through Yukawa interaction, the flavor-changing process $K^+\rightarrow\pi^++A_i$ is induced by the flavored-axions $A_i$.
Then, from Eqs.\,(\ref{lagrangian_qd}) and (\ref{Ch2}) the flavored-axion interactions with the flavor violating coupling to the $s$- and $d$-quark is given by
\begin{eqnarray}
-{\cal L}^{A_isd}_Y\simeq \frac{i}{2}\Big(\frac{|X_1|\,A_1}{f_{a_1}}-\frac{|X_2|\,A_2}{f_{a_2}}\Big)\bar{s}d\,(m_s-m_d)\lambda\Big(1-\frac{\lambda^2}{2}\Big)\,,
  \label{Gkp}
\end{eqnarray}
where\,\footnote{The flavor violating coupling to the $s$- and $d$-quark induced by the mixing matrix $V^d_R$  is absent due to $(V^d_R\,{\rm Diag.}(5\frac{A_2}{v_{g}}, 5\frac{A_2}{v_{g}}, 2\frac{A_1}{v_{\cal F}}+5\frac{A_2}{v_{g}})\,V^{d\dag}_R)_{12}=0$, see Eq.\,(\ref{c_num}).} $V^{d\dag}_L\simeq V_{\rm CKM}$, $f_{a_1}=|X_1|v_{\cal F}$, and $f_{a_2}=|X_2|v_g$ are used. Then the decay width of $K^+\rightarrow\pi^++A_i$ is given by\,\cite{Wilczek:1982rv, raredecay} $\Gamma(K^+\rightarrow\pi^++A_i)=\frac{m^3_K}{16\pi}\big(1-\frac{m^2_{\pi}}{m^2_{K}}\big)^3\big|{\cal M}_{dsi}\big|^2$
 where $m_{K^{\pm}}=493.677\pm0.013$ MeV, $m_{\pi^{\pm}}=139.57061\pm0.00024$ MeV\,\cite{PDG}, and
 \begin{eqnarray}
   \big|{\cal M}_{dsi}\big|^2=\Big|\frac{X_i}{2\sqrt{2}\delta^{\rm G}_i\,F_{A}}\lambda\Big(1-\frac{\lambda^2}{2}\Big)\Big|^2\qquad\text{with}~i=1,2\,,
    \label{}
 \end{eqnarray}
with $F_A=f_{a_i}/(\delta^{\rm G}_i\sqrt{2})$.
From the present experimental upper bound ${\rm Br}(K^+\rightarrow\pi^+A_i)<7.3\times10^{-11}$ at $90\%$ CL\,\cite{Artamonov:2008qb} with ${\rm Br}(K^+\rightarrow\pi^+\nu\bar{\nu})=1.73^{+1.15}_{-1.05}\times10^{-10}$\,\cite{Artamonov:2009sz},
we obtain the lower limit on the QCD axion decay constant,
 \begin{eqnarray}
  F_{A}\gtrsim2.04\times10^{10}\,{\rm GeV}\,.
 \label{cons_2}
 \end{eqnarray}

\subsection{QCD axion interactions with photons and electrons}
\noindent The QCD axion mass $m_a$ in terms of the pion mass and pion decay constant reads\,\cite{Ahn:2014gva, Ahn:2016hbn}
 \begin{eqnarray}
 m^{2}_{a}F^{2}_{A}=m^{2}_{\pi^0}f^{2}_{\pi}F(z,w)\,,
\label{axiMass2}
 \end{eqnarray}
where $f_\pi\simeq92.1$ MeV\,\cite{PDG} and $F(z,w)=z/(1+z)(1+z+w)$ with $\omega=0.315\,z$.
Recalling that the Weinberg value lies in $0.40<z<0.53$ in Eq.\,(\ref{axipra}).
With the model predicted value of Eq.\,(\ref{cas}) the axion-photon coupling is expressed in terms of the QCD axion mass, pion mass, pion decay constant, $z$ and $w$:
 \begin{eqnarray}
 g_{a\gamma\gamma}=\frac{\alpha_{\rm em}}{2\pi}\frac{m_a}{f_{\pi}m_{\pi^0}}\frac{1}{\sqrt{F(z,w)}}\left(\frac{E}{N_C}-\frac{2}{3}\,\frac{4+z+w}{1+z+w}\right)\,.
 \label{gagg}
 \end{eqnarray}
The upper bound on the axion-photon coupling is derived from the recent analysis of the horizontal branch stars in galactic globular clusters\,\cite{Ayala:2014pea}, $|g_{a\gamma\gamma}|<6.6\times10^{-11}\,{\rm GeV}^{-1}\,(95\%\,{\rm CL})$, which translates into the lower bound of decay constant through Eq.\,(\ref{axiMass2}) 
\begin{eqnarray}
  F_A\gtrsim1.34\times10^{7}\,{\rm GeV}\,,
 \label{axph}
\end{eqnarray}
 where $z=0.47$ is used. In the model the axion to photon coupling of Eq.\,(\ref{gagg}) can be linked to the active neutrino mass at seesaw scale via Eq.\,(\ref{mnu0}). Thus, using the values of $m_0(M)$ in Table-\ref{inp_1} and -\ref{inp_2} and $\kappa, \nabla_\Psi, \tan\beta$ in Eqs.\,(\ref{tanpara}) and (\ref{quarkvalue}), we obtain 
\begin{eqnarray}
  1.74\times10^{-14}\lesssim g_{a\gamma\gamma}{\rm [GeV^{-1}]}\lesssim1.74\times10^{-12}\,,
 \label{nCon}
\end{eqnarray}
where $1/\sqrt{10}\lesssim|\hat{y}_\nu|\lesssim\sqrt{10}$ is taken into account, and which subsequently translates into the QCD axion decay constant
\begin{eqnarray}
  5.09\times10^{8}\lesssim F_A{\rm [GeV]}\lesssim5.09\times10^{10}\,.
 \label{nCon1}
\end{eqnarray}
A narrow range of the QCD axion decay constant is derived from the bounds Eqs.\,(\ref{cons_2}, \ref{nCon1}):
 \begin{eqnarray}
 F_{A}=3.57^{\,+1.52}_{\,-1.53}\times10^{10}\,{\rm GeV}\,.
  \label{k_bound}
 \end{eqnarray}
The value of this range should satisfy the constraints coming from axion cooling of stars via bremsstrahlung off electrons $e+Ze\rightarrow Ze+e+A_i$\,\cite{Giannotti:2017hny}. 
From Eq.\,(\ref{lagrangian_chL}) the flavored-axion $A_2$ coupling to electrons in the model reads 
\begin{eqnarray}
 g_{Aee}=\frac{|X_e|\,m_e}{\sqrt{2}\,|\delta^{\rm G}_2|\,F_A}\qquad\text{with}~X_e=-13X_2
 \label{ax_e}
\end{eqnarray}
where $m_e=0.511$ MeV and $F_A=f_{a_i}/\sqrt{2}\,\delta^{\rm G}_i$ with Eq.\,(\ref{dGi1}). We stress that the electron quantum number of $U(1)_X$ in Eq.\,(\ref{ax_e}) and the value of $\delta^{\rm G}_2$ in Eq.\,(\ref{dGi1}) have been determined via the anomaly cancellation conditions of both $U(1)_Y\times[U(1)_X]^2$ and $U(1)_X\times[gravity]^2$, as shown in Table-\ref{reps_q} and -\ref{reps_l}.
Indeed, the longstanding anomaly in the cooling of white dwarfs (WDs)\,\cite{Corsico:2019nmr} and red giants branch (RGB) stars in globular clusters where bremsstrahlung off electrons is mainly efficient\,\cite{Raffelt:1985nj} could be explained by the flavored axions with the fine-structure constant of axion to electrons $\alpha_{Aee}=(0.29-2.30)\times10^{-27}$\,\cite{WD01} and  $\alpha_{Aee}=(0.41-3.70)\times10^{-27}$\,\cite{Bertolami:2014wua, wd_recent2} with $\alpha_{Aee}=g^2_{Aee}/4\pi$, indicating the clear systematic tendency of stars to cool faster than  predicted. A recent reexamination of the WD Luminosity Function and its combination with hints from the WD pulsation and RGB stars has provided bounds of the axion-electron coupling in Refs.\,\cite{Giannotti:2017hny, DiLuzio:2020wdo}. 
Taking the constraint of Eq.\,(\ref{k_bound}) into account,  the axion-electron coupling of Eq.\,(\ref{ax_e}) reads
 \begin{eqnarray}
 g_{Aee}=3.29^{+2.47}_{-0.98}\times10^{-14}\,,
  \label{a_ee}
 \end{eqnarray}
 which lies in the $3\sigma$ bound of Ref.\,\cite{DiLuzio:2020wdo}, where the central value corresponds to that of Eq.\,(\ref{k_bound}).
Clearly, the strongest bound on the QCD axion decay constant comes from the seesaw scale as the common origin of active neutrino mass Eq.\,(\ref{MR2}) as well as the flavor-changing process $K^+\rightarrow\pi^++A_i$ induced by the flavored-axions in Eq.\,(\ref{cons_2}).

 \subsection{The correlated predictions via flavored-axion's network}
 \label{cpvf}
Since the flavored-axion $A_i$ couples directly to SM matter fields, direct correlations between axion interactions to the SM fields and QCD axion mass can be generically derived in flavored axion models\,\cite{Ahn:2014gva, Ahn:2016hbn, Ahn:2018nfb, Ahn:2018cau}. Then we see that the QCD axion decay constant or the scale of $U(1)_X$ group breakdown has tight correlations with the below terminology in quotes:
  \begin{eqnarray}
 && F_A=\Lambda\sqrt{2}\,\nabla_\Psi\frac{|X_2|}{|\delta^{\rm G}_2|}\qquad\qquad\qquad\qquad\qquad\qquad\text{``FN scale"}\nonumber\\
  &&=\frac{m_{\pi^0}f_\pi}{m_a}\sqrt{F(z,w)}\qquad\qquad\qquad\qquad\qquad\text{``QCD axion mass"}\nonumber\\
  &&=\frac{1}{g_{Aee}}\frac{m_e\,|X_e|}{\sqrt{2}\,|\delta^{\rm G}_i|}\qquad\qquad\qquad\qquad\qquad~\text{``axion-electron coupling"}\nonumber\\
  &&=\frac{1}{g_{a\gamma\gamma}}\frac{\alpha_{\rm em}}{2\pi}\Big(\frac{E}{N_C}-\frac{2}{3}\frac{4+z+w}{1+z+w}\Big)\quad\qquad\text{``axion-photon coupling"}\nonumber\\
  &&=\frac{|X_n|\,m_n}{g_{Ann}}\qquad\qquad\qquad\qquad\qquad\qquad\text{``axion-neutron coupling"}\nonumber\\
  &&=\Big|\frac{\sum_{C=L,R}(V^d_C\,C_i\,V^{d\dag}_C)_{12}}{2\sqrt{2}\,\delta^{\rm G}_i}\Big|\left\{\Big(1-\frac{m^2_\pi}{m^2_K}\Big)^3\frac{m^3_K}{16\pi\Gamma(K^+\rightarrow\pi^+\nu\bar{\nu})}\frac{{\rm Br}(K^+\rightarrow\pi^+\nu\bar{\nu})}{{\rm Br}(K^+\rightarrow\pi^+A_i)}\right\}^{\frac{1}{2}}\nonumber\\
  &&\qquad\qquad\qquad\qquad\qquad\qquad\qquad\qquad\quad{\text{``axion-kaon coupling"}}\nonumber\\
   &&=M\,\sqrt{1+\kappa^2}\,\Big|\frac{X_1}{\hat{y}_{\Theta}\,\delta^{\rm G}_1}\Big|\,\quad\qquad\qquad\qquad\text{``see-saw scale"~(see Eq.}\,(\ref{MR2}))\nonumber\\
   &&=|\hat{y}_\nu|^2\,\nabla^{14}_\Psi\,v^2_u\frac{\sqrt{1+\kappa^2}|X_1|}{m_0(M)\,|\hat{y}_\Theta\,\delta^{\rm G}_1|}\,\qquad\text{``active neutrino mass scale"~(see Eq.}\,(\ref{mnu0}))
  \label{emg}
 \end{eqnarray}
 where $C_i=X_i\,{\rm Diag}.(C_{id}, C_{is}, C_{ib})$ is given by Eq.\,(\ref{c_num}), $g_{Ann}$ is the QCD axion coupling to the neutron with the neutron mass $m_n=939.6$ MeV\,\cite{PDG} and the axion to neutron coupling $X_n=-0.02(3)+0.88(3)\frac{\tilde{X}_d}{N_C}-0.39(2)\frac{\tilde{X}_u}{N_C}-0.038(5)\frac{\tilde{X}_s}{N_C}-0.012(5)\frac{\tilde{X}_c}{N_C}-0.009(2)\frac{\tilde{X}_b}{N_C}-0.0035(4)\frac{\tilde{X}_t}{N_C}$
 with $\tilde{X}_q=\delta^{\rm G}_2X_{1q}+\delta^{\rm G}_1X_{2q}$ ($q=u,c,t,d,s,b$) which can be extracted at high precision\,\cite{diCortona:2015ldu}, and the active neutrino mass scale $m_0(M)$ is defined at the seesaw scale $M$. In contrast with the scenario in Refs.\,\cite{Ahn:2018nfb, Ahn:2018cau}, the above correlated relations
mean that QCD axion mass (or the scale of PQ symmetry breakdown $F_A$, or equivalently seesaw scale $M$, or FN scale $\Lambda$) is automatically predicted due the assigned quantum numbers in Table-\ref{DrivingRef}, -\ref{reps_q}, and -\ref{reps_l} once either of axion-electron coupling, axion-photon coupling, axion-neutron coupling, ${\rm Br}(K^+\rightarrow\pi^+A_i)$, and active neutrino mass scale $m_0(M)$ is fixed by experiments and/or observations. And the value of $g_{a\gamma\gamma}$ is fixed by the value of $E/N_C$ in the model via the anomaly-free conditions of both $U(1)_Y\times[U(1)_{\tilde{X}}]^2$ and $U(1)_{\tilde{X}}\times[gravity]^2$, see Eq.\,(\ref{cas}). 
\begin{figure}[h]
\includegraphics[width=11.0cm]{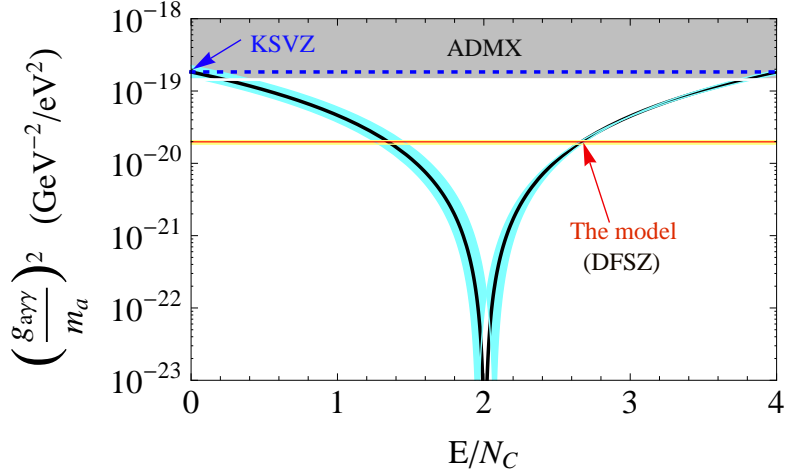}
\caption{\label{Fig2} Plot of $(g_{a\gamma\gamma}/m_{a})^2$ versus $E/N_C$ for $z=0.47$ (black curve) and $0.40<z<0.53$ (cyon-band curve). The gray-band represents the experimentally excluded  bound $(g_{a\gamma\gamma}/m_{a})^2\leq1.44\times10^{-19}\,{\rm GeV}^{-2}\,{\rm eV}^{-2}$ from ADMX\,\cite{Asztalos:2003px, Asztalos:2009yp}. The horizontal red line, and blue dashed line stand for $(g_{a\gamma\gamma}/m_{a})^2=1.98\times10^{-20}\,{\rm GeV}^{-2}\,{\rm eV}^{-2}$ for $E/N_C=8/3$ and $1.84\times10^{-19}\,{\rm GeV}^{-2}\,{\rm eV}^{-2}$ for $E/N_C=0$, respectively, for $z=0.47$. Yellow band represents the model prediction of $(g_{a\gamma\gamma}/m_{a})^2=1.98^{+0.08}_{-0.14}\times10^{-20}\,{\rm GeV}^{-2}\,{\rm eV}^{-2}$ for $E/N_C=8/3$ and $0.40<z<0.53$.}
\end{figure}
 The QCD axion coupling to photon $g_{a\gamma\gamma}$ divided by the QCD axion mass $m_{a}$ is dependent on $E/N_C$.
 Fig.\,\ref{Fig2} shows the $E/N_C$ dependence of $(g_{a\gamma\gamma}/m_{a})^2$ so that the experimental limit is independent of the axion mass $m_{a}$\,\cite{Ahn:2014gva}, where the cyon-band curve stands for the uncertainties of the value of $z$ in Eq.\,(\ref{axipra}) while the solid black curves for $z=0.47$. The gray-band represents the experimentally excluded ADMX (Axion Dark Matter eXperiment) bound $(g_{a\gamma\gamma}/m_{a})^2_{\rm ADMX}\leq1.44\times10^{-19}\,{\rm GeV}^{-2}\,{\rm eV}^{-2}$\,\cite{Asztalos:2003px}. The horizontal blue dashed and red lines stand for $(g_{a\gamma\gamma}/m_{a})^2=1.84\times10^{-19}\,{\rm GeV}^{-2}\,{\rm eV}^{-2}$ for $E/N_C=0$ (KSVZ model\,\cite{KSVZ}), and $1.98\times10^{-20}\,{\rm GeV}^{-2}\,{\rm eV}^{-2}$ for $E/N_C=8/3$ (DFSZ model\,\cite{DFSZ} and Eq.\,(\ref{cas}) predicted in the present model), respectively, for $z=0.47$.
For the {\it updated} Weinberg value Eq.\,(\ref{axipra}),  clearly, the value of $(g_{a\gamma\gamma}/m_{a})^2=1.98^{+0.08}_{-0.14}\times10^{-20}\,{\rm GeV}^{-2}\,{\rm eV}^{-2}$ for the present model of $E/N_C=8/3$ (yellow-band) is located much lower than that of the ADMX bound, while the KSVZ model of $E/N_C=0$ (blue dotted line) lies in $(g_{a\gamma\gamma}/m_{a})^2=(1.55-2.24)\times10^{-20}\,{\rm GeV}^{-2}\,{\rm eV}^{-2}$ showing that its model is excluded by the ADMX bound (gray band). 
 
\begin{figure}[h]
\includegraphics[width=11.0cm]{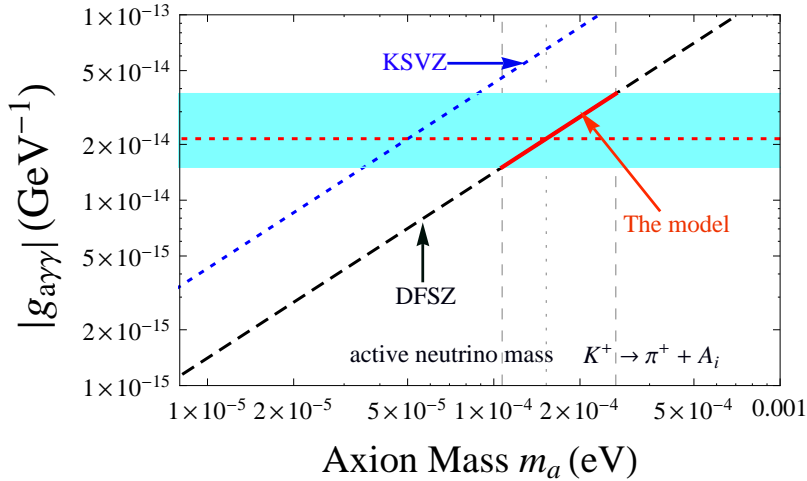}
\caption{\label{Fig3} Plot of $|g_{a\gamma\gamma}|$ versus $m_{a}$ for KSVZ (slanted blue-dotted line), DFSZ (slanted black-dashed line), and the model (``localized" slanted red line) in terms of $E/N_C=$ $0$, $8/3$, and $8/3$, respectively. Especially, the QCD axion mass $m_{a}=1.52^{+1.14}_{-0.46}\times10^{-4}\,\text{eV}$ (vertical black-dashed lines) corresponds to the axion photon coupling $|g_{a\gamma\gamma}|=2.15^{+1.61}_{-0.64}\times10^{-14}\,\text{GeV}^{-1}$ (cyan band), predicted in Eq.\,(\ref{Amass_pre}). Vertical black-dotted and horizontal red-dotted lines correspond to the central values of $m_a$ and $|g_{a\gamma\gamma}|$, respectively.}
\end{figure}

From Eqs.\,(\ref{quarkvalue}) and (\ref{k_bound}) the scale $\Lambda$ responsible for the FN mechanism is obtained through the flavor structure parameters in Eq.\,(\ref{expan_1}) (see also Eq.\,(\ref{m_para})),
 \begin{eqnarray}
\Lambda=1.94^{\,+0.83}_{\,-0.83}\times10^{11}\,{\rm GeV}\,.
  \label{lamScale}
 \end{eqnarray}
Plugging Eq.\,(\ref{k_bound}) into Eq.\,(\ref{emg}) the QCD axion mass is predicted for $z=0.47$ (whose Weinberg value is well in good agreement with the result in Sec.\,\ref{num1}), and subsequently its corresponding couplings of axion-photon and -neutron are obtained:
  \begin{eqnarray}
  &m_a=1.52^{+1.14}_{-0.46}\times10^{-4}\,\text{eV}\,\Leftrightarrow\,|g_{a\gamma\gamma}|=2.15^{+1.61}_{-0.64}\times10^{-14}\,\text{GeV}^{-1}\nonumber\\
 & \Leftrightarrow|g_{Ann}|=1.87^{+1.41}_{-0.56}\times10^{-11}\,.
  \label{Amass_pre}
 \end{eqnarray}
The new plasma-based axion detector suggested by Ref.\,\cite{Lawson:2019brd} can test this prediction for axion mass.
 Its corresponding Compton wavelength of axion oscillation is $\lambda_a=(2{\pi\!\!\not\!h}/m_a)c$ with $c\simeq2.997\times10^{8}\,{\rm m/s}$ and ${\!\!\not\!h}\simeq1.055\times10^{-34}\,{\rm J}\cdot{\rm s}$:
 \begin{eqnarray}
  \lambda_a=8.16^{+3.54}_{-3.50}\,{\rm mm}\,.
 \end{eqnarray}
 Note here that, considering $0.40<z<0.53$ for the given axion decay constant in Eq.\,(\ref{k_bound}), the range of $|g_{a\gamma\gamma}|$ ({\it e.g.} Eq.\,(\ref{Amass_pre})) becomes wider than that for $z=0.47$, see Fig.\,\ref{Fig3}. 
The value of  axion-neutron coupling in Eq.\,(\ref{Amass_pre}) is located in one order of magnitude lower than that of the hint for extra cooling from the neutron star in the supernova remnant ``Cassiopeia A" by axion neutron bremsstrahlung, $g_{Ann}=3.74^{+0.62}_{-0.74}\times10^{-10}$\,\cite{Leinson:2014ioa}. This discrepancy may be explained by considering other means in the cooling of the superfluid core in the neutron star\,\cite{Leinson:2014cja}.
Interestingly, the NA62 experiment is expected to reach the sensitivity of ${\rm Br}(K^+\rightarrow\pi^++A_i)<1.0\times10^{-12}$ in the near future\,\cite{Fantechi:2014hqa}, which is interpreted as the flavored-axion decay constants $F_{ai}$ with $i=1,2$ (see Eq.\,(\ref{agc})) and its corresponding QCD axion decay constant $F_{A}>1.74\times10^{11}$ GeV.
Hence, it is clear that the NA62 experiment will probe the flavored-axions or exclude the present model.

Fig.\,\ref{Fig3} shows the plot for the axion-photon coupling $|g_{a\gamma\gamma}|$ as a function of the axion mass $m_{a}$ in terms of anomaly values $E/N_C=8/3$ for our model (and DFSZ) and $0$ for KSVZ, respectively. Unlike the prediction of DFSZ model on the plot of $|g_{a\gamma\gamma}|$ versus $m_a$, our model prediction on the plot has a localized line, see the red-line of Fig.\,\ref{Fig3}. Especially, in the model with $F_A=3.57^{+1.52}_{-1.53}\times10^{10}$ GeV in Eq.\,(\ref{k_bound}) we obtain the QCD axion mass $m_{a}=1.52^{+1.14}_{-0.46}\times10^{-4}\,\text{eV}$ and the axion photon coupling $|g_{a\gamma\gamma}|=2.15^{+1.61}_{-0.64}\times10^{-14}\,\text{GeV}^{-1}$ (cyan band) for $z=0.47$. 
As the upper bound on ${\rm Br}(K^+\rightarrow\pi^++A_i)$ gets tighter, the range of the QCD axion mass gets narrower, and consequently the corresponding bandwidth on $|g_{a\gamma\gamma}|$ in Fig.\,\ref{Fig3} is getting narrower. In Fig.\,\ref{Fig3}, the top of the band comes from the upper bound on ${\rm Br}(K^+\rightarrow\pi^++A_i)$, while the bottom of the band is from the constraints of active neutrino mass shown in Table-\ref{inp_1} and -\ref{inp_2} induced by the seesaw mechanism.

\section{conclusion}
Motivated by the unsolved theoretical and cosmological issues under the SM, we have started with an undisputed ansatz ``all elementary fields that make up the Universe are charged under the gauge group $G_0=G_f\times G_{\rm SM}$", where $G_f$ stands for a newly introduced gauge group containing (non)-Abelian symmetry as a copy of SM gauge system $G_{\rm SM}=SU(3)_C\times SU(2)_L\times U(1)_Y$. Finding a new mathematical structure leads to the development of a new physical framework and where that framework describes the Universe.

In a flavored-axion framework where $U(1)$ flavored-PQ symmetry is embedded in non-Abelian finite group\,\cite{Ahn:2014gva, Ahn:2016hbn, Ahn:2018nfb, Ahn:2018cau}, we have studied the reasonable requirements of two anomalous $U(1)$s for the anomaly cancellations of both $U(1)$-mixed gravity and $U(1)_Y\times[U(1)]^2$ which in turn determine the $U(1)_Y$ charges uniquely. And we have argued that axion-induced topology in symmetry-broken phases plays crucial roles in describing how quarks and leptons are organized at a fundamental level and make deep connections with each other. Since the particle content of low energy theories is constrained under the $G_F\times G_{\rm SM}$ group where $G_F$ is a subset of the gauge group $G_f$, the anomaly cancellations of both $U(1)$-mixed gravity and $U(1)_Y\times[U(1)]^2$ should be canceled by enlarging particle content together with another $U(1)$ contribution.
In such a flavored-axion framework, the seesaw and PQ mechanisms are naturally triggered. 

We then have proposed an explicit model sewed by the flavor symmetry group $G_F=SL_2(F_3)\times U(1)_X$, where $U(1)_X\equiv U(1)_{X_1}\times U(1)_{X_2}$, by taking the anomaly cancellations of both $U(1)$-mixed gravity and $U(1)_Y\times[U(1)]^2$ into account. In modeling both the ADMX experimental bound and updated Weinberg value of Eq.\,(\ref{axipra}) are considered, leading to a constraint $0<E/N_C<4$ in Eq.\,(\ref{axipra}).
 Flavored-axion interactions with normal matter and the masses and mixings of fermions emerge from the spontaneous breaking of the flavor symmetry.
We have shown a whole spectrum of quarks and leptons in the new parameterization way via Eqs.\,(\ref{m_para}, \ref{expan_1}), see also Refs.\,\cite{Ahn:2018cau, Ahn:2018nfb}, and the model prediction $E/N_C=8/3$ with four axionic domain-walls. 
Consecutively, we have shown how the neutrino sector could be tightly connected to the quark and charged-lepton sectors where the flavored-seesaw is referred to as the seesaw that makes several parameters compact in the flavored-axion framework, showing that the higher-order corrections do not change its flavor structure. 
We have explored numerically, by taking into account the quantum corrections of the Yukawa matrix between the cut-off scale $\Lambda$ and the weak scale, what values of CP phase $\delta_{CP}$, atmospheric mixing angle $\theta_{23}$, and effective Majorana neutrino mass $0\nu\beta\beta${\it-decay rate} can be predicted through RG running effects, as shown in Table-\ref{inp_1} and -\ref{inp_2} for normal mass ordering and inverted one, respectively. 

Finally, we have shown that the fundamental physical parameters, QCD axion decay constant $F_A$, FN scale $\Lambda$, and see-saw scale $M$, can be predicted due to Eq.\,(\ref{AFN}) and the quantum numbers of Table-\ref{DrivingRef}, -\ref{reps_q}, and -\ref{reps_l} since they are complementary to each other, see Eq.\,(\ref{emg}):
 Once a scale of active neutrino mass defined at a seesaw scale is fixed by the commensurate $U(1)$ flavored-PQ charges of fermions, that of $F_A$ is determined, and vice versa. 
Moreover, their related physical observables are precisely determined by precision flavor experiments and astrophysical observations.
The quantum numbers in Table-\ref{DrivingRef}, -\ref{reps_q}, and  -\ref{reps_l} predicts a well-constrained QCD axion mass with the help of both flavor changing process $K^+\rightarrow\pi^++A_i$ induced by the flavored-axions and the active neutrino mass induced by the seesaw mechanism in particle physics. 
Model predictions are shown on the characteristics of flavored-axions: $F_A=3.57^{+1.52}_{-1.53}\times10^{10}$ GeV corresponds to $\Lambda=1.94^{\,+0.83}_{\,-0.83}\times10^{11}\,{\rm GeV}$ (consequently, QCD axion mass $m_a=1.52^{+1.14}_{-0.46}\times10^{-4}$ eV, axion to electron coupling $g_{Aee}=3.29^{+2.47}_{-0.98}\times10^{-14}$, axion to photon coupling $|g_{a\gamma\gamma}|=2.15^{+1.61}_{-0.64}\times10^{-14}\,\text{GeV}^{-1}$, axion to neutron coupling $|g_{Ann}|=1.87^{+1.41}_{-0.56}\times10^{-11}$, and its Compton wavelength $\lambda_a=8.16^{+3.54}_{-3.50}\,{\rm mm}$, where we have used $z=0.47$ whose value is consistent with the numerical one calculated in the quark sector, see Sec.\,\ref{num1}.).

Future searches of $0\nu\beta\beta$ decay and/or flavor-changing rare processes such as $K^+\rightarrow\pi^++A_i$, as well as precise measurements of $\theta_{23}$ and/or axion to electron coupling, will exclude or favor the example model.
\acknowledgments{Ahn thanks Sue subin An for fruitful comments. This work is partly supported by the Chung-Ang University Research Grants in 2019 and partly by Basic Science Research Program through the National Research Foundation of Korea (NRF) funded by the Ministry of Education, Science and Technology (NRF-2019R1A2C2003738).
}

\appendix
\section{Flavon and driving fields configurations and vacuum configuration at leading order}
\label{vacu_0}
Let us consider how the desired vacuum configuration for a compact description of the quark and lepton mixings and masses could be derived.  
\begin{table}[h]
\caption{\label{DrivingRef} Representations of the driving, flavon, and Higgs fields, respectively, under $SL_2(F_{3})\times U(1)_{X}\times U(1)_R$ where $U(1)_X\equiv U(1)_{X_1}\times U(1)_{X_2}$.}
\begin{ruledtabular}
\begin{tabular}{cccccccccccccccccc}
Field &$\Phi^{T}_{0}$&$\Phi^{S}_{0}$&$\Theta_{0}$&$\Psi_{0}$&$\eta_{0}$&\vline\vline&$\Phi_{S}$&$\Phi_{T}$&$\Theta$&$\tilde{\Theta}$&$\Psi$&$\tilde{\Psi}$&$\eta$&\vline\vline&$H_{d}$&$H_{u}$\\
\hline
$SL_2(F_{3})$&$\mathbf{3}$&$\mathbf{3}$&$\mathbf{1}$&$\mathbf{1}$&$\mathbf{2}''$&\vline\vline&$\mathbf{3}$&$\mathbf{3}$&$\mathbf{1}$&$\mathbf{1}$&$\mathbf{1}$&$\mathbf{1}$&$\mathbf{2}'$&\vline\vline&$\mathbf{1}$&$\mathbf{1}$\\
$U(1)_{X}$&$0$&$-2X_1$&$-2X_1$&$0$&$0$&\vline\vline&$X_1$&$0$&$X_1$&$X_1$&$X_2$&$-X_2$&$0$&\vline\vline&$0$&$0$\\
$U(1)_R$&$2$&$2$&$2$&$2$&$2$&\vline\vline&$0$&$0$&$0$&$0$&$0$&$0$&$0$&\vline\vline&$0$&$0$\\
$U(1)_Y$&$0$&$0$&$0$&$0$&$0$&\vline\vline&$0$&$0$&$0$&$0$&$0$&$0$&$0$&\vline\vline&$-\frac{1}{2}$&$\frac{1}{2}$\\
\end{tabular}
\end{ruledtabular}
\end{table}
The representations of the field contents, summarized in Table-\ref{DrivingRef}, under $G_F=SL_2(F_3)\times U(1)_X\times U(1)_R$ are as follows.
Apart from the usual two Higgs doublets $H_{u,d}$ responsible for electroweak symmetry breaking, which are invariant under $SL_2(F_3)$ ({\it i.e.} flavor singlets $\mathbf{1}$), the scalar sector is extended via two types of new scalar multiplets, flavon fields $\Phi_{T},\Phi_{S},\Theta,\tilde{\Theta}, \eta, \Psi, \tilde{\Psi}$ that are SM gauge singlets and driving fields $\Phi^{T}_{0},\Phi^S_{0},\eta_0, \Theta_{0},\Psi_{0}$ that are to break the flavor group along required VEV directions and to allow the flavons to get VEVs, which couple only to the flavons: we take the flavon fields $\Phi_{T},\Phi_{S}$ to be $SL_2(F_3)$ triplets, $\eta$ to be a $SL_2(F_3)$ doublet ($\mathbf{2}'$ representation), and $\Theta,\tilde{\Theta},\Psi,\tilde{\Psi}$ to be $SL_2(F_3)$ singlets ($\mathbf{1}$ representation), respectively, that are SM gauge singlets, and driving fields $\Phi_{0}^{T},\Phi_{0}^{S}$ to be $SL_2(F_3)$ triplets, $\eta_0$ to be a $SL_2(F_3)$ doublet ($\mathbf{2}''$ representation) and $\Theta_{0}, \Psi_{0}$ to be $SL_2(F_3)$ singlets.

 The flavon fields $\{\Phi_S,\Theta,\tilde{\Theta}\}$ have $X_1$ charge, and $\{\Phi^S_0, \Theta_0\}$ have $ -2X_1$ charge under $U(1)_{X_1}$; the field $\Psi$ ($\tilde{\Psi})$ is $X_2 (-X_2)$ charged under $U(1)_{X_2}$. For vacuum stability and a desired vacuum alignment solution, we enforce $\{\Phi_T, \eta\}$ to be neutral under $U(1)_X$. And the others $H_{u,d}$, $\Phi^T_0, \eta_0$, and $\Psi_0$ are neutral under $U(1)_X$.
 Moreover, the superpotential $W$ in the theory is uniquely determined by the $U(1)_R$ symmetry, containing the usual $R$-parity as a subgroup: $\{matter\,fields\rightarrow e^{i\xi/2}\,matter\,fields\}$ and $\{driving\,fields\rightarrow e^{i\xi}\,driving\,fields\}$, with $W\rightarrow e^{i\xi}W$, whereas flavon and Higgs fields remain invariant under an $U(1)_R$ symmetry. As a consequence of the $R$ symmetry, the other superpotential term $\kappa_{\alpha}L_{\alpha}H_{u}$ and the terms violating the lepton and baryon number symmetries are not allowed. Besides, dimension 6 supersymmetric operators like $Q_{i}Q_{j}Q_{k}L_{l}$ ($i,j,k$ must not all be the same) are not allowed either, and stabilizing proton. 
 Here the global $U(1)$ symmetry is a remnant of the broken $U(1)$ gauge symmetry which can connect string theory with flavor physics\,\cite{Ahn:2016typ, Ahn:2016hbn}.

The superpotential dependent on the driving fields having $U(1)_R$ charge $2$, which is invariant under  $G_{\rm SM}\times U(1)_{X}\times SL_2(F_{3})$, is given at leading order by
\begin{eqnarray}
 W_v&=&\Phi^T_0(\mu_T\Phi_T+g_{T_1}\Phi_T\Phi_T+g_{T_2}\eta\eta)+\Phi^S_0(g_{S_1}\Phi_S\Phi_S+g_{S_2}\tilde{\Theta}\Phi_S)\nonumber\\
 &+&\Theta_0(g_{\Theta_1}\Phi_S\Phi_S+g_{\Theta_2}\Theta\Theta+g_{\Theta_3}\Theta\tilde{\Theta}+g_{\Theta_4}\tilde{\Theta}\tilde{\Theta})\nonumber\\
 &+&\Psi_0\{g_{\Psi_1}\Psi\tilde{\Psi}+g_{\Psi_2}\Phi_T\Phi_T-\mu^2_\Psi\}
 +\eta_0(\mu_\eta\eta+g_\eta\eta\Phi_T)+...\,,
  \label{super_d}
\end{eqnarray}
where dots stand for the higher-dimensional operators of Eq.\,(\ref{vah}), $\mu_{i=T,\Psi, \eta}$ are dimensional parameters, and $g_{T_i, S_i, \Theta_i, \Psi_i, \eta}$ are dimensionless coupling constants. At this level there is no fundamental distinction between the $\Theta$ field and $\tilde{\Theta}$ in the superpotential $W_v$ so that the field $\tilde{\Theta}$ is defined as the combination that couples to $(\Phi^S_0\Phi_S)$\,\cite{Altarelli:2005yx}.

It is important to show a non-trivial vacuum configuration at leading order for flavor physics.
In supersymmetric (SUSY) limit, the vacuum configuration is obtained by the $F$-terms of all fields being required to vanish. The vacuum alignments of the flavons $\Phi_{T}$ and $\eta$ are determined by
 \begin{eqnarray}
 \frac{\partial W_{v}}{\partial\Phi^{T}_{01}}&=&\mu_T\,\Phi_{T1}+\frac{2g_{T_1}}{3}\left(\Phi^{2}_{T1}-\Phi_{T2}\Phi_{T3}\right)+ig_{T_2}\,\eta^2_1=0\,,\nonumber\\
 \frac{\partial W_{v}}{\partial\Phi^{T}_{02}}&=&\mu_T\,\Phi_{T3}+\frac{2g_{T_1}}{3}\left(\Phi^{2}_{T2}-\Phi_{T1}\Phi_{T3}\right)+g_{T_2}\,(1-i)\eta_1\eta_2=0\,,\nonumber\\
 \frac{\partial W_{v}}{\partial\Phi^{T}_{03}}&=&\mu_T\,\Phi_{T2}+\frac{2g_{T_1}}{3}\left(\Phi^{2}_{T3}-\Phi_{T1}\Phi_{T2}\right)+g_{T_2}\,\eta^2_2=0\,, \label{potential1}
  \end{eqnarray}
  \begin{eqnarray}
 \frac{\partial W_{v}}{\partial\eta_{01}}&=&\mu_\eta\,\eta_{2}+\frac{5g_\eta}{6}\Big(\frac{1-i}{2}\eta_{2}\Phi_{T1}+i\eta_1\Phi_{T3}\Big)=0\,,\nonumber\\
 \frac{\partial W_{v}}{\partial\eta_{02}}&=&-\mu_\eta\,\eta_{1}+\frac{5g_\eta}{6}\Big(\frac{1-i}{2}\eta_{1}\Phi_{T1}+i\eta_2\Phi_{T2}\Big)=0\,.
 \label{potential01}
 \end{eqnarray}
From this set of five equations, we can obtain the supersymmetric vacua for $\Phi_{T}$ and $\eta$
 \begin{eqnarray}
\langle\Phi_{T}\rangle&=&\Big(\frac{v_{T}}{\sqrt{2}},\,0,\,0\Big)\,,\qquad \text{with}\,\, \mu_{T}=-g_{T_1}\frac{\sqrt{2}}{3}v_T-i\frac{g_{T_2}}{\sqrt{2}}\frac{v^2_\eta}{v_T}\,,\nonumber\\
\langle\eta\rangle&=&\Big(\pm\frac{v_{\eta}}{\sqrt{2}},\,0\Big)\,,\qquad~\, \text{with}\,\, \mu_{\eta}=g_\eta\,\frac{v_T}{\sqrt{2}}\frac{5(1-i)}{12}\,,
 \label{vevdirection1}
 \end{eqnarray}
where $g_{T_{1, 2}}$ and $g_\eta$ are dimensionless couplings, and $v_T$ and $v_\eta$ are not determined.
And the minimization equations for the vacuum configuration of $\Phi_{S}$ and $(\Theta,\tilde{\Theta})$ are given by
 \begin{eqnarray}
 \frac{\partial W_{v}}{\partial\Phi^{S}_{01}}&=&\frac{2g_{S_1}}{3}\left(\Phi_{S1}\Phi_{S1}-\Phi_{S2}\Phi_{S3}\right)+g_{S_2}\Phi_{S1}\tilde{\Theta}=0\,,\nonumber\\
 \frac{\partial W_{v}}{\partial\Phi^{S}_{02}}&=&\frac{2g_{S_1}}{3}\left(\Phi_{S2}\Phi_{S2}-\Phi_{S1}\Phi_{S3}\right)+g_{S_2}\Phi_{S3}\tilde{\Theta}=0\,,\nonumber\\
 \frac{\partial W_{v}}{\partial\Phi^{S}_{03}}&=&\frac{2g_{S_1}}{3}\left(\Phi_{S3}\Phi_{S3}-\Phi_{1}\Phi_{S2}\right)+g_{S_2}\Phi_{S2}\tilde{\Theta}=0\,,\nonumber\\
 \frac{\partial W_{v}}{\partial\Theta_{0}}&=&g_{\Theta_1}\left(\Phi_{S1}\Phi_{S1}+2\Phi_{S2}\Phi_{S3}\right)+g_{\Theta_2}\Theta^{2}+g_{\Theta_3}\Theta\tilde{\Theta}+g_{\Theta_4}\tilde{\Theta}^{2}=0\,.
 \label{potential2}
 \end{eqnarray}
From the above set of four equations we can get the supersymmetric vacua for the fields $\Phi_{S},\Theta,\tilde{\Theta}$:
 \begin{eqnarray}
 \langle\Phi_{S}\rangle=\frac{1}{\sqrt{2}}\left(v_{S},v_{S},v_{S}\right)\,,\quad\langle\Theta\rangle=\frac{v_{\Theta}}{\sqrt{2}}\,,\quad\langle\tilde{\Theta}\rangle=0\,,\qquad\text{with}\,\,v_{\Theta}=v_{S}\sqrt{-3\frac{g_{\Theta_1}}{g_{\Theta_2}}}\,,
 \label{vevdirection2}
 \end{eqnarray}
where $v_{\Theta}$ is undetermined. The flat direction $\langle\tilde{\Theta}\rangle=0$ is guaranteed by adding to the scalar potential a soft SUSY breaking mass term
for the scalar field $\tilde{\Theta}$ with $m^2_{\tilde{\Theta}}>0$. The VEVs $v_{\Theta}$ and $v_{S}$ are naturally of the same order of magnitude (the dimensionless parameters $g_{\Theta_1}$ and $g_{\Theta_2}$ are the same order of magnitude) and its flat direction is simply parametrized as $v_S/v_\Theta=\kappa$.
Finally, the minimization equation for the vacuum configuration of $\Psi(\tilde{\Psi}$) is given by
 \begin{eqnarray}
 \frac{\partial W_{v}}{\partial\Psi_{0}}=g_{\Psi_1}\Psi\tilde{\Psi}+g_{\Psi_2}(\Phi^2_{T_1}+2\Phi_{T_2}\Phi_{T_3})-\mu^{2}_{\Psi}=0\,,
 \label{potential3}
 \end{eqnarray}
where $g_{\Psi_i}$ are dimensionless couplings.
From the above equation together with Eq.\,(\ref{vevdirection1}) we can get the supersymmetric vacua for the fields $\Psi,\tilde{\Psi}$:
 \begin{eqnarray}
 \langle\Psi\rangle=\langle\tilde{\Psi}\rangle=\frac{v_{\Psi}}{\sqrt{2}}\,,\quad\text{with}~\mu_\Psi=\sqrt{\frac{g_{\Psi_1}v^2_\Psi+g_{\Psi_2}v^2_T}{2}}\,.
 \label{vevdirection3}
 \end{eqnarray}
The scale of dimensionful parameters $\mu_T, \mu_\eta$ and $\mu_\Psi$ roughly lies in the same order of scale of the $G_F$ symmetry breakdown due to the associated dimensionless parameters being order of unity.

\section{Higher order corrections}
\label{vacu_l}
Considering higher-dimensional operators induced by $\Phi_{T},\Phi_{S},\Theta, \Psi, \tilde{\Psi}, \eta$ invariant under $SL_{2}(F_3)\times U(1)_X$ in the driving superpotential $W_{v}$, which are suppressed by additional powers of the cut-off scale $\Lambda$, they can lead to small deviations from the leading order vacuum configurations of Eq.\,(\ref{vev}).
The next leading order superpotential $\delta W_{v}$, which is linear in the driving fields and invariant under $SL_{2}(F_3)\times U(1)_{X}\times U(1)_R$, is given by
\begin{eqnarray}
\delta W_v=\frac{1}{\Lambda}\Big\{\sum^3_{i=1}a_iI^T_i+\sum^{6}_{i=1}b_iI^S_i+\sum^2_{i=1}c_iI^{\Theta}_i+\sum^2_{i=1}d_iI^\Psi_i+\sum^3_{i=1}e_iI^\eta_i\Big\}
\label{vah}
\end{eqnarray}
with the dimensionless parameters $a_i, b_i, c_i, d_i, e_i$ and 
\begin{eqnarray}
I^T_1&=&(\Phi^{T}_{0}\Phi_{T}\Phi_{T}\Phi_{T})_{{\bf 1}}\,,\quad\qquad I^T_2=(\Phi^{T}_{0}\Phi_{T}\Psi\tilde{\Psi})_{{\bf 1}}\,,~\quad\qquad I^T_3=(\Phi^{T}_0\Phi_{T}\eta\eta)_{{\bf 1}}\,,\\
I^S_1&=&(\Phi^{S}_{0}\Phi_{T}\Phi_{S}\Phi_{S})_{{\bf 1}}\,,\quad\qquad
 I^S_2=(\Phi^{S}_{0}\Phi_{T}\Phi_{S}\Theta)_{{\bf 1}}\,,~\quad\qquad
  I^S_3=(\Phi^{S}_{0}\Phi_{T}\Phi_{S}\tilde{\Theta})_{{\bf 1}}\,,\nonumber\\
I^S_{4}&=&(\Phi^{S}_{0}\Phi_{T}\Theta\Theta)_{{\bf 1}}\,,\qquad\qquad I^S_{5}=(\Phi^{S}_{0}\Phi_{T}\Theta\tilde{\Theta})_{{\bf 1}}\,,\qquad\qquad I^S_{6}=(\Phi^{S}_{0}\Phi_{T}\tilde{\Theta}\tilde{\Theta})_{{\bf 1}}\,,\\
I^\Theta_1&=&(\Theta_{0}\Phi_{T}\Phi_{S}\Phi_{S})_{{\bf 1}}\,,\quad\qquad I^\Theta_2=(\Theta_{0}\Phi_{T}\Phi_{S}\tilde{\Theta})_{{\bf 1}}\,,\\
I^\Psi_1&=&(\Psi_{0}\Phi_{T}\Phi_{T}\Phi_{T})_{{\bf 1}}\,,\quad\qquad I^\Psi_2=(\Psi_{0}\Phi_{T}\eta\eta)_{{\bf 1}}\,,\\
I^\eta_1&=&(\eta_{0}\eta\eta\eta)_{{\bf 1}}\,,~\,\quad\qquad\qquad I^\eta_2=(\eta_{0}\eta\Phi_{T}\Phi_{T})_{{\bf 1}}\,,\qquad\qquad
I^\eta_3=(\eta_{0}\eta\Psi\tilde{\Psi})_{{\bf 1}}\,.
\end{eqnarray}
The corrections to the VEVs, Eq.\,(\ref{vev}), are of relative order $1/\Lambda$ and affect the flavon fields $\Phi_{S}$, $\Phi_{T}$, $\Theta$, $\tilde{\Theta}$, $\eta$ and $\Psi$, and the vacuum configuration can be modified with relations among the dimensionless parameters ($a_{1}...a_{3}$, $b_{1}...b_{6}$, $c_{1}, c_{2}$, $d_{1}$, $d_{2}$, $e_{1}...e_{3}$).
The corrected minimum for the flavon fields $\Phi_{T}$, $\Theta$, $\tilde{\Theta}$, $\eta$, $\Psi$, and $\Phi_{S}$ is obtained by searching for the vanishing of the $F$ terms, the first derivatives of $W_v+\delta W_v$, associated to the driving fields $\Phi^T_0, \Theta_0, \Psi_0$, $\eta_0$, and $\Phi^S_0$:
  \begin{eqnarray}
  &\langle\Phi_T\rangle=\left(\frac{v_T}{\sqrt{2}}+\delta v_{T_1}, \delta v_{T_2}, \delta v_{T_3}\right),\quad \langle \eta\rangle=\left(\frac{v_\eta}{\sqrt{2}}+\delta v_{\eta_1}, \delta v_{\eta_2}\right),\quad\langle \Psi\rangle=\frac{v_\Psi}{\sqrt{2}}+\delta v_{\Psi},\nonumber\\
   &\langle \Phi_S\rangle= \left(\frac{v_S}{\sqrt{2}}+\delta v_{S_1}, \frac{v_S}{\sqrt{2}}+\delta v_{S_2}, \frac{v_S}{\sqrt{2}}+\delta v_{S_3}\right),\quad \langle \Theta\rangle= \frac{v_\Theta}{\sqrt{2}}+\delta v_{\Theta},\quad \langle \tilde{\Theta}\rangle= \delta v_{\tilde{\Theta}}
 \label{}
 \end{eqnarray} 
 with $\mu_{T}=-g_{T_1}\frac{\sqrt{2}}{3}v_T-i\frac{g_{T_2}}{\sqrt{2}}\frac{v^2_\eta}{v_T}$, $\mu_{\eta}=g_\eta\,\frac{v_T}{\sqrt{2}}\frac{5(1-i)}{12}$, $v_{\Theta}=v_{S}\sqrt{-3\frac{g_{\Theta_1}}{g_{\Theta_2}}}$, and $\mu_\Psi=\sqrt{\frac{g_{\Psi_1}v^2_\Psi+g_{\Psi_2}v^2_T}{2}}$ in Sec.\,\ref{vac_a}.
By keeping only the first order in the expansion, we obtain the minimization equations: For $\Phi_T$, $\Theta$, $\Psi$ and $\eta$ fields,
 \begin{eqnarray}
  &\delta v_{T_1}\,T_{11}+\delta v_{\eta_1}\,T_{12}+\delta v_\Psi\,T_{13}=T\,v_T\,,\nonumber\\
 &\delta v_{T_3}\,T_{21}+\delta v_{\eta_2}\,T_{22}=0\,,\nonumber\\
 &\delta v_{T_2}=0\,, \label{Npo04}\\
 &\delta v_{T_1}P_{11}+\delta v_{\eta_1} P_{12}+\delta v_{\Psi}P_{13}=P\,v_T\,, \label{Npo01}\\
 &\delta v_{T_3}E_{11}+\delta v_{\eta_2}E_{12}=0\,,\nonumber\\
 &\delta v_{T_1}E_{21}+\delta v_{\eta_1}E_{22}+\delta v_{\Psi}E_{23}=E\,v_\eta\,, \label{Npo02}
 \end{eqnarray}
 where the parameters $T, T_{ij}, P, P_{ij}, E, E_{ij}$ are dimensionless:
 \begin{eqnarray}
  & T_{11}=\frac{\sqrt{2}}{3}g_{T_1}-\frac{ig_{T_2}}{\sqrt{2}}(\frac{\nabla_\eta}{\nabla_T})^2+\frac{3a_1}{\sqrt{2}}\nabla_T+\frac{a_2}{\sqrt{2}}\frac{\nabla^2_\Psi}{\nabla_T}+\frac{a_3(1-i)}{\sqrt{2}}\frac{\nabla^2_\eta}{\nabla_T}\,,\nonumber\\
 &T_{12}=i\sqrt{2}g_{T_2}\frac{\nabla_\eta}{\nabla_T}+\sqrt{2}(1-i)a_3\nabla_\eta\,, \qquad  T_{13}=a_2\sqrt{2}\nabla_\Psi\,,\nonumber\\
 &T_{21}=-\frac{2\sqrt{2}}{3}g_{T_1}-\frac{ig_{T_2}}{\sqrt{2}}(\frac{\nabla_\eta}{\nabla_T})^2+\frac{a_17}{3\sqrt{2}}\nabla_T+\frac{a_2}{\sqrt{2}}\frac{\nabla^2_\Psi}{\nabla_T}\,,\qquad T_{22}=\frac{g_{T_2}(1-i)}{\sqrt{2}}\frac{\nabla_\eta}{\nabla_T}+\sqrt{2}a_3\nabla_\eta\,,\nonumber\\
 &T=-\frac{1}{2}\Big\{a_1\nabla_T+a_2\frac{\nabla^2_\Psi}{\nabla_T}+(1-i)a_3\frac{\nabla^2_\eta}{\nabla_T}\Big\}\,,\\
 &P_{11}=\sqrt{2}g_{\Psi_2}+\sqrt{2}d_1\nabla_T+\frac{i}{\sqrt{2}}d_2\frac{\nabla^2_\eta}{\nabla_T}\,,\quad P_{12}=i\sqrt{2}d_2\nabla_\eta\,,\quad P=-\frac{d_1}{3}\nabla_T-\frac{id_2}{2}\frac{\nabla^2_\eta}{\nabla_T}\,,\\
 &E_{11}=g_\eta\frac{5i}{6\sqrt{2}}\frac{\nabla_\eta}{\nabla_T}-e_2\frac{i\sqrt{2}}{3}\nabla_\eta\,,\qquad E_{12}=g_\eta\frac{5(1-i)}{6\sqrt{2}}-e_2\frac{\sqrt{2}(2+i)}{6}\nabla_T+e_1\frac{3(1+i)\sqrt{2}}{4}\frac{\nabla^2_\eta}{\nabla_T}\,,\nonumber\\
 &E_{21}=g_\eta\frac{5(1-i)}{12\sqrt{2}}+e_2\frac{(4-i)\sqrt{2}}{3}\nabla_T\,,\quad E_{22}=e_1\frac{3\sqrt{2}(1+i)}{4}\nabla_\eta+e_2\frac{(4-i)\sqrt{2}}{6}\frac{\nabla^2_T}{\nabla_\eta}+e_3\frac{1}{\sqrt{2}}\frac{\nabla^2_\Psi}{\nabla_T}\,,\nonumber\\
 &E_{23}=e_3\sqrt{2}\nabla_\Psi\,,\qquad E=-e_1\frac{1+i}{4}\nabla_\eta-e_2\frac{3-2i}{2}\frac{\nabla^2_T}{\nabla_\eta}-e_3\frac{\nabla^2_\Psi}{2\nabla_\eta}\,.
  \label{Npo2}
 \end{eqnarray}
 From the above set of equations we obtain $\delta v_{T_2}=\delta v_{T_3}=0$ and $\delta v_{\eta_2}=0$, and there remain three equations for three unknown parameters $\delta v_{T_1}, \delta v_{\eta_1}$, and $\delta v_{\Psi}$. One can obtain the modified vacuum configurations
  \begin{eqnarray}
  \langle\Phi_T\rangle\rightarrow\left(\frac{v_T}{\sqrt{2}}+\delta v_{T_1}, 0,0\right)\,,\quad \langle \eta\rangle\rightarrow\left(\pm\frac{v_\eta}{\sqrt{2}}+\delta v_{\eta_1}, 0\right)\,,\quad\langle \Psi\rangle\rightarrow \frac{v_\Psi}{\sqrt{2}}+\delta v_{\Psi}\,.
 \label{Npo4}
 \end{eqnarray} 

 For $\Theta$ and $\Phi_S$ fields,
  \begin{eqnarray}
   &\delta v_{S_1}\Theta_{11}+(\delta v_{S_2}+\delta v_{S_3})\Theta_{12}+\delta v_\Theta\Theta_{13}+\delta v_{\tilde{\Theta}}\Theta_{14}=0\,, \label{Npo03}\\
 &\delta v_{S_1}\,S_{11}+(\delta v_{S_2}+\delta v_{S_3})S_{12}+\delta v_\Theta\,S_{17}+\delta v_{\tilde{\Theta}}\,S_{18}=S_{a}\,v_S-\delta v_{T_1}\,S_{14}\,,\nonumber\\
 &\delta v_{S_1}\,S_{21}+\delta v_{S_2}\,S_{22}+\delta v_{S_3}\,S_{23}+\delta v_\Theta\,S_{27}+\delta v_{\tilde{\Theta}}\,S_{28}=S_b\,v_S-\delta v_{T_1}\,S_{24}\,,\nonumber\\
  &\delta v_{S_1}\,S_{31}+\delta v_{S_2}\,S_{32}+\delta v_{S_3}\,S_{33}+\delta v_\Theta\,S_{37}+\delta v_{\tilde{\Theta}}\,S_{38}=S_c\,v_S-\delta v_{T_1}\,S_{34}\,, \label{Npo05}
 \end{eqnarray}
 where the dimensionless parameters $\Theta_{ij}$ are given by
  \begin{eqnarray}
   &\Theta_{11}=g_{\Theta_1}+c_1\frac{2\sqrt{2}}{3}\nabla_T\,,\qquad \Theta_{12}=g_{\Theta_1}-c_1\frac{\sqrt{2}}{3}\nabla_T\,,\nonumber\\
   & \Theta_{13}=g_{\Theta_2}\frac{\sqrt{2}}{\kappa}\,,\qquad \Theta_{14}=g_{\Theta_3}\frac{1}{\kappa\sqrt{2}}\,,\\
   &S_{11}=\frac{2\sqrt{2}}{3}g_{S_1}+g_{S_2}\frac{1}{\kappa\sqrt{2}}+b_1\frac{5\sqrt{2}}{3}\nabla_T+b_2\frac{\sqrt{2}}{3\kappa}\nabla_T\,,\nonumber\\
 &S_{12}=-\frac{\sqrt{2}}{3}g_{S_1}+b_1\frac{2\sqrt{2}}{3}\nabla_T\,,\qquad S_{17}=g_{S_2}\frac{1}{\sqrt{2}}+b_2\frac{\sqrt{2}}{3}\nabla_T+b_4\frac{\sqrt{2}}{\kappa}\nabla_T\,,\nonumber\\
 &S_{18}=g_{S_2}\frac{1}{\sqrt{2}}+b_3\frac{\sqrt{2}}{3}\nabla_T+b_5\frac{1}{\kappa\sqrt{2}}\nabla_T\,,\quad S_{14}=\sqrt{2}\nabla_S\Big(b_1\frac{3}{2}+b_2\frac{1}{3\kappa}+b_4\frac{1}{2\kappa^2}\Big)\,,\nonumber\\
 &S_{a}=-g_{S_2}\frac{1}{2\kappa}-b_1\frac{3}{2}\nabla_T-b_2\frac{1}{3\kappa}\nabla_T-b_4\frac{1}{2\kappa^2}\nabla_T\,,\\
 & S_{21}=-\frac{\sqrt{2}}{3}g_{S_1}+b_1\sqrt{2}\nabla_T\,,\qquad~~ S_{22}=\frac{2\sqrt{2}}{3}g_{S_1}+b_1\sqrt{2}\nabla_T\,,\nonumber\\
 &S_{23}=-\frac{\sqrt{2}}{3}g_{S_1}+g_{S_2}\frac{1}{\kappa\sqrt{2}}+b_1\sqrt{2}\nabla_T+b_{2}\frac{1}{6\sqrt{2}\kappa}\nabla_T\,,\nonumber\\
  & S_{27}=\frac{1}{\sqrt{2}}g_{S_2}+b_2\frac{\sqrt{2}}{12}\nabla_T\,,\quad S_{28}=\frac{1}{\sqrt{2}}g_{S_2}-b_3\frac{5\sqrt{2}}{12}\nabla_T\,,\quad S_{24}=\nabla_S\Big(b_1\frac{3}{\sqrt{2}}+b_2\frac{1}{6\sqrt{2}\kappa}\Big)\,,\nonumber\\
  &S_{b}=-g_{S_2}\frac{1}{2\kappa}-b_1\frac{3}{2}\nabla_T-b_2\frac{1}{12\kappa}\nabla_T\,,\\
 &S_{31}=-\frac{\sqrt{2}}{3}g_{S_1}+b_1\frac{4\sqrt{2}}{3}\nabla_T\,,\nonumber\\
 &S_{32}=-\frac{\sqrt{2}}{3}g_{S_1}+g_{S_2}\frac{1}{\sqrt{2}\kappa}+b_1\frac{4\sqrt{2}}{3}\nabla_T-b_{2}\frac{5}{6\sqrt{2}\kappa}\nabla_T\,,\nonumber\\
 &S_{33}=\frac{2\sqrt{2}}{3}g_{S_1}+b_1\frac{\sqrt{2}}{3}\nabla_T\,,\quad S_{37}=\frac{1}{\sqrt{2}}g_{S_2}-b_2\frac{5\sqrt{2}}{12}\nabla_T\,,\quad S_{37}=\frac{1}{\sqrt{2}}g_{S_2}+b_2\frac{\sqrt{2}}{12}\nabla_T\,,\nonumber\\
  &S_{34}=\nabla_S\Big(b_1\frac{3}{\sqrt{2}}-b_2\frac{5}{6\sqrt{2}\kappa}\Big)\,,\nonumber\\
    &S_{c}=-g_{S_2}\frac{1}{2\kappa}-b_1\frac{3}{2}\nabla_T-b_2\frac{5}{12\kappa}\nabla_T\,,
 \label{Npo3}
 \end{eqnarray}
where $1/\kappa=\sqrt{-3g_{\Theta_1}/g_{\Theta_2}}$ is used.
Taking $\delta v_{S_2}=-\delta v_{S_3}$, four equations in Eqs.\,(\ref{Npo03}, \ref{Npo05}) have four unknown parameters $\delta v_{S_{1,2,3}}, \delta v_\Theta$ and $\delta v_{\tilde{\Theta}}$. Then one obtains
  \begin{eqnarray}
  \langle \Phi_S\rangle\rightarrow \left(\frac{v_S}{\sqrt{2}}+\delta v_{S_1}, \frac{v_S}{\sqrt{2}}+\delta v_{S_2}, \frac{v_S}{\sqrt{2}}-\delta v_{S_2}\right)\,,\quad \langle \Theta\rangle\rightarrow \frac{v_\Theta}{\sqrt{2}}+\delta v_{\Theta}\,,\quad \langle \tilde{\Theta}\rangle\rightarrow \delta v_{\tilde{\Theta}}\,.
 \label{Npo5}
 \end{eqnarray} 
Given the values for $\nabla_Q$ with $Q=\eta,S,T,\Theta,\Psi$ in Eq.\,(\ref{quarkvalue}), one can expect that the shifts $|\delta v_{\tilde{\Theta}}|/v_{\Theta}, |\delta v_\Theta|/v_{\Theta},|\delta v_{S_i}|/v_S,|\delta v_{T_i}|/v_T, |\delta v_{\eta_i}|/v_\eta, |\delta v_{\Psi}|/v_\Psi$ can be kept small enough, below a few percent level. The crucial thing about the new VEVs of Eqs.\,(\ref{Npo4}, \ref{Npo5}) is that they do not change the flavor structures of quark and lepton in Eqs.\,(\ref{Ch2}, \ref{Ch1}, \ref{ChL1}, \ref{MR1}, \ref{Ynu1}).

\section{Quark masses and Mixings}
\label{quark_l}
Form the up-type quark mass matrix of Eq.\,(\ref{Ch1}) its left-handed mixing matrix $V^u_L$ reads at leading order
 \begin{eqnarray}
 V^{u}_L={\left(\begin{array}{ccc}
 1-\frac{1}{2}\theta^2_u & \theta_u\,e^{i\phi_u} & 0  \\
-\theta_u\,e^{-i\phi_u} & 1-\frac{1}{2}\theta^2_u & 0   \\
 0 & 0  & 1
 \end{array}\right)}+{\cal O}(\theta^3_u)\,,
 \label{u_qm}
 \end{eqnarray}
where $\theta_u\simeq|({\cal M}_u)_{21}|/|({\cal M}_u)_{22}|= \frac{\kappa}{\sqrt{2}}\frac{|\hat{y}^\ast_u|}{|\hat{y}_c+\frac{1-i}{2}\hat{y}_{\tilde{c}}\nabla_T|}\nabla_\Theta\,\nabla_\Psi$ and $\phi_u=\pi/8+\frac{1}{2}\arg(\hat{y}^\ast_u)+\frac{1}{2}\arg(\hat{y}_c+\frac{1-i}{2}\hat{y}_{\tilde{c}}\nabla_T)$.
Then the up-type quark masses are simply written in a good approximation by
\begin{eqnarray}
m_t&=&|\hat{y}_t|v_u\,,\nonumber\\
 m_c&=&\big|\hat{y}_c+\hat{y}_{\tilde{c}}\frac{1-i}{2}\nabla_T\big|(1+2\theta^2_u)\,\nabla_\eta\nabla^7_\Psi\,v_u+{\cal O}(\theta^3_u)v_u\,,\nonumber\\
 m_u&=&\big|i\hat{y}_u\kappa+i\hat{y}_{\tilde{u}}\nabla_T-\hat{y}_{\bar{u}}\nabla^2_\eta\nabla_\Theta\,\nabla^8_\Psi\nabla_\eta\,v_u+{\cal O}(\theta^3_u)v_u\,.
  \label{qum}
\end{eqnarray}
 Form the down-type quark mass matrix of Eq.\,(\ref{Ch2}) its left-handed mixing matrix $V^d_L$ reads in the parametrization of Ref.\,\cite{Ahn:2011yj} at leading order
  \begin{eqnarray}
 V^{d}_L={\left(\begin{array}{ccc}
 1-\frac{1}{2}\tilde{\lambda}^2 & \tilde{\lambda}\,e^{i\phi^d_3} & B_d\tilde{\lambda}^3\,e^{i\phi^d_2}  \\
-\tilde{\lambda}\,e^{-i\phi^d_3}& 1-\frac{1}{2}\tilde{\lambda}^2 &  A_d\tilde{\lambda}^2\,e^{i\phi^d_1}    \\
 \tilde{\lambda}^3(A_d\,e^{-i(\phi^d_1+\phi^d_3)}-B_d\,e^{-i\phi^d_2}) & -A_d\tilde{\lambda}^2\,e^{-i\phi^d_1}  & 1
 \end{array}\right)}+{\cal O}(\lambda^4)\,,
 \label{d_qm}
 \end{eqnarray}
 where we have set $\theta^d_{12}\equiv\tilde{\lambda}$, $\theta^d_{13}\equiv B_d\tilde{\lambda}^3$ and $\theta^d_{23}\equiv A_d\tilde{\lambda}^2$ with $\tilde{\lambda}\approx\lambda$, and its corresponding mixing angles are expressed in a good approximation as
 \begin{eqnarray}
\theta^d_{12}&=&\frac{\kappa}{\sqrt{2}} \Big|\frac{\hat{y}^\ast_d}{\hat{y}_s}\Big|\,\Big|\frac{X_2\delta^{\rm G}_1}{X_1\delta^{\rm G}_2}\sqrt{\frac{2}{1+\kappa^2}}\Big|\,,\nonumber\\
 \theta^d_{23}&=&\kappa\frac{|3\kappa\hat{Y}_{s}+\hat{Y}_{\tilde{s}}\nabla_T|}{|\hat{y}_b|}\Big|\frac{X_2\delta^{\rm G}_1}{X_1\delta^{\rm G}_2}\sqrt{\frac{2}{1+\kappa^2}}\Big|^2\,\nabla^5_\Psi\,,\nonumber\\
\theta^d_{13}&=&\kappa\frac{|3\kappa\hat{Y}^\ast_{d}+\hat{Y}^\ast_{\tilde{d}}\nabla_T|}{|\hat{y}_b|}\,\Big|\frac{X_2\delta^{\rm G}_1}{X_1\delta^{\rm G}_2}\sqrt{\frac{2}{1+\kappa^2}}\Big|^3\,\nabla^5_\Psi\,,
\label{qckm}
 \end{eqnarray}
 recalling that $\hat{Y}_{\tilde{d}}=\hat{Y}_{d1}+3\kappa^2\hat{Y}_{d2}$ with $|\hat{Y}_{di}|\sim{\cal O}(1)$.
Subsequently, the down-type quark masses are obtained in a good approximation as
 \begin{eqnarray}
 m_b=|\hat{y}_b|\,\nabla^2_\Psi\,v_d\,,\quad
 m_s=|\hat{y}_s|\,\nabla^5_\Psi\nabla_\eta\,v_d\,,\quad
 m_d=2\kappa\,|\hat{y}_d\cos\phi_d|\nabla_\Theta\,\nabla^4_\Psi\nabla_\eta\,v_d\,,
 \label{qm}
 \end{eqnarray}
 where
$\nabla_T=\kappa\big|\hat{y}_d/\hat{y}_{\tilde{d}}\big|$ and $\phi_{\tilde{d}}=-\phi_d$
with $\arg(\hat{y}_i)=\phi_i$ ($i=d, \tilde{d}$) are used for fitting the empirical results in Ref\,\cite{PDG}. 
And the ratio of electroweak Higgs field VEVs $\langle H_u\rangle/\langle H_d\rangle$ is approximately given in terms of the PDG value by $\tan\beta\simeq(m_t/m_b)_{\rm PDG}|\hat{y}_b/\hat{y}_t|\,\nabla^2_\Psi$. Setting $\phi_u=\phi^d_3$ and redefining the quark fields $c\rightarrow e^{i(\phi^d_3+\pi)}\,c$, $t\rightarrow e^{i(\phi^d_1+\phi^d_3)}\,t$, $s\rightarrow e^{i(\phi^d_3+\pi)}\,s$, and $b\rightarrow e^{i(\phi^d_1+\phi^d_3)}\,b$, we obtain
   \begin{eqnarray}
 V_{\rm CKM}={\left(\begin{array}{ccc}
 1-\frac{1}{2}\tilde{\lambda}^2+\theta_u\tilde{\lambda}  &\tilde{\lambda}- \theta_u &  \tilde{\lambda}^3(A_d-B_d\,e^{-i(\phi^d_1-\phi^d_2+\phi^d_3)})  \\
 \theta_u-\tilde{\lambda} & 1-\frac{1}{2}\tilde{\lambda}^2+\theta_u\tilde{\lambda} &  A_d\tilde{\lambda}^2+B_d\tilde{\lambda}^4\,e^{-i(\phi^d_1-\phi^d_2+\phi^d_3)}    \\
 B_d\tilde{\lambda}^3\,e^{i(\phi^d_1-\phi^d_2+\phi^d_3)}& -A_d\tilde{\lambda}^2  & 1
 \end{array}\right)}
 +{\cal O}(\theta^2_u;\theta_u\tilde{\lambda}^2),
 \label{ckm_1}
 \end{eqnarray}
 where $\theta_u$ can be rewritten with the help of $\theta^d_{12}$ in Eq.\,(\ref{qckm}) as $\theta_u\simeq\tilde{\lambda}\,x_u$ with $x_u=\nabla^2_\Psi\,(|\hat{y}^\ast_u|/|\hat{y}_c+\frac{1-i}{2}\hat{y}_{\tilde{c}}\nabla_T|)(|\hat{y}_s|/|\hat{y}^\ast_d|)$.
In order to obtain the CKM matrix in the Wolfenstein parametrization\,\cite{Wolfenstein:1983yz} we redefine $A_d\equiv A$, $B_d\equiv A\sqrt{\rho^2+\eta^2}$, and $\tilde{\lambda}-\theta_u\equiv\lambda$.
Then the CKM $CP$ phase is given by
 \begin{eqnarray}
 \delta^{q}_{CP}\equiv\phi^{d}_{1}-\phi^{d}_{2}+\phi^{d}_{3}=\tan^{-1}\left(\eta/\rho\right)\,,
  \label{cpDirac}
 \end{eqnarray}
 where each $\phi^d_i$ is expressed in terms of the components of Yukawa matrices Eqs.\,(\ref{Ch2}, \ref{Ch1})
  \begin{eqnarray}
  \phi^{d}_{1}&=&\frac{1}{2}\arg(3\hat{Y}^\ast_{s}\kappa+\hat{Y}^\ast_{\tilde{s}}\nabla_T)\,,\nonumber\\
 \phi^{d}_{2}&\simeq&\frac{1}{2}\arg(3\hat{Y}^\ast_{d}\kappa+\hat{Y}^\ast_{\tilde{d}}\nabla_T)-\phi^{d}_{1}/2\,, \nonumber\\
 \phi^{d}_{3}&\simeq&\frac{1}{2}\arg(\hat{y}^\ast_d\hat{y}_s)+\phi^{d}_{1}/2-\phi^{d}_{2}/2+\pi/8\,,
   \label{cpDirac0}
 \end{eqnarray}
with\,\footnote{The condition $\phi^d_3=\phi_u$ connects the phases of up- and down-type quark Yukawa couplings, {\it i.e.}, $\arg(\hat{y}_c+\frac{1-i}{2}\hat{y}_{\tilde{c}}\nabla_T)+\arg(\hat{y}^\ast_{u})-\arg(\hat{y}^\ast_d\hat{y}_s)=\frac{3}{4}\arg(3\hat{Y}^\ast_s\kappa+\hat{Y}^\ast_{\tilde{s}}\nabla_T)-\frac{1}{2}\arg(3\hat{Y}^\ast_d\kappa+\hat{Y}^\ast_{\tilde{d}}\nabla_T)$ in a good approximation, reproducing the empirical results of quark masses and their mixing angles and a CP phase. See Sec.\,\ref{num1}.} $\phi^d_3=\phi_u$ in Eq.\,(\ref{u_qm}); without loss of generality, we have set $\arg(\hat{y}_b)=0$.  
For $x_u\lesssim\lambda^2$ the CKM matrix reads
 \begin{eqnarray}
 V_{\rm CKM}={\left(\begin{array}{ccc}
 1-\frac{1}{2}\lambda^2 & \lambda & A\lambda^3(\rho-i\eta)  \\
-\lambda & 1-\frac{1}{2}\lambda^2 & A\lambda^2   \\
 A\lambda^3(1-\rho-i\eta) & -A\lambda^2  & 1
 \end{array}\right)}+{\cal O}(\lambda^4)\,,
 \label{ckm0}
 \end{eqnarray}
in the Wolfenstein parametrization\,\footnote{Equivalently, it could be expressed as the Qin-Ma parametrization\,\cite{Qin:2011ub}}\,\cite{Wolfenstein:1983yz} and at higher precision\,\cite{Ahn:2011fg}, where $\lambda=0.22475^{+0.00106}_{-0.00018}$, $A=0.840^{+0.016}_{-0.043}$, $\bar{\rho}=\rho/(1-\lambda^2/2)=0.158^{+0.036}_{-0.020}$, and $\bar{\eta}=\eta/(1-\lambda^2/2)=0.349^{+0.029}_{-0.025}$ with $3\sigma$ errors\,\cite{ckm}. 
Their corresponding current best-fit values in the standard parameterization\,\cite{Chau:1984fp} read in the $3\sigma$ range\,\cite{ckm}
 \begin{eqnarray}
  \theta^q_{23}[^\circ]=2.430^{+0.054}_{-0.118}\,,\quad\theta^q_{13}[^\circ]=0.215^{+0.017}_{-0.012}\,,\quad\theta^q_{12}[^\circ]=12.988^{+0.068}_{-0.011}\,,\quad\delta^q_{CP}[^\circ]=65.8^{+2.9}_{-5.5}\,.
 \label{ckmmixing}
 \end{eqnarray}
 
\newpage

\end{document}